\documentclass[a4paper,11pt]{article}
\pdfoutput=1

\usepackage[utf8]{inputenc}
\usepackage{amsmath,amssymb,mathtools}
\usepackage{color}
\usepackage{cite}
\usepackage[colorlinks=true,linkcolor=blue, citecolor=red, urlcolor=blue, bookmarks]{hyperref}
\usepackage{graphics}
\definecolor{comments}{RGB}{220,20,60}
\usepackage[text={17.1cm,24.6cm},centering]{geometry} 
\usepackage{braket}
\usepackage{comment}

\usepackage[normalem]{ulem}

\usepackage{authblk}
\usepackage{cleveref}
\usepackage{multirow,makecell}
\usepackage{textcomp}
\usepackage{wasysym}
\numberwithin{equation}{section}
\usepackage[dvipsnames]{xcolor}

\newcommand{\beq}{\begin{equation}}
\newcommand{\eeq}{\end{equation}}
\newcommand{\bea}{\begin{eqnarray}}
\newcommand{\eea}{\end{eqnarray}}

\newcommand{\nc}{\newcommand}
\nc{\ir}{\mathrm{i}}
\nc{\dd}{\mathrm{d}} 
\nc{\eE}{\mathsf{e}}
\nc{\Tr}{\text{Tr}}
\nc{\id}{\mathbb{I}}
\nc{\tdet}{\tilde{\det}}
\nc{\J}{\mathcal{J}}

\begin{document} 

\title{\bf Exact quench dynamics of symmetry resolved entanglement in a free fermion chain}

\author[1]{Gilles Parez}
\affil[1]{\it Universit\'e catholique de Louvain, Institut de Recherche en Math\'ematique et Physique, Chemin du Cyclotron 2, 1348 Louvain-la-Neuve, Belgium
}

\author[2]{Riccarda Bonsignori}
\affil[2]{\it International School for Advanced Studies (SISSA) and INFN, Via Bonomea 265, 34136 Trieste, Italy}

\author[2,3]{Pasquale Calabrese}
\affil[3]{\it International Centre for Theoretical Physics (ICTP), Strada Costiera 11, 34151 Trieste, Italy}

\date{}

\maketitle

\begin{abstract}
The study of the entanglement dynamics plays a fundamental role in understanding the behaviour of many-body quantum systems out of equilibrium. 
In the presence of a globally conserved charge, further insights are provided by the knowledge of the resolution of entanglement in the various symmetry sectors. 
Here, we carry on the program we initiated in \href{https://doi.org/10.1103/PhysRevB.103.L041104}{[Phys. Rev. B 103, L041104 (2021)]}, for the study of the time evolution of the symmetry resolved entanglement in free fermion systems. 
We complete and extend our derivations also by defining and quantifying a symmetry resolved mutual information.
The entanglement entropies display a time delay that depends on the charge sector that we characterise exactly. 
Both entanglement entropies and mutual information show effective equipartition in the scaling limit of large time and subsystem size. 
Furthermore, we argue that the behaviour of the charged entropies can be quantitatively understood in the framework of the quasiparticle picture for the spreading of entanglement, 
and hence we expect that a proper adaptation of our results should apply to a large class of integrable systems.
We also find that the number entropy grows logarithmically with time before saturating to a value proportional to the logarithm of the subsystem size. 

\end{abstract}

\baselineskip 18pt
\thispagestyle{empty}
\newpage

\tableofcontents

\section{Introduction}

Understanding the dynamics of entanglement in out-of-equilibrium quantum many-body systems is a prominent challenge in many problems of contemporary physics, such as the equilibration and thermalisation of isolated many-body systems \cite{PolkonikovRMP11, GE15, DKPR15, SI,EF16}, the emergence of thermodynamics in quantum systems \cite{C:18, DLS:13, SPR:11,ckc-14} or the effectiveness of classical computers to simulate quantum dynamics \cite{SWVC:PRL, SWVC:NJP, PV:08, HCTDL:12, D:17}. 

Let us consider a system in a pure state $|\psi\rangle$ described by the density matrix $\rho=|\psi\rangle \langle \psi|$. 
The entanglement between a subsystem $A$ and its complement $B$, is encoded in the reduced density matrix $\rho_A$ of system $A$, defined as $\rho_A=\Tr_{B}\rho$. 
The \textit{R\'enyi entropies} are entanglement measures labelled by a parameter $n$ (which is integer in the replica approach \cite{cc-04}), and are defined as
\begin{equation}
    \label{RenyiEnt}
S_n \equiv\frac{1}{1-n}\log \Tr \rho_A^n.
\end{equation}
The knowledge of the R\'enyi entropies for all $n>0$ gives access to the full spectrum of $\rho_A$ \cite{cl-08}. The limit $n \to 1$ of Eq. \eqref{RenyiEnt} defines the \textit{entanglement entropy} $S_1$. It is the von Neumann entropy of the reduced density matrix of system $A$, i.e.
\begin{equation}
  \label{EntEnt}
S_{1} \equiv\lim_{n \to 1}S_n=-\Tr \rho_A \log \rho_A.
\end{equation}
  
The entanglement entropy, and more generally the R\'enyi entropies $S_n$, quantify the entanglement between a subsystem $A$ and its complement. 
This is independent of the topology of $A$. For example, when $A$ consists of two or more disconnected subsets, $S_n$ is still a measure of entanglement between $A$
and the remainder. However, by no means does $S_n$ quantify the entanglement between two subsets of $A$. 
It can anyway be used to construct a measure of the total correlations between the subsets. 
Let us denote two subsystems $A_1$ and $A_2$ with $A=A_1\cup A_2$. 
The  \textit{mutual information}\begin{equation}
  \label{MutualInfo}  
  I_1^{A_1:A_2} \equiv S_{1}^{A_1} +  S_{1}^{A_2} -S_{1}^{A_1\cup A_2},
\end{equation}
is a very useful measure of the total correlation between $A_1$ and $A_2$  \cite{wvhc-08}.
This definition can readily be generalised to a R\'enyi mutual information
\begin{equation}
    \label{RenyiMutInfo}
    I_{n}^{A_1:A_2}= S^{A_1}_{n} +  S^{A_2}_{n} -S^{A_1\cup A_2}_{n}=\frac{1}{n-1}\log  \left(\frac{\Tr\rho_{A_1\cup A_2}^n}{\Tr\rho_{A_1}^n\Tr\rho_{A_2}^n} \right).
\end{equation}
However, as a very important difference compared to the von Neumann mutual information \eqref{MutualInfo}, the R\'enyi one is not a measure of the 
correlations (e.g., it can be negative for some states \cite{kz-17}).   
The definition of a proper and calculable R\'enyi mutual information is a long standing problem, see, e.g., the discussion in the very recent manuscript \cite{sasc-21}. 
We mention that a proper measure to quantify the entanglement between non-complementary subsystems is instead the entanglement negativity \cite{vw-02}, 
that is much more difficult to deal with, especially out of equilibrium \cite{ctc-14,vez-14,hd-15,ac-19,gh-19,kkr-21,wkp-20}, and will be investigated 
in a forthcoming publication \cite{pbc-prep}.  

The last decade witnessed an intense theoretical effort aimed at better understanding the dynamics of quantum many-body systems after a \textit{quantum quench}, the simplest and most broadly studied protocol to drive a quantum system out of equilibrium \cite{cc-06,cc-07}. In integrable systems, the entanglement dynamics after a quench is well described and understood in terms of the \textit{quasiparticle~picture}~\cite{cc-05, ac-17,ac-18,c-20}. 
These intense theoretical progress moved together with groundbreaking cold-atom and ion-trap experiments. Most notably, it has been possible to measure the many-body entanglement of non-equilibrium states \cite{kaufman-2016,exp-lukin,brydges-2018,ekh-20}.
In particular, in the experimental paper \cite{exp-lukin}, the authors pointed out that a refined understanding of how the entanglement arises from the various symmetry sectors 
is necessary to better grasp the full many-body dynamics. 
Even though the study of \textit{the symmetry resolution of entanglement} became a very active research area of the last years \cite{GS,equi-sierra,bons,bons-20,Luca,pbc-21,lr-14,cgs-18,fg-19,mdc-20b,fg-20,mdc-20,ccdm-20,tr-20,mrc-20,trac-20,Topology,Anyons,hc-20,as-20,eimd-21,mbc-21,vecd-21,ncv-21,hcc-21,znm-20,fg-21,c-21,cc-21,hcc-21b}, only few results are available for out-of-equilibrium situations \cite{fg-19,pbc-21,vecd-21,fg-21}, see also \cite{cmr-13,cnn-16} as earlier works on the charged moments.

This article is a completion and an extension of our initial Letter \cite{pbc-21}.
On the one hand, we report many details of calculations in a system of free fermions after a global quantum quench
that were only sketched in Ref. \cite{pbc-21} due to the lack of space. 
On the other hand, we also present various new results that we obtained in the meantime, in particular concerning a proposal for the definition 
of a symmetry resolved mutual information and for the number entropy. 
We focus on exact lattice calculations for two distinct quenches and compare our results with the quasiparticle picture predictions.

This paper is organised as follows. In Sec. \ref{Sect2} we provide the definitions of the symmetry resolved entanglement entropies and mutual information. We then introduce the model of free fermions that we consider in Sec. \ref{Sect3}. Sections \ref{Sect4} and \ref{Sect5} are devoted to the analytical study of the dynamics of the symmetry resolved quantities after a quench from two distinct initial states: the N\'eel state and the Majumdar-Ghosh dimer state, respectively. In Sec. \ref{Sect6} we give an interpretation of our results in the framework of the quasiparticle picture for the dynamics of entanglement in integrable models. Finally, we give a summary of the results and future outlook of this work in Sec. \ref{Sect7}.

\section{Symmetry resolution and flux insertion}
\label{Sect2}

In this section we give the general definitions of the main quantities of interest, namely the symmetry resolved entanglement entropies and mutual information. 
We consider an extended quantum system with an internal  $U(1)$ symmetry generated by a local operator $Q$. 
If the symmetry is not  broken, we have $[\rho,Q]=0$.
For a bipartition in two complementary subsystems $A$ and $B$, by locality the charge $Q$ splits as sum of operators that act on the 
degrees of freedom of the two parts, $Q=Q_A+Q_B$. Hence, the trace over the degrees of freedom of subsystem $B$ in the commutation relation $[\rho,Q]=0$ yields 
$[\rho_A,Q_A]=0$. 
This relation implies that the reduced density matrix $\rho_A$ acquires a block diagonal form, where each block corresponds to a given eigenvalue $q$ of $Q_A$. 
We thus have
\begin{equation}
    \label{reducedblock}
    \rho_A=\oplus_q \Pi_q\rho_A= \oplus_q[p(q)\rho_A(q)]
\end{equation}
 where $\Pi_q$ is the projector on the eigenspace of the eigenvalue $q$ and $p(q)=\Tr \left(\Pi_q \rho_A \right)$ is the probability that a measurement of $Q_A$ 
 gives $q$ as outcome.
The reduced density matrix in the sector $q$ is normalised so that $\Tr{\rho_A}(q)=1$.

\subsection{Symmetry resolved entropies}

The block structure of the reduced density matrix induces a decomposition of the entropy into contributions associated to each charge sector. 
For example, plugging Eq. \eqref{reducedblock} into the definition of the von Neumann entropy \eqref{EntEnt}, we obtain 
\begin{equation}
\label{decompositionSvN}
S_{1}=\sum_q p(q)S_{1}(q)-\sum_q p(q)\log (p(q))\equiv S^c+S^n
\end{equation}
where $S_{1}(q)$ is the \textit{symmetry resolved entanglement entropy} associated to the charge sector $q$, i.e.
\begin{equation}
    \label{SymResEnt}
    S_{1}(q)\equiv -\Tr [\rho_A(q)\log \rho_A(q)].
\end{equation}
The two terms in Eq. \eqref{decompositionSvN} are the  \textit{configurational entropy} ($S^c$)\cite{exp-lukin,wv-03,bhd-18}, which quantifies the average of the entanglement in each charge sector, and \textit{number entropy} ($S^n$) \cite{exp-lukin,kusf,kusf2,kufs3,kufs,ctd-19,zfh-21}, which measures the entropy due to the fluctuations of the value of the charge within subsystem $A$.

Similarly, we define the \textit{symmetry resolved R\'enyi entropies} as
\begin{equation}
    \label{SymResReny}
    S_n(q)\equiv\frac{1}{1-n}\log \Tr [\rho_A(q)]^n.
\end{equation} 
A decomposition of the total R\'enyi entropies into symmetry sectors can also be written down \cite{Luca}, but it is less informative and clear than 
Eq. \eqref{decompositionSvN} for the von Neumann entropy. 

In principle, the evaluation of the symmetry resolved contributions requires the knowledge of the resolution of the spectrum of $\rho_A$ in $Q_A$. This is a highly non-trivial problem, mainly because of the non-local nature of the projector $\Pi_q$. A different strategy, put forward in Ref. \cite{GS, equi-sierra}, is based on the evaluation of the \textit{charged moments} 
\begin{equation}
\label{ChargedMom}
Z_n({\alpha})\equiv\mbox{Tr}[\rho_A^n \eE^{\ir\alpha Q_A}]
\end{equation}
and their Fourier transform
\begin{equation}
\label{eq:ZnqFT}
\mathcal{Z}_n(q)= \int_{-\pi}^{\pi}\frac{d \alpha}{2 \pi} \eE^{-\ir q \alpha} {Z}_n(\alpha)\equiv \mbox{Tr}[\Pi_q\rho_A^n].
\end{equation}
From these quantities, the symmetry resolved entropies are simply given by
\begin{equation}
S_n(q)=\frac{1}{1-n}\log \left[\frac{\mathcal{Z}_n(q)}{\mathcal{Z}_1(q)^n} \right], \quad \quad S_{1}(q)=-\partial_n\left[\frac{\mathcal{Z}_n(q)}{\mathcal{Z}_1(q)^n} \right]_{n=1},
\label{SvsZ}
\end{equation}
and the probability $p(q)$ is 
\begin{equation}
\label{eq:pq}
    p(q)=\mathcal{Z}_1(q).
\end{equation}

\subsection{Symmetry resolved mutual information}

The mutual information defined in Eq. \eqref{MutualInfo} is constructed from three different entanglement entropies and so from three different reduced density matrices 
$\rho_{A_1}$, $\rho_{A_2}$, and $\rho_{A_1 \cup A_2}$. Each of them admits its independent symmetry decomposition, and it is not obvious how to combine them 
to construct a symmetry resolved quantity, and whether it is possible at all.
We propose to define the \textit{symmetry resolved mutual information} as follows:
 \begin{equation}
 \label{eq:SymResMut}
     I^{A_1 : A_2}_1(q) =\sum_{q_1 = 0}^{q} p(q_1,q-q_1)\Big(S^{A_1}_1(q_1)+ S^{A_2}_1(q-q_1)\Big) - S^{A_1 \cup A_2}_1(q),
 \end{equation}
where the superscripts $A_1$, $A_2$ or $A_1 \cup A_2$ indicate to which system the various quantities pertain to. 
The idea of Eq. \eqref{eq:SymResMut} is that for a given charge sector $q$ of the whole subsystem $A$, we take a weighted sum over the contributions from charge sectors $q_1$ and $q_2$ of $A_1$ and $A_2$ such that $q_1+q_2=q$. The weight $p(q_1,q-q_1)$ is the probability that a simultaneous measurement of the charges $Q_{A_1}$ and  $Q_{A_2}$ yields $q_1$ and $q-q_1$, respectively, while the charge sector of the whole system $A$ is fixed to $q$. It is given by
\begin{subequations}
\label{eq:pqq1}
\begin{equation}
 p(q_1,q-q_1) =  \frac{\mathcal{Z}_1^{A_1:A_2}(q_1,q-q_1)}{\mathcal{Z}_1^{A_1 \cup A_2}(q)}
\end{equation}
where $\mathcal{Z}_1^{A_1:A_2}(q_1,q-q_1)$ is the probability of obtaining $q_1$ and $q-q_1$ by simultaneous measure of the charge 
in $A_1$ and $A_2$, respectively, and satisfies
\begin{equation}
\sum_{q_1=0}^q p(q_1,q-q_1)=1.
\end{equation}
\end{subequations}
The above definition of symmetry resolved mutual information is sort of natural, but does not represent a measure of the total correlations within the symmetry sector 
(e.g., we shall see that it can be negative) and does not account for many processes (e.g., the total charge $q$ of $A$ should not be split as a direct sum of eigenvalues $q_1,q_2$).
It is very likely that a more appropriate symmetry resolved mutual information can be defined and we strongly hope that this manuscript will boost the research in this direction.
In spite of all these caveats, we will see that the symmetry resolved mutual information defined in \eqref{eq:SymResMut} satisfies an equipartition for small 
$q-\langle Q_A\rangle$ after a quench.

In full analogy with $\mathcal{Z}_1^A(q)$, $\mathcal{Z}_1^{A_1:A_2}(q_1,q-q_1)$ can be reconstructed from the knowledge of 
a more complicated charged moment given by
\begin{equation}
\label{eq:Z1ab}
Z_1^{A_1:A_2} (\alpha,\beta) = \Tr  [\rho_A \eE^{\ir\alpha Q_{A_1}+\ir \beta Q_{A_2}}]
\end{equation}
and performing the double Fourier transform
\begin{equation}
\label{eq:Z1q1q2}
\mathcal{Z}_1^{A_1:A_2}(q_1,q_2) = \int_{-\pi}^{\pi}\frac{d \alpha}{2 \pi}\frac{d \beta}{2 \pi} \eE^{-\ir q_1 \alpha} \eE^{-\ir q_2 \beta} Z_1^{A_1:A_2} (\alpha,\beta).
\end{equation}
The charged moment $Z_1^{A_1:A_2} (\alpha,\beta)$ is a new quantity that, to the best of our knowledge, has not yet been studied in the literature, even in equilibrium.

Similarly to the entanglement entropy, it is possible to reconstruct the total mutual information from the symmetry resolved ones. We have
\begin{equation}
\label{eq:IqTot}
 I^{A_1:A_2}_1 = \sum_{q} p(q) I^{A_1 : A_2}_1(q) + S^{A_1,n}+S^{A_2,n}-S^{A_1\cup A_2,n}=
 \sum_{q} p(q) I^{A_1 : A_2}_1(q) + I^{A_1:A_2,n},
\end{equation}
where the notation $S^{s,n}$ stands for the number entropy associated with system $s$, and $p(q)=\mathcal{Z}_1^{A_1\cup A_2}(q)$.
The last equality defines the number mutual information $I^{A_1:A_2,n}\equiv S^{A_1,n}+S^{A_2,n}-S^{A_1\cup A_2,n}$.

\section{Free fermions}\label{sec:FF}
\label{Sect3}

In this paper, we study the time evolution of the symmetry resolved entanglement after a global quench in the tight-binding model. It is a model of free fermions constrained on a one-dimensional lattice with $L$ sites and endowed with periodic boundary conditions. The Hamiltonian is
\begin{equation}
\label{eq:Hfree}
H=\sum_{j=1}^{L}(c_j^{\dagger}c_{j+1}+c_{j+1}^{\dagger}c_{j}),
\end{equation}
where $c_j$ and $c_j^\dagger$ are the canonical spinless fermionic annihilation and creation operators on site $j$, that satisfy the anticommutation relations $\{c_i,c_j^\dagger \} = \delta_{i,j}$ and $\{c_i,c_j \} = \{c_i^\dagger,c_j^{\dagger}\}=0$. 
The model is equivalent to the spin-$1/2$ XX spin chain via the Jordan-Wigner transformation. The conserved charge is the fermion number 
\begin{equation}
    \label{eq:Qff}
    Q=\sum \limits_{j=1}^L c_j^\dagger c_j,
\end{equation}
i.e. the $z$-component of the spin in the spin chain.

We consider a subsystem $A$ of length $\ell$ and its complement $B$, of length $L-\ell$. 
Because of locality, the global conserved charge $Q$ trivially splits as a sum over $A$ and $B$, 
\begin{equation}
Q = \sum_{j \in A}c_j^\dagger c_j +\sum_{j \in B}c_j^\dagger c_j  = Q_A+Q_B.
\end{equation}
We will study both the cases in which $A$ is a block of $\ell$ consecutive sites or the union of two disjoint blocks $A=A_1 \cup A_2$ where 
the subsystems have respective lengths $\ell_1$ and $\ell_2=\ell-\ell_1$ and are separated by $d$ lattice sites. 
The reduced density matrix of $A$ is always written in terms of the correlations matrix $C_A=\langle c_x^\dagger c_{x'}\rangle$ with $x,x' \in A$ \cite{cp-01,p-03,pe-09}, 
using the property that the many-body time-dependent state is a Slater determinant and so $\rho_A$ is a Gaussian operator at any time. 
Using the standard algebra of Gaussian operators, the charged moments $Z_n(\alpha)$ defined in Eq. \eqref{ChargedMom} can be expressed as \cite{GS}
\begin{equation}
\label{eq:Znalpha}
\log Z_n(\alpha)= \Tr \log \left[ (C_A)^n \eE^{\ir \alpha}+ ( 1- C_A)^n \right].
\end{equation}
For later convenience, we introduce the $\ell \times \ell$ matrix $J_A$ defined as $C_A = \frac 12 (\id_\ell + J_A)$, with eigenvalues $\nu_i$.
In terms of the latter, the charged moments are recast as
\begin{equation}
\label{eq:chargedZ}
\log Z_n(\alpha)=\sum_{i=1}^{\ell} \log \left[ \left(\frac{1+\nu_i}{2} \right)^n \eE^{\ir \alpha}+ \left( \frac{1-\nu_i}{2}\right)^n \right].
\end{equation}
To perform analytical calculations, it is useful to express the charged moments as a Taylor series in $\Tr J_A^m$. We thus introduce the function $h_{n,\alpha}(x)$ and the coefficients $c_{n,\alpha}(m)$ as
\begin{equation}
\label{eq:hna}
h_{n,\alpha}(x) = \log \left[ \left(\frac{1+x}{2} \right)^n \eE^{\ir \alpha}+ \left( \frac{1 - x}{2}\right)^n \right] \equiv \sum_{m=0}^{\infty}c_{n,\alpha}(m) x^m
\end{equation}
and conclude
\begin{equation}
\label{eq:chargedZTrK}
\log Z_n(\alpha)=  \sum_{m=0}^{\infty}c_{n,\alpha}(m) \Tr J_A^m.
\end{equation}
As we shall see in the following sections, the precise form of the coefficients $c_{n,\alpha}(m)$ is never needed and hence will not be reported.

In the case of disjoint intervals, the matrix $J_{A_1\cup A_2}$  has a structure  
\begin{equation}
\label{eq:CorrMat}
J_{A_1\cup A_2} = \begin{pmatrix}
J_{11} & J_{12} \\
J_{21} & J_{22}\end{pmatrix} ,
\end{equation}
in which the notation $J_{a b}$, $a,b=1,2$, refers to correlations between sites in $A_{a}$ and $A_b$. 
We recall that  for disjoint blocks, fermionic and spin entanglement are not equal \cite{atc-09,ip-09,fc-10} and here we focus only on the fermionic one. 
The computation of $Z_n^A(\alpha)$ is not affected by the fact that $A$ is not a connected block.
Instead, to compute the symmetry resolved mutual information, we need to evaluate $Z_1^{A_1:A_2} (\alpha,\beta)$ defined in Eq. \eqref{eq:Z1ab}. 
Since the reduced density matrix $\rho_A$ does not commute with the charges $Q_{A_1}$ and $Q_{A_2}$ of the two disjoint subsystems, the method used to compute 
$Z_n(\alpha)$ and find Eq. \eqref{eq:chargedZ} does not apply. 
To circumvent this problem, we adapt a technique devised in \cite{gec-18} to compute the full counting statistics of the magnetisation in the transverse field Ising chain. 
The idea is to express $Z_1^{A_1:A_2} (\alpha,\beta)$ in terms of a trace of two non-commuting putative density matrices. We recast Eq. \eqref{eq:Z1ab} as 
\begin{equation}
\label{eq:Z1abTwoRhos}
Z_1^{A_1:A_2} (\alpha,\beta) = \tilde Z_A \Tr_A(\rho_A \tilde \rho_A)
\end{equation}
were
\begin{equation}
\tilde \rho_A=\frac{1}{\tilde Z_A}\eE^{\ir\alpha Q_{A_1}+\ir \beta Q_{A_2}}, \quad \tilde Z_A=\Tr_A \Big(\eE^{\ir\alpha Q_{A_1}+\ir \beta Q_{A_2}}\Big)
\end{equation}
and the normalisation $\tilde Z_A$ ensures that $\Tr_A \tilde \rho_A=1$. It is simply given by 
\begin{equation}
\label{eq:tildeZA}
\tilde Z_A = (1+\eE^{\ir \alpha})^{\ell_1}(1+\eE^{\ir \beta})^{\ell_2}.
\end{equation}
Obviously, $\tilde \rho_A$ is not a true density matrix (it is not even hermitian), but it is a very simple Gaussian operator and can be 
associated to a two-point correlation matrix. We can thus use the rules for composition of Gaussian operators. In particular, the trace of a product of two $\ell \times \ell$ density matrices $\rho_1$ and $\rho_2$ is \cite{fc-10,bb-69} 
\begin{equation}
\label{eq:Trr1r2}
\Tr(\rho_1 \rho_2) = \det \Big(\frac{\mathbb{I}_\ell+J_1J_2}{2} \Big)
\end{equation}
where 
\begin{equation}
J_j=2C_j-\mathbb{I}_\ell, \quad [C_j]_{x,x'}=\Tr(\rho_j c^\dagger_x c_{x'}).
\end{equation}
To ease the comparison with Ref. \cite{gec-18} we mention that for free fermionic systems without particle number conservation (i.e. the Ising and XY spin chain), one 
also has to consider correlations of the form $\Tr(\rho_j c^\dagger_x c^\dagger_{x'})$ and $\Tr(\rho_j c_x c_{x'})$. 
For that reason, the size of the correlation matrices are $2 \ell$ instead of $\ell$, and the right-hand side of Eq. \eqref{eq:Trr1r2} is then written with 
an overall square root in \cite{gec-18}.

For the case of interest here, the two correlation matrices are
\begin{equation}
\begin{split}
[C_A]_{x,x'} = \Tr_A(\rho_A c^\dagger_x c_{x'}), \quad x,x' \in A, \\
[\tilde C_A]_{x,x'} = \Tr_A(\tilde \rho_A c^\dagger_x c_{x'}), \quad x,x' \in A.
\end{split}
\end{equation}
Of course, $C_A$ is the usual correlation matrix related to $J_{A_1 \cup A_2}$ given in Eq. \eqref{eq:CorrMat}, by definition. 
The second one, $\tilde C_A$, is very easy to compute because $Q_{A_1}$ and $Q_{A_2}$ are diagonal operators in terms of the fermion operators $c_j$. 
We then find 
\begin{equation}
\label{eq:tildeCA}
\begin{split}
 \Tr_A(\tilde \rho_A c^\dagger_x c_{x'}) &= \delta_{x,x'} \left\{\begin{array}{cl}
\frac{1}{\tilde Z_A}\eE^{\ir \alpha}  (1+\eE^{\ir \alpha})^{\ell_1-1}(1+\eE^{\ir \beta})^{\ell_2}& x \in A_1, \\[0.15cm]
\frac{1}{\tilde Z_A}\eE^{\ir \beta}  (1+\eE^{\ir \alpha})^{\ell_1}(1+\eE^{\ir \beta})^{\ell_2-1}& x \in A_2,
\end{array}\right. 
\\[.3cm]&
= \delta_{x,x'} \left\{\begin{array}{ll}
\frac{\eE^{\ir \alpha}}{1+\eE^{\ir \alpha}}& x \in A_1, \\[0.15cm]
\frac{\eE^{\ir \beta}}{1+\eE^{\ir \beta}}& x \in A_2.
\end{array}\right. 
\end{split}
\end{equation}
Finally, combining Eqs. \eqref{eq:Z1abTwoRhos}, \eqref{eq:tildeZA}, \eqref{eq:Trr1r2} and \eqref{eq:tildeCA}, we have
\begin{equation}
Z_1^{A_1:A_2} (\alpha,\beta)=(1+\eE^{\ir \alpha})^{\ell_1}(1+\eE^{\ir \beta})^{\ell_2}\det \Big( \frac{\mathbb{I}_\ell+ J_{\alpha \beta}}{2}\Big)
\label{Z1abff}
\end{equation}
where
\begin{equation}
\label{eq:TildeJA}
 J_{\alpha \beta} = \begin{pmatrix}
J_{11} & J_{12} \\
J_{21} & J_{22}\end{pmatrix}  \cdot \begin{pmatrix}
\frac{\eE^{\ir \alpha}-1}{\eE^{\ir \alpha}+1}\mathbb{I}_{\ell_1} & 0\\
0&  \frac{\eE^{\ir \beta}-1}{\eE^{\ir \beta}+1}\mathbb{I}_{\ell_2}\end{pmatrix}  .
\end{equation}

Similarly to the charged entropies, we write $Z_1^{A_1:A_2} (\alpha,\beta)$ as an expansion in terms of the traces of the powers of $J_{\alpha \beta}$, namely
\begin{equation}
\label{eq:Z1abSum}
\log Z_1^{A_1:A_2} (\alpha,\beta) = \ell_1 \log(1+\eE^{\ir \alpha})+\ell_2 \log(1+\eE^{\ir \beta}) + \sum_{m=0}^{\infty}\tilde c(m) \Tr J_{\alpha \beta} ^m
\end{equation}
where $\tilde c(m)$ are the Taylor coefficient of the function $\tilde h (x) = \log [ (1+x)/2]$.

In this work, we focus on the quench dynamics starting from two low-entangled initial states $|\psi_0\rangle$ that are invariant under $U(1)$ symmetry 
and with a resulting time-dependent correlation matrix $C_A(t)=\langle \psi_0| \eE^{\ir t H}c_x^\dagger c_{x'}\eE^{-\ir t H}|\psi_0\rangle$ which is exactly known 
when $H$ is the tight-binding Hamiltonian \eqref{eq:Hfree}. 
Those are the N\'eel and the dimer states, denoted by $|N \rangle$ and $|D\rangle$, respectively, defined as
\begin{equation}
    \label{eq:InitialStates}
  \begin{split}
        |N\rangle&=\prod_{j=1}^{L/2}c_{2j}^{\dagger}|0\rangle, \\
        |D\rangle &=\prod_{j=1}^{L/2}\frac{c_{2j}^{\dagger}-c_{2j-1}^{\dagger}}{\sqrt{2}}|0\rangle.
    \end{split}
\end{equation}

\section{Quench from the N\'eel state }
\label{Sect4}

The correlation function after a quench from the N\'eel state $|N\rangle$ in the tight-binding model \eqref{eq:Hfree} is known in the thermodynamic limit and reads 
(see, e.g., \cite{ac-19})
\begin{equation}
\label{eq:CNeel}
[C(t)]_{x,x'}=\frac{\delta_{x,x'}}2 +\frac{(-1)^{x'}}2\int_{-\pi}^{\pi}\frac{\dd k}{2\pi}\eE^{\ir k(x-x')+4\ir t\cos k}.
\end{equation}
As explained in Sec. \ref{sec:FF}, the expression of the time-dependent correlation matrix is the starting point to compute the charged moments, and hence symmetry resolved entanglement measures. 
In the following subsection, we separately consider the situations where the subsystem $A$ is a single interval of adjacent sites and where $A=A_1\cup A_2$ 
is made of two disjoint intervals.

\subsection{Single interval}
In this subsection we study the case where $A$ is a connected interval of $\ell$ neighbouring sites in an infinite chain. The correlation matrix is then simply $C_A = \frac 12 (\id_\ell + J_A)$ with
\begin{equation}
\label{eq:JAExactNeel}
[J_A(t)]_{x,x'} =(-1)^{x'}\int_{-\pi}^{\pi}\frac{\dd k}{2\pi}\eE^{\ir k(x-x')+4\ir t\cos k}, \quad x,x' \in \{1,2,\dots,\ell\}.
\end{equation}
Because of the structure of the initial state, we only work with $\ell$ even. 

\subsubsection{Charged moments}\label{sec:ZnaNeel}

\begin{figure}[t]
    \centering
    \includegraphics[scale=0.25]{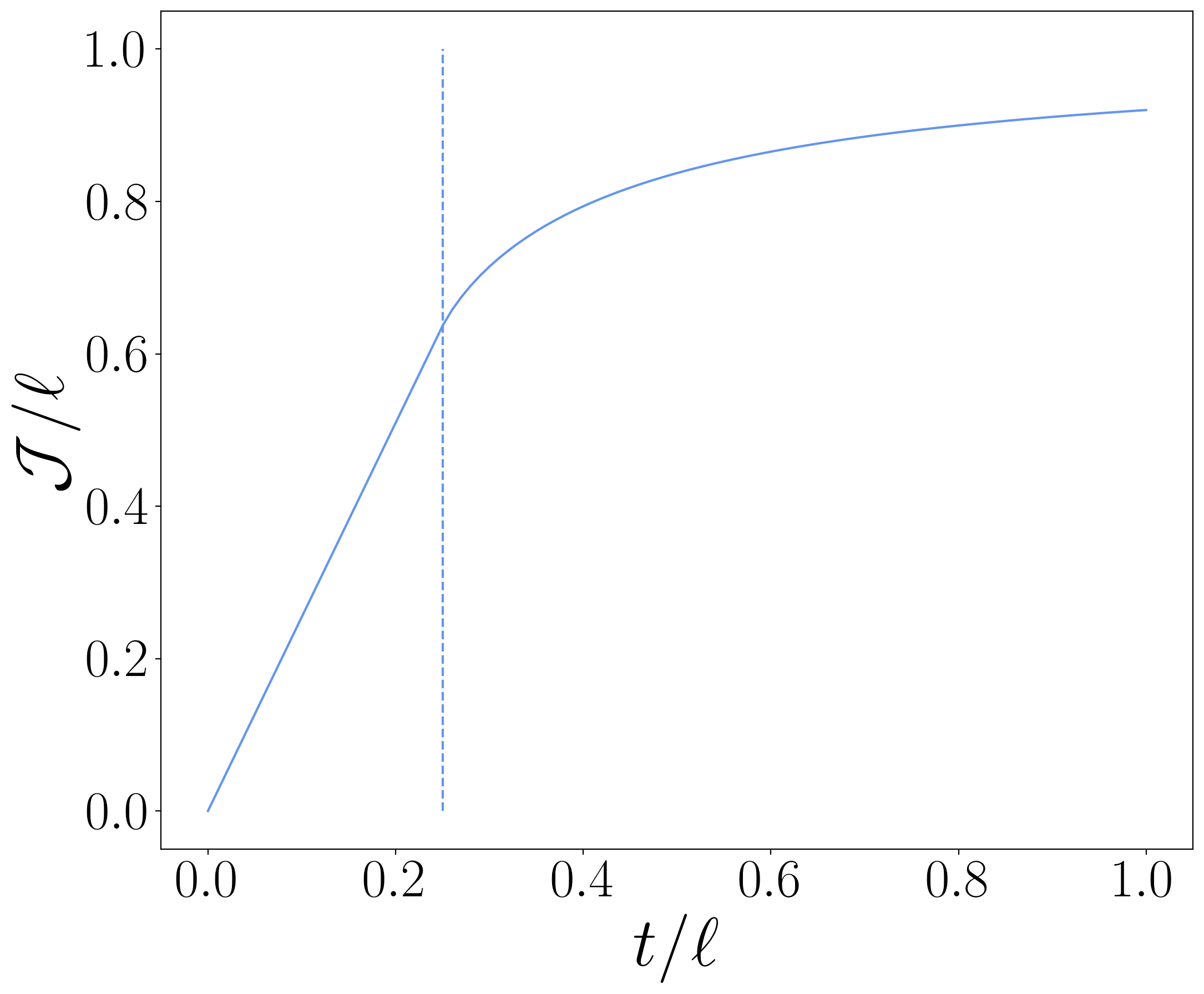}
    \includegraphics[scale=0.25]{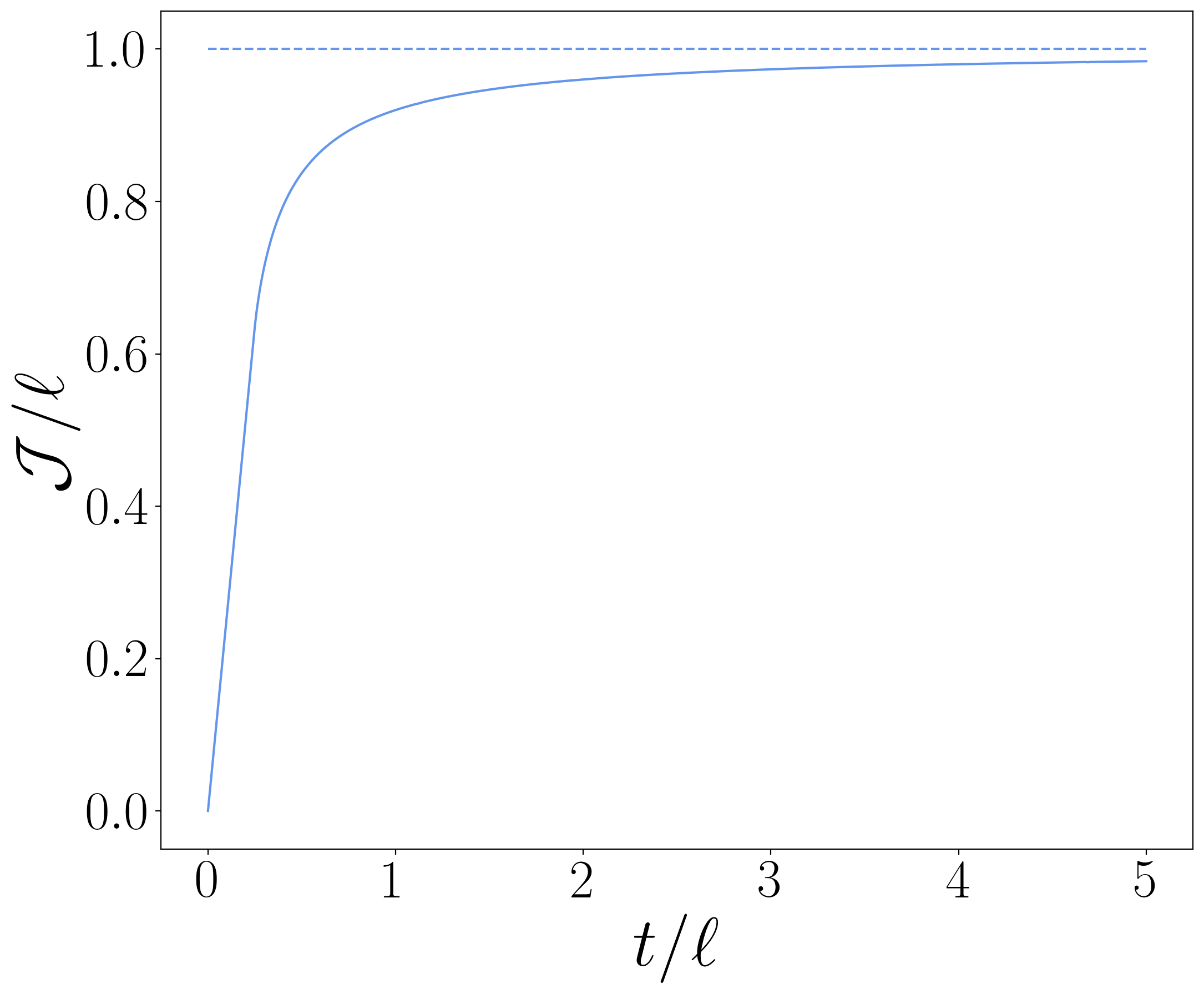}
    \caption{Time evolution of $\J/\ell$ in Eq. \eqref{eq:JNeel} as a function of $t/\ell$. 
    \textit{Left:} The crossover between the linear growth and the saturation; the vertical dashed line indicates the time at which the transition occurs, 
    i.e. $t/\ell=1/4$ (in agreement with  $t/\ell=1/(2v_M)$ and $v_M=2$). 
    \textit{Right:} For large time, $\J/\ell$ saturates to $1$ (horizontal dashed line).}
    \label{Fig:JNeel}
\end{figure}

The first step to compute the charged moments $Z_n(\alpha)$ is the evaluation of the trace of powers of $J_A$. 
With the exact expression  \eqref{eq:JAExactNeel}, it is possible to perform this calculation analytically using the multidimentional stationary phase approximation 
in the scaling limit where $t,\ell~\to~\infty$ with a fixed finite ratio $t/\ell$. 
Actually, the computation of $\Tr J_{A}(t)^{j}$ is a simple adaptation of the calculations reported in Ref. \cite{fc-08}, and the final result can be written as 
\begin{equation}
\label{eq:TrJNeeld0}
\Tr J_{A}(t)^{2j} = \ell - \int_{-\pi}^\pi \frac{\dd k}{2 \pi}\min(\ell,2 v_k t), \quad \Tr J_{A}(t)^{2j+1} =0,
\end{equation}
with $v_k = 2 |\sin k|$. We note that, with our normalisation, the maximal velocity $v_M=\textrm{max}_k v_k$ is $v_M=2$. To compute the charged moments $Z_n(\alpha)$, we evaluate the sum \eqref{eq:chargedZTrK} with \eqref{eq:TrJNeeld0}. We use the fact that $\Tr J_{A}(t)^{m}$ only depends on the parity of the exponent for $m>0$, and $\Tr J_{A}(t)^{0}=\ell$. We then find
\begin{equation}
\label{eq:SumNeel1}
\log Z_n(\alpha) =\ell  c_{n,\alpha}(0) + \Tr J_A(t)^2 \sum_{m=1}^{\infty} c_{n,\alpha}(2m),
\end{equation}
where the coefficients $c_{n,\alpha}(k)$ are defined in \eqref{eq:hna}. To proceed, we notice that the coefficient $ c_{n,\alpha}(0)$, as well as the sum $\sum_{m=1}^{\infty} c_{n,\alpha}(2m)$ in fact correspond to the function $h_{n,\alpha}(x)$ from Eq. \eqref{eq:hna} evaluated in certain simple points. Indeed, we have 
\begin{equation}
\begin{split}
c_{n,\alpha}(0) &= h_{n,\alpha}(0)=(1-n) \log 2 + \log \cos \frac{\alpha}{2}+ \ir\frac{\alpha}{2}, \\
\sum_{m=1}^{\infty} c_{n,\alpha}(2m) &= \left(\frac{h_{n,\alpha}(1)+h_{n,\alpha}(-1)  }{2}-c_{n,\alpha}(0)\right) = (n-1)\log2 - \log\cos \frac{\alpha}{2}.
\end{split}
\end{equation}
It immediately follows that
\begin{equation}
\label{eq:logZnaExactNeel}
\log Z_n(\alpha) = \ir \ell \frac{\alpha}{2}+\int_{-\pi}^\pi  \frac{\dd k}{2 \pi}\mathrm{Re}[h_{n,\alpha}(0)]\min(\ell,2 v_k t), \quad
\end{equation}
or
\begin{figure}[t]
\begin{center}
\includegraphics[scale=0.24]{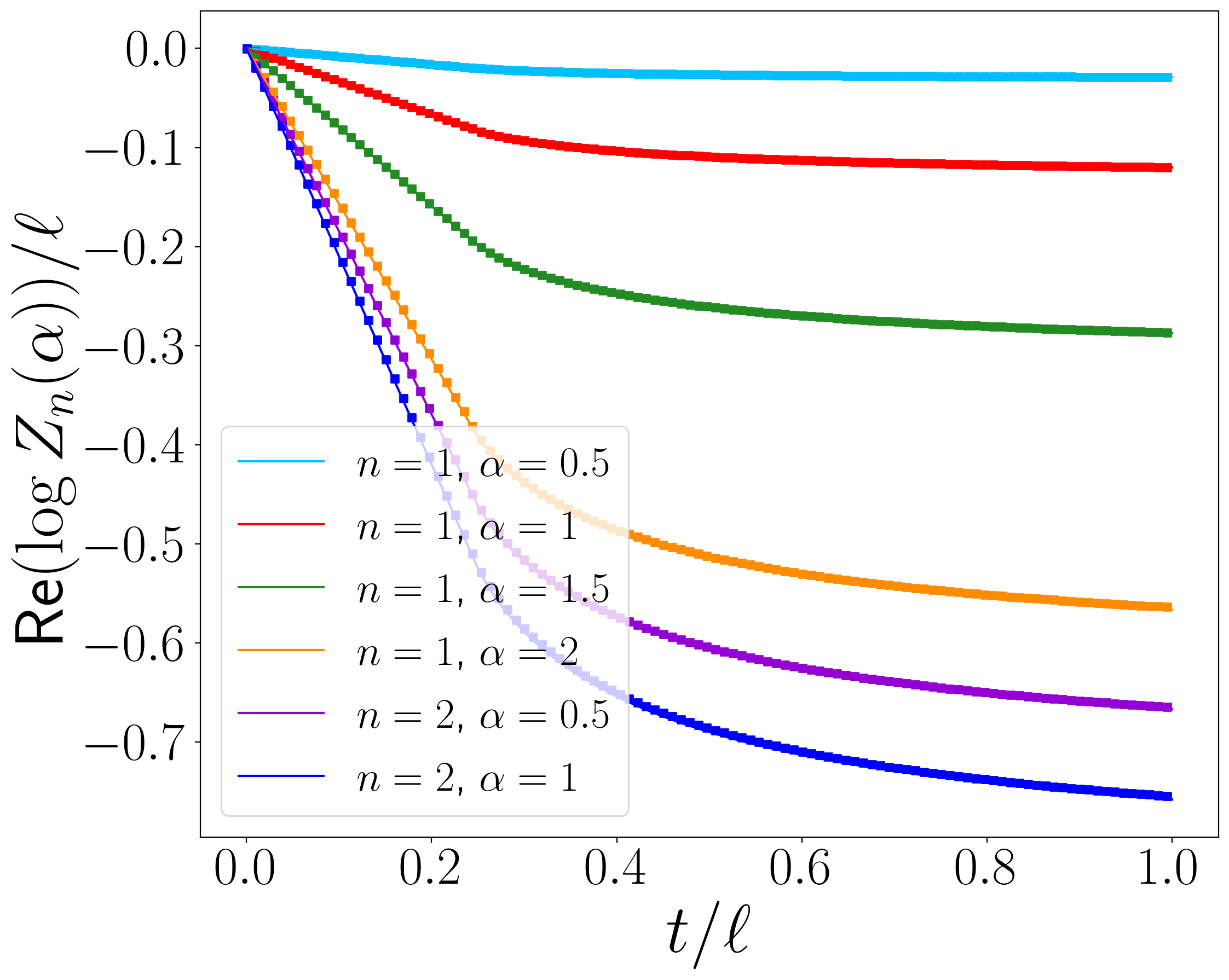}\quad
\includegraphics[scale=0.24]{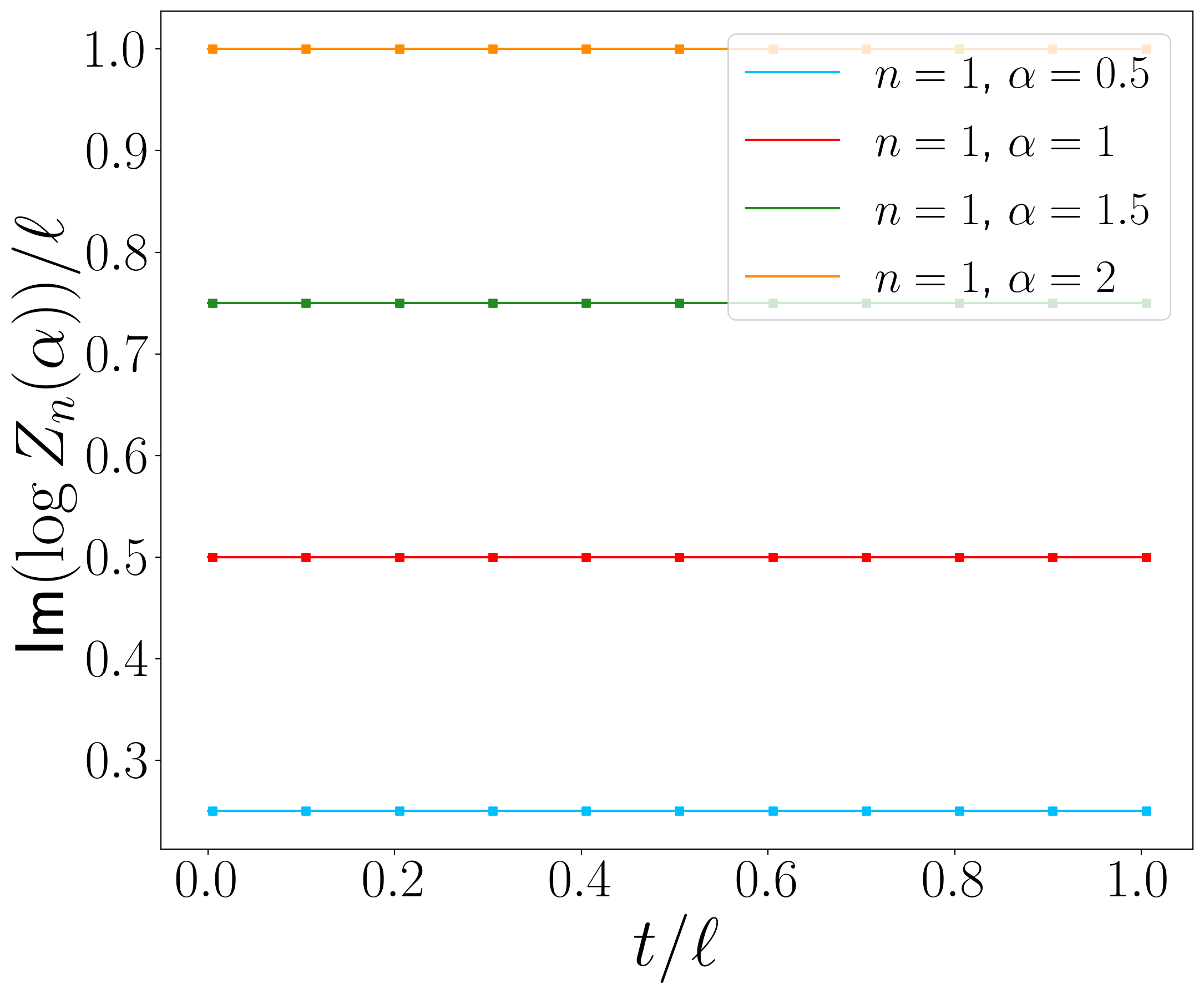} \\
\includegraphics[scale=0.24]{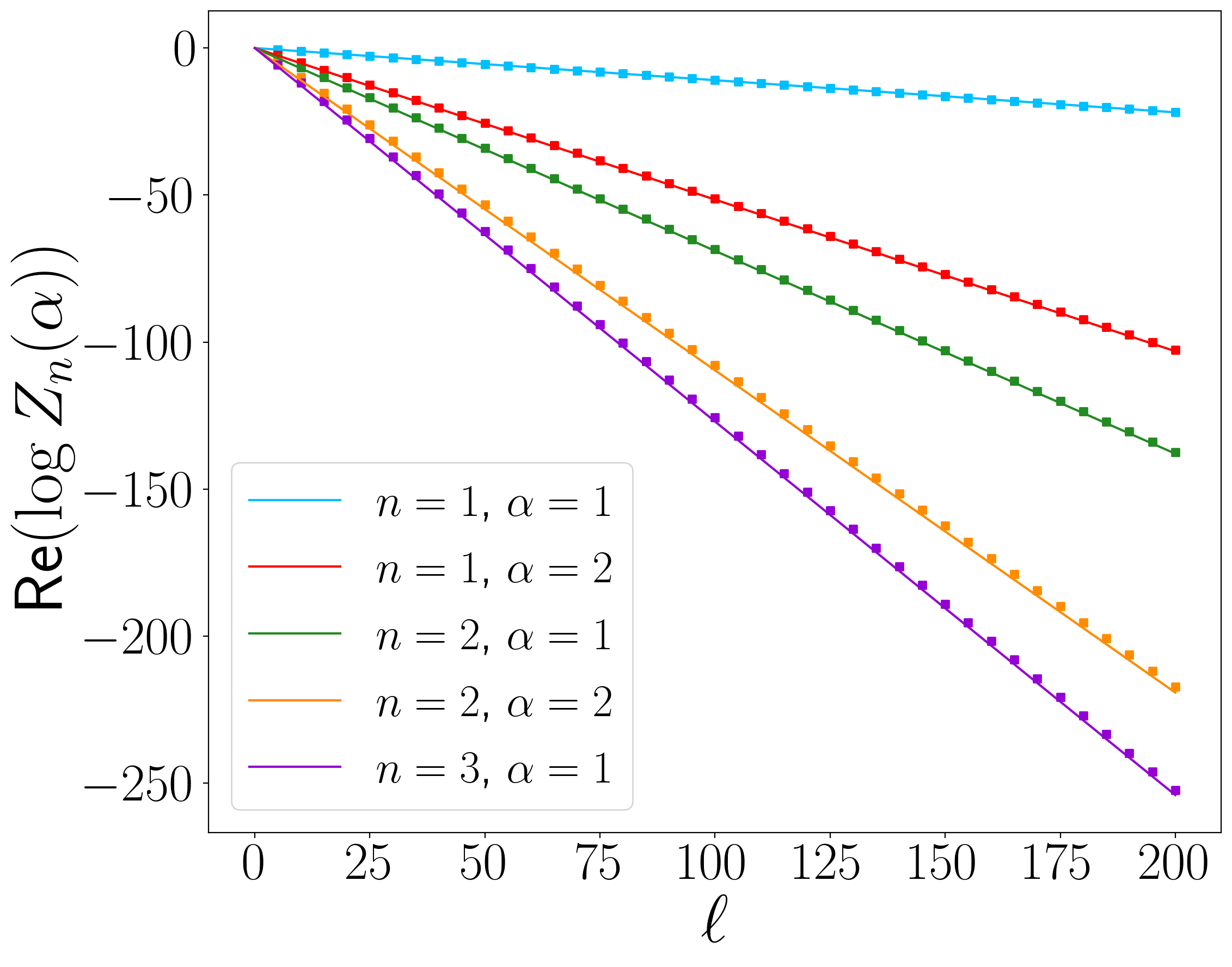} \quad 
\includegraphics[scale=0.24]{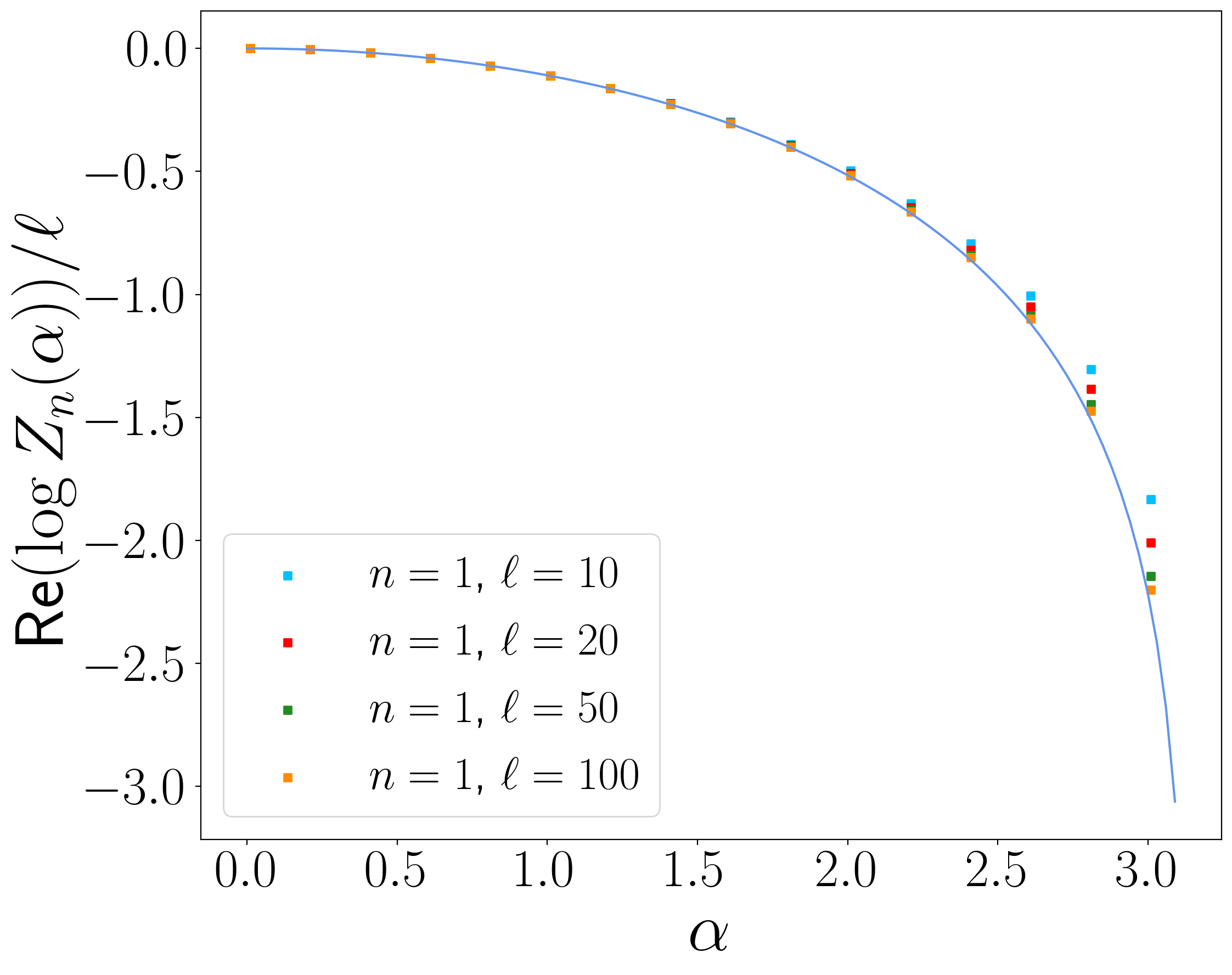}
\end{center}
\caption{Comparison between the asymptotic result \eqref{eq:ZnaExactNeel} (solid lines) and exact numerical data (symbols) for $\log Z_n(\alpha)$ after a quench from the N\'eel state in the free fermion model \eqref{eq:Hfree}. \textit{Top left:} Time evolution of the real part of $\log Z_n(\alpha)$ as a function of $t/\ell$ with $\ell=120$. \textit{Top right}: Time evolution of the imaginary part of $\log Z_n(\alpha)$ as a function of $t/\ell$ with $\ell=20$ and $n=1$. \textit{Bottom left:} Real part of $\log Z_n(\alpha)$ as a function of $\ell$ with $t/\ell=0.5$. \textit{Bottom right:} Real part of $\log Z_n(\alpha)$ as a function of $\alpha$ with $t/\ell=0.5$ and $n=1$.
}
\label{fig:Znalpha}
\end{figure}
\begin{equation}
\label{eq:ZnaExactNeel}
Z_n(\alpha) = \left(\frac{\cos \frac{\alpha}{2}}{2^{n-1}} \right)^{\J} \eE^{\ir \ell \frac{\alpha}{2}}
\end{equation}
where we introduced the quantity 
\begin{equation}
\label{eq:JNeel}
\J \equiv \ell -\text{Tr}J_A(t)^2 = \int_{-\pi}^\pi  \frac{\dd k}{2 \pi}\min(\ell,2 v_k t).
\end{equation}

The function $\J(t)$ is an increasing function of time, going from $\J(t=0)=0$ to $\J(t=\infty)=\ell$. Its time evolution is separated in two distinct regimes. 
For short times $t<\ell/(2v_M)$, $\J(t) \propto t$ increases linearly with time. On the other hand, for $t>\ell/(2v_M)$, $\J(t)$ slowly saturates to its asymptotic value $\ell$. 
These behaviours are illustrated in Fig. \ref{Fig:JNeel}. The ratio $\J(t)/\ell$ is a scaling function of $t/\ell$, a fact that we will use repeatedly in what follows.

\begin{figure}[t]
\begin{center}
\includegraphics[scale=0.3]{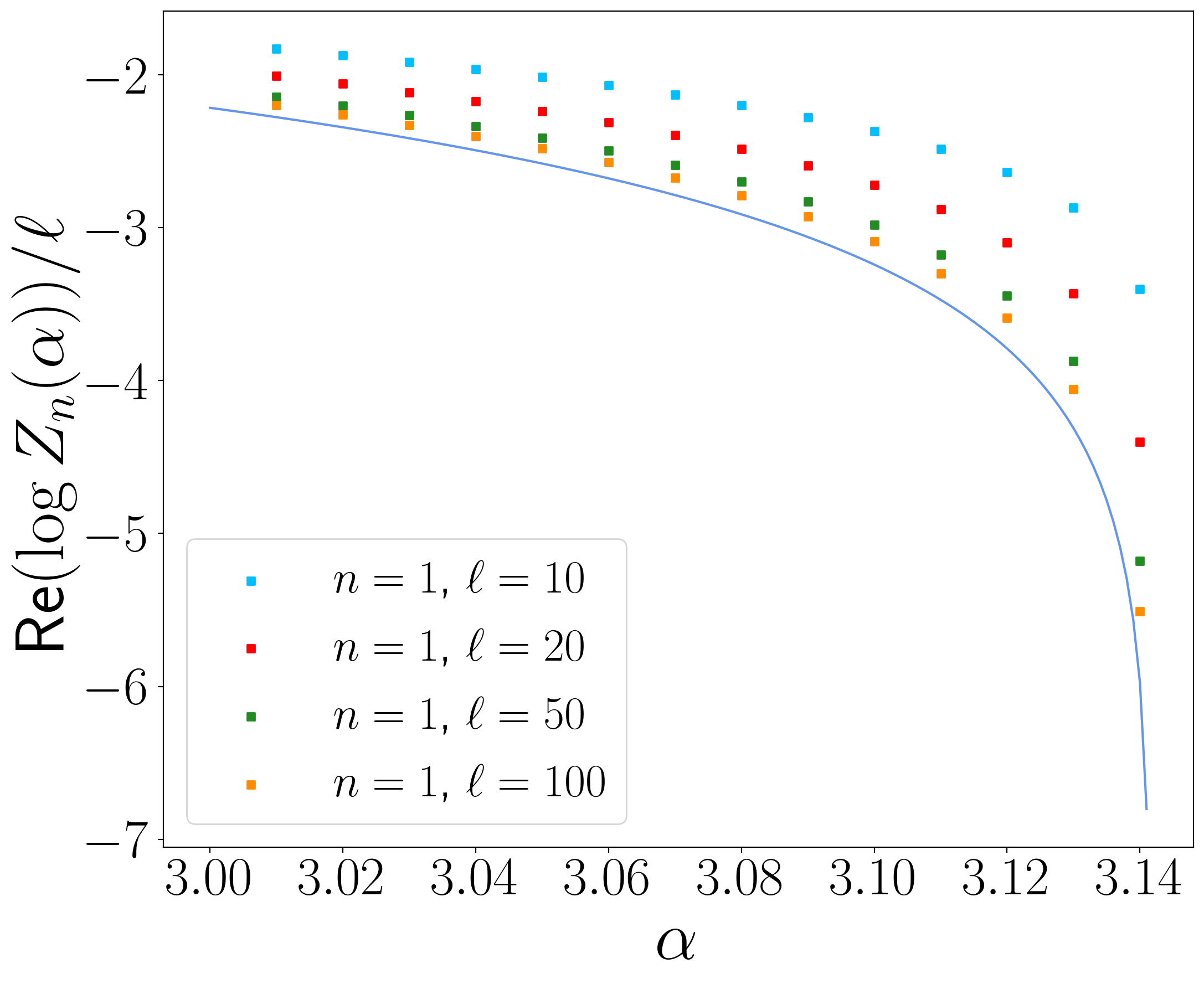}
\end{center}
\caption{Real part of $\log Z_n(\alpha)$ after a quench from the N\'eel state in the free fermion model \eqref{eq:Hfree} as a function of $\alpha$ with $t/\ell=0.5$ and $n=1$ near the value $\alpha=\pi$.
}
\label{fig:ZnalphaAlpha}
\end{figure}

In Fig. \ref{fig:Znalpha} we compare the prediction \eqref{eq:ZnaExactNeel} for the charged moments $Z_n(\alpha)$ with ab-initio computations. The dependence of $Z_n(\alpha)$ in $t$ and $\ell$ is perfectly reproduced by the exact result, as can be seen from top left, right and bottom left panels. 
For this quench, the imaginary part is trivial since it is given by the average conserved charge. Concerning the $\alpha$-dependence, we observe that away from $\alpha=\pi$, the analytic prediction and numerical data match perfectly, even for relatively small $\ell$. 
However, as $\alpha$ gets closer to $\pm \pi$, the difference between the numerical results and the exact prediction gets larger. 
To highlight this phenomenon, we zoom on the region in the vicinity of $\alpha=\pi$ in Fig. \ref{fig:ZnalphaAlpha}.
It is clear that for large $\ell$, the data approach the asymptotic result for any $\alpha\neq\pi$. 
As is well known also in equilibrium, see e.g. \cite{bons,fg-20}, as $\alpha$ gets closer to $\pm\pi$, the match between the numerical data and the analytical predictions only holds for very large values of $\ell$. Additional subleading terms (vanishing in the scaling limit) in the expansion of $\J$ should be considered to correctly describe the large corrections in finite size.
However, this point will not affect the asymptotic behaviour of the symmetry resolved quantities which are calculated by a saddle point integral 
close to $\alpha=0$ and will not be further discussed.

\subsubsection{Fourier transform}
\label{sec:FTNeel}

The symmetry resolved moments $\mathcal{Z}_n(q)$ with $q=\Delta q+ \langle Q_A \rangle$ are defined as the Fourier transform of the charged moments $Z_n(\alpha)$. 
Using Eq. \eqref{eq:ZnqFT}, the asymptotic formula \eqref{eq:ZnaExactNeel}, and $\langle Q_A \rangle=\ell /2$, we have
\begin{equation}
\label{eq:ZnqNeel}
\mathcal{Z}_n(q)\simeq2^{(1-n)\J} \int_{-\pi}^{\pi}\frac{\dd \alpha}{2\pi}  \left(\cos \frac{\alpha}{2} \right)^{\J} \eE^{-\ir\alpha\Delta q}.
\end{equation}
First of all, we stress that this formula is valid for arbitrary $\Delta q$, but only for asymptotically large $\J$ because we used the asymptotic behaviour of $Z_n(\alpha)$.
Indeed, as we shall see, the integral in Eq. \eqref{eq:ZnqNeel} can be also negative for small $\J$.
Furthermore, in the scaling limit $\Delta q$ should be taken proportional to $\ell$ (or, equivalently, $t$), otherwise one obtains trivial results. 

Since the $n$-dependence in Eq. \eqref{eq:ZnqNeel} is trivial, we focus on $n=1$. 
The integral in Eq. \eqref{eq:ZnqNeel} can be evaluated in closed form in terms of the Euler Beta function $B(x,y)= \frac{\Gamma(x)\Gamma(y)}{\Gamma(x+y)}$, 
where $\Gamma(z)$ is the Gamma function. 
The result reads \cite{GR}
\begin{equation}
\label{eq:Z1Gamma}
\zeta_1(q) = 2^{-\J} \frac{\Gamma(\J+1)}{\Gamma\big(\frac{\J+2 \Delta q+2}{2}\big)\Gamma\big(\frac{\J-2 \Delta q+2}{2}\big)},
\end{equation}
and we recall 
\begin{equation}
{\cal Z}_1(q)\simeq \zeta_1(q) , \qquad \rm {for }\; \J\gg1.
\end{equation}

\begin{figure}[t]
    \centering
    \includegraphics[width=0.44\textwidth]{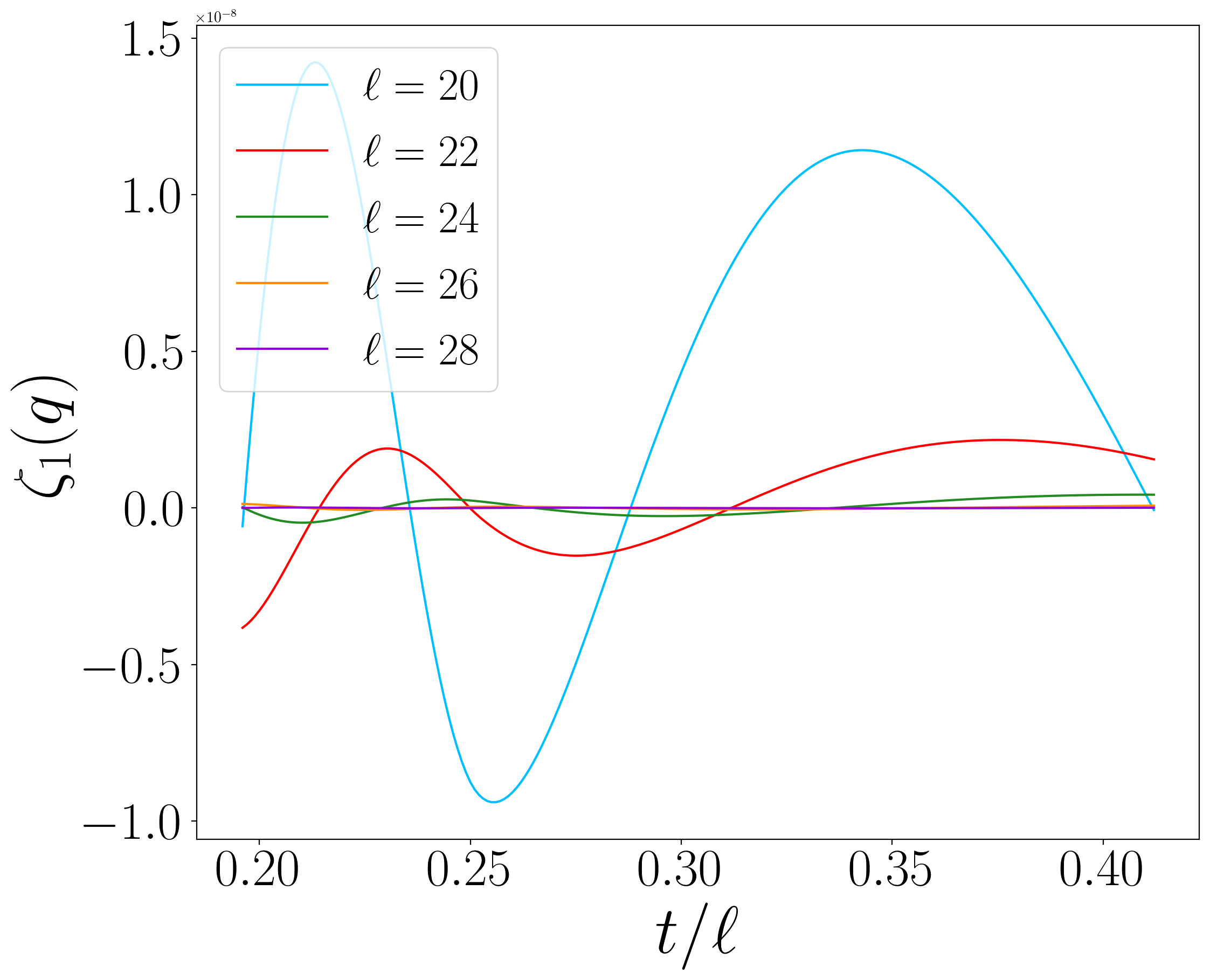} 
    \includegraphics[width=0.44\textwidth]{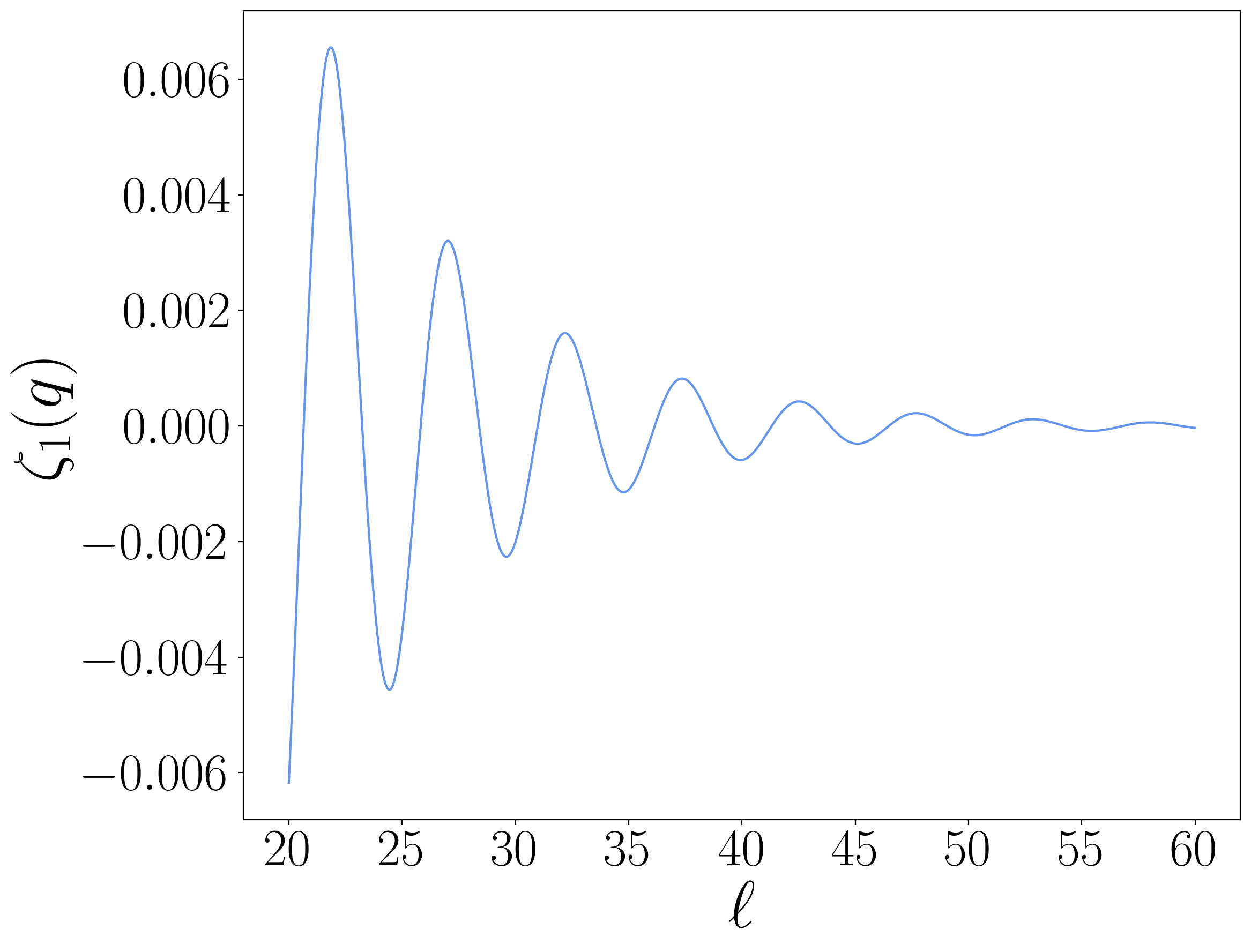} \\
    \includegraphics[width=0.44\textwidth]{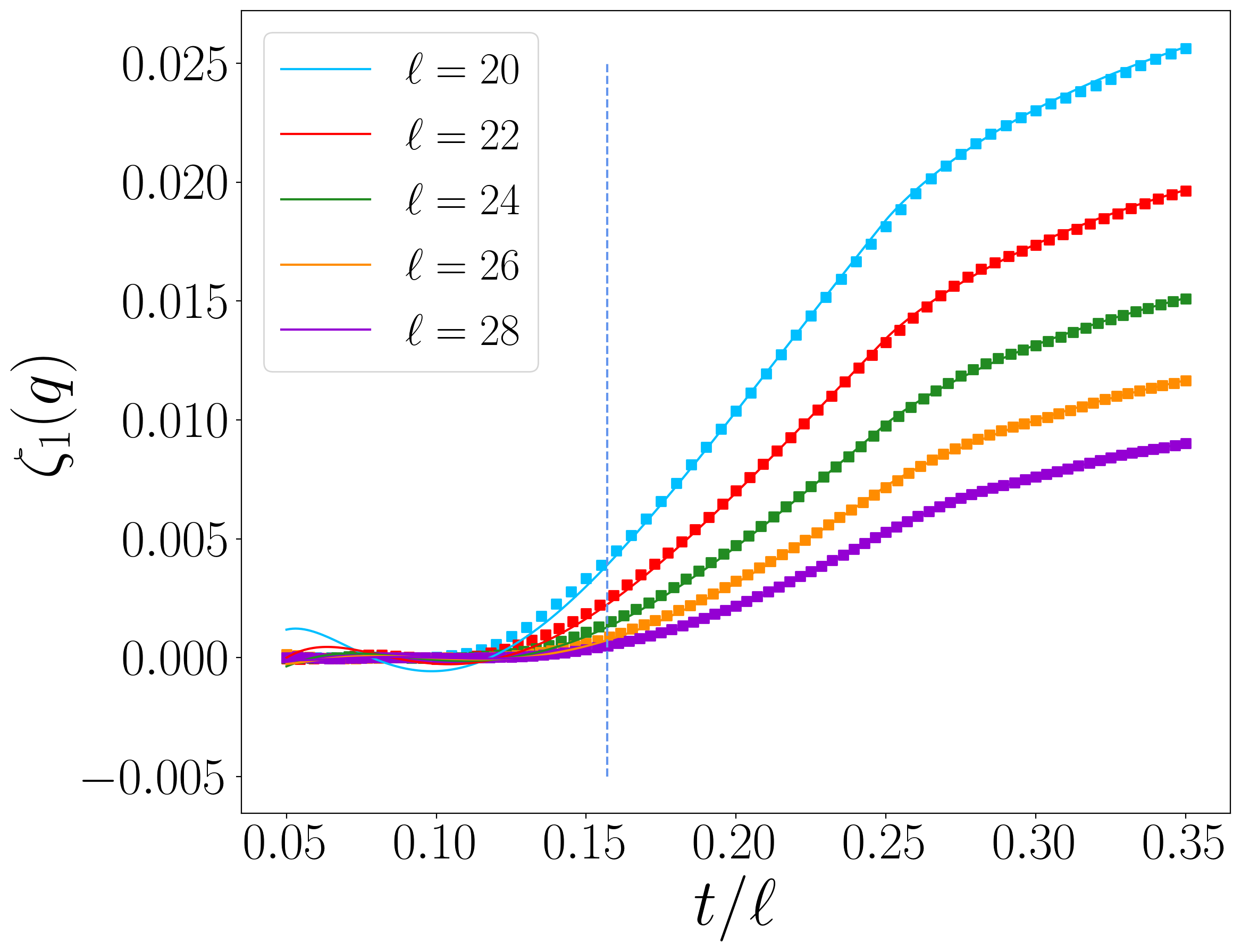} 
    \includegraphics[width=0.44\textwidth]{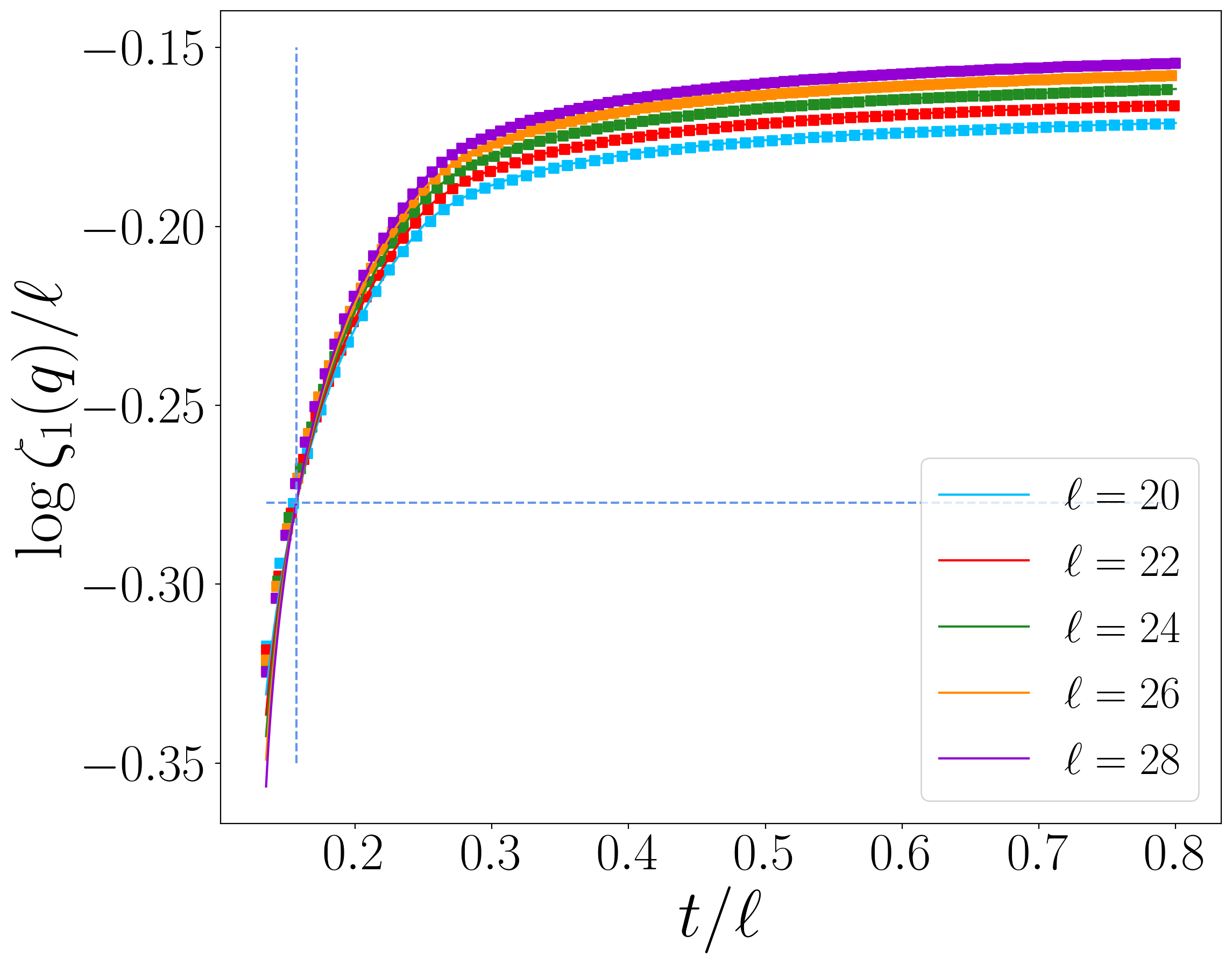} 
     \caption{Study of the function $\zeta_1(q)$ in Eq. \eqref{eq:Z1Gamma} that tends to the physical probability $\mathcal{Z}_1(q)$ for large $\J$ after a quench from the N\'eel state in the tight-biding model \eqref{eq:Hfree}.
   {\bf Top}: Short time regime $\J<2|\Delta q|$. \textit{Left:} $\zeta_1(q)$ as a function of $\J/\ell$ for $q=\ell$ and various values of $\ell$. 
     As $\ell$ increases, $\zeta_1(q)$ quickly averages to zero.
     \textit{Right:} $\zeta_1(q)$ as a function of $\ell$ with $q=9 \ell/10$ and fixed $\J=\ell/100$.
     {\bf Bottom}: $\zeta_1(q)$ (left) and $(\log \zeta_1(q))/\ell$ (right) as a function of $\J/\ell$ with $q=7 \ell/10$. We compare the exact formula of Eq. \eqref{eq:Z1Gamma} for $\zeta_1(q)$ (solid lines) with numerical data for $\mathcal{Z}_1(q)$ (symbols).
     The vertical dashed lines are placed at $\J/\ell = 2 |\Delta q|/\ell=2/5$ in both panels. 
     \textit{Left:} $\zeta_1(q)$ starts growing monotonically at $\J=2|\Delta q|-2$.
     \textit{Right:} The horizontal dashed line is at position $\log\zeta_1(q)	 =  -2 |\Delta q| \log 2$. All the curves collapse on the point $\zeta_1(q)=2^{-\J}$ at $\J=2 |\Delta q|$.}
    \label{fig:Z1}
\end{figure}

We now discuss the properties of the function $\zeta_1(q)$ and in particular its limit for $\J\gg1$ when it tends to the physical probability ${\cal Z}_1(q)$.  
The function $\Gamma (z)$ has poles when $z$ is a negative integer. 
Since $\J \geq 0$, it follows that $\zeta_1(q)$ vanishes when $\J<2 |\Delta q|$ is an even integer. For other values of $\J$ in the regime $\J<2 |\Delta q|$, the function $\zeta_1(q)$ oscillates in $\ell$ around zero, as can be seen in Fig.~\ref{fig:Z1} (top panels). The amplitude of these oscillations decays quickly to zero and, in the scaling limit, averages to $\zeta_1(q)\simeq\mathcal{Z}_1(q)\simeq0$ for all the relevant physics. 
Actually, the last zero of $\zeta_1(q)$ is at position $\J=2|\Delta q|-2$ and afterward it monotonically increases with $\J$ 
(see Fig.  \ref{fig:Z1}, bottom panels).
At $\J=2|\Delta q|$, we have $\zeta_1(q)=2^{-\J}=2^{-2|\Delta q|}$. 
Thus, in the scaling limit, when $\Delta q$ also grows with $\ell$, the entire window $2|\Delta q|-2<\J < 2|\Delta q|$ shrinks to zero length, 
and $\mathcal{Z}_1(q)$ can be considered zero up to $\J=2|\Delta q|$.
Anyhow, for finite $q$, ${\cal Z}_1(q)$ is expected to grow from $\J=2|\Delta q|-2$. 
We illustrate these behaviours in Fig. \ref{fig:Z1} for relatively small values of $\ell$.

In the regime  $\J>2 |\Delta q|\gg1$, we use Stirling formula to derive the asymptotic behaviour of $\zeta_1(q)$ and hence of  $\mathcal{Z}_1(q)$. 
We  find
\begin{multline}
\label{eq:Z1qNeelbigL}
    \log \mathcal{Z}_1(q) = - \Big(\frac{\J}{2}+ |\Delta q|\Big) \log \Big(1+\frac{2 |\Delta q|}{\J}\Big) - \Big(\frac{\J}{2}- |\Delta q|\Big) \log \Big(1-\frac{2 |\Delta q|}{\J}\Big) \\
    +\frac{1}{2} \log \Big( \frac{4 \J}{(\J+2 \Delta q)(\J-2 \Delta q)}\Big)- \frac{1}{2} \log 2 \pi.
\end{multline}
The first line of Eq. \eqref{eq:Z1qNeelbigL} contains the leading terms, the second one gives the first corrections, and sub-leading terms are omitted. 
We stress that the corrections due to the subleading terms in $Z_n(\alpha)$ enter at an order higher than the ones in Eq. \eqref{eq:Z1qNeelbigL}.

\begin{figure}
\begin{center}
\includegraphics[width=0.5\textwidth]{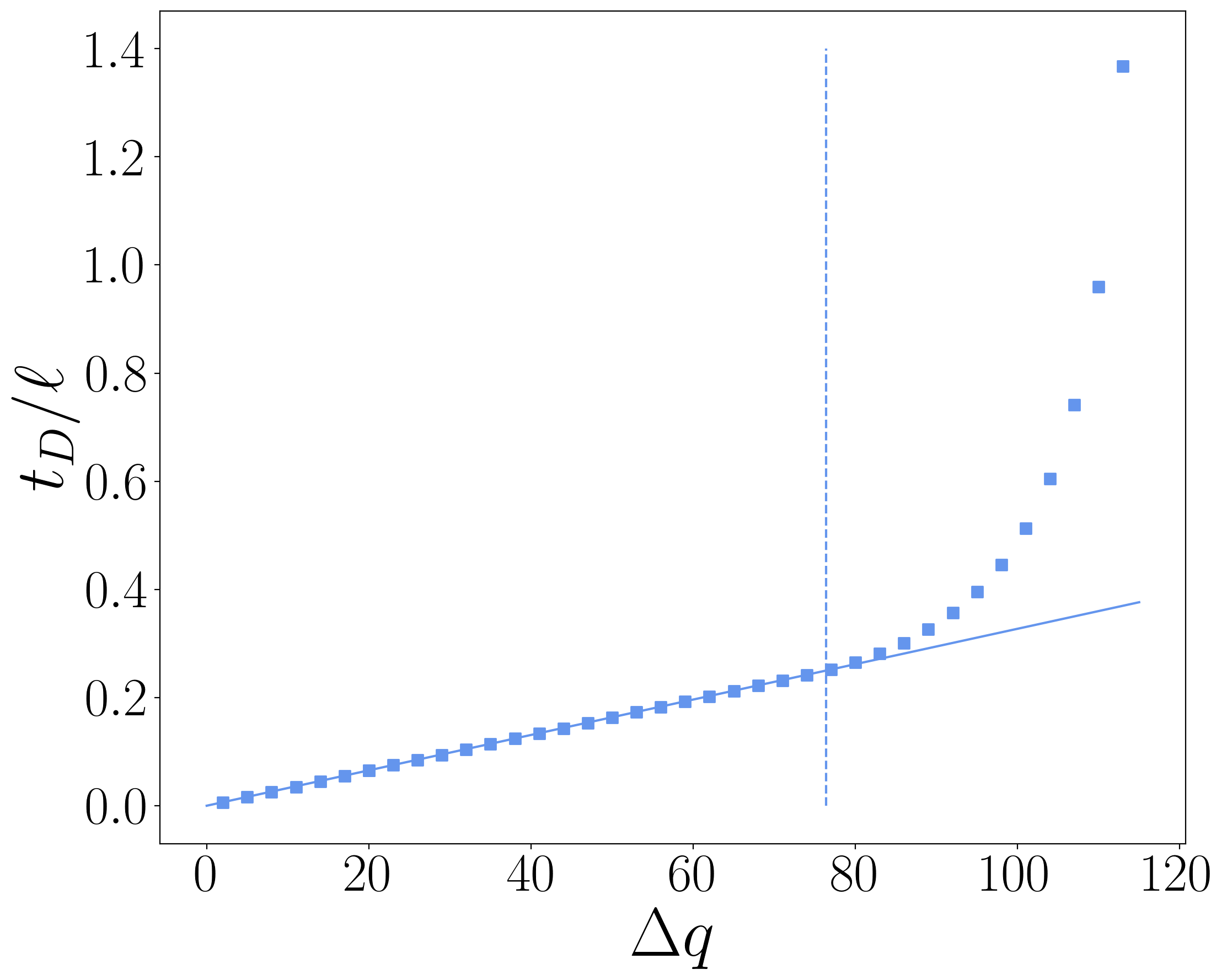}
\end{center}
\caption{Delay $t_D$ obtained from the numerical solution of $\J(t_D)=2 |\Delta q|$ for $\ell=240$ (symbols) compared with the linear result $t_D = \pi \frac{|\Delta q|}{4}$ (solid line), 
valid up to $\Delta q= \frac{\ell}{\pi}$ (dashed vertical line).} 
\label{Fig:DelayNeel}
\end{figure}

We have seen that in the scaling regime $\mathcal{Z}_1(q)$ is zero for short $\J$ and starts growing at $\J=2\Delta q$. 
These relations can be transferred from the auxiliary variable $\J$ to the physical time $t$.  
The crossover between the zero probability regime and the growing one takes place at a time $t_D$ that we dub \textit{delay time}. 
The latter  is defined by the relation $\J(t_D)=2 |\Delta q|$. 
In general, this relation is not explicit. However, from Eq.~\eqref{eq:JNeel}, we have that for $2v_M t<\ell$ 
\begin{equation}
    \J = 2 t \int_{-\pi}^{\pi} \frac{\dd k}{2 \pi} v_k, \quad 2v_M t<\ell.
\end{equation}
Since $v_k =2|\sin k|$ and $v_M=2$, we find 
\begin{equation}
    4 t_D \int_{-\pi}^{\pi} \frac{\dd k}{2 \pi}|\sin k| = 2| \Delta q|, \quad 4t_D<\ell,
\end{equation}
and conclude
\begin{equation}
\label{eq:tDNeel}
    t_D = \pi \frac{|\Delta q|}{4}, \quad {\rm for}\quad  |\Delta q|< \frac{\ell}{\pi}. 
\end{equation}
For larger $|\Delta q|$, there is no closed form for $t_D$. In Fig. \ref{Fig:DelayNeel} we compare the numerical solution of $\J(t_D)=2 |\Delta q|$ (symbols) with the linear result $t_D=\pi \frac{|\Delta q|}{4}$ (solid line). For $ |\Delta q|< \frac{\ell}{\pi}$ the match is perfect, but for larger $\Delta q$ the delay grows very quickly. 

Even though the scaling behaviour of $\mathcal{Z}_1(q)$ is completely encoded in Eq. \eqref{eq:Z1Gamma}, we also analyse it with the saddle point approximation. The reason is that this method is useful in the cases where there are no analytical formulas like Eq. \eqref{eq:Z1Gamma}. We recast Eq. \eqref{eq:ZnqNeel} as
\begin{equation}
\label{eq:Z1qInt}
\mathcal{Z}_1(q)  =  \int_{-\pi}^{\pi} \frac{\dd \alpha}{2 \pi} \eE^{ \ell h(\alpha)}, \quad 
h(\alpha) = -\ir \alpha \frac{\Delta q}{\ell} + \frac{\mathcal{J}}{\ell} \log \cos \frac{\alpha}{2},
\end{equation}
so that the saddle point $\alpha^*$, defined as $h'(\alpha^*)=0$, is 
\begin{equation}
\label{eq:alphaStar}
\alpha^* = - 2\ir \, {\rm arctanh} \left( \frac{2  \Delta q}{\mathcal{J}}\right) \\
= \ir \log \left(\frac{\mathcal{J}-2\Delta q}{\mathcal{J}+2\Delta q}\right).
\end{equation}
We have two regimes, (i) $\J<2 |\Delta q|$, where $\alpha^*$ has a real part, and (ii) $\J>2 |\Delta q|$, where $\alpha^*$ is purely imaginary. In the first regime, the real part of $\alpha^*$ leads to a non-zero imaginary part of $h(\alpha^*)$ and $\mathcal{Z}_1(q)$ oscillates quickly in $\ell$ around zero. This behaviour is the same as the one we found by analyzing the exact formula \eqref{eq:Z1Gamma}. In the scaling limit, these oscillations average to zero for all the relevant physics.

Conversely, for $t>t_D$ (i.e. $\J>2 |\Delta q|$), we deform the integration contour to pass through $\alpha^*$ while remaining in the analyticity region of the integrand. The saddle point method gives 
\begin{equation}
\label{eq:Z1qSPA}
\mathcal{Z}_1(q) = \eE^{ \ell h(\alpha^*)} \sqrt{\frac{1}{2 \pi \ell |h''(\alpha^*)|}}.
\end{equation}
The logarithm of this equation precisely gives back Eq. \eqref{eq:Z1qNeelbigL}, as expected. In the limit where $\J \gg |\Delta q|$ we have 
\begin{equation}
\label{eq:Z1qSPAAsympt}
\mathcal{Z}_1(q) \simeq  \sqrt{\frac{2}{\J \pi}} \eE^{-\frac{2 \Delta q^2}{\J} }
\end{equation}
at leading order. 

\subsubsection{Symmetry resolved R\'enyi entropies}\label{sec:SnqNeel}
We now turn to the evaluation of the symmetry resolved R\'enyi entropies. Plugging Eq.~\eqref{eq:ZnqNeel} into  Eq.~\eqref{SvsZ},  $S_n(q)$ reads
\begin{equation}
\label{eq:SnqIndep}
\begin{split}
    S_n(q) &= \J \log 2 + \log \mathcal{Z}_1(q) \\
\end{split}
\end{equation}
and does not depend on $n$. 
In the scaling regime  $\J>2 |\Delta q|\gg1$,  $\log \mathcal{Z}_1(q)$ is given by \eqref{eq:Z1qNeelbigL}.

\begin{figure}[t]
\begin{center}
\includegraphics[width=0.45\textwidth]{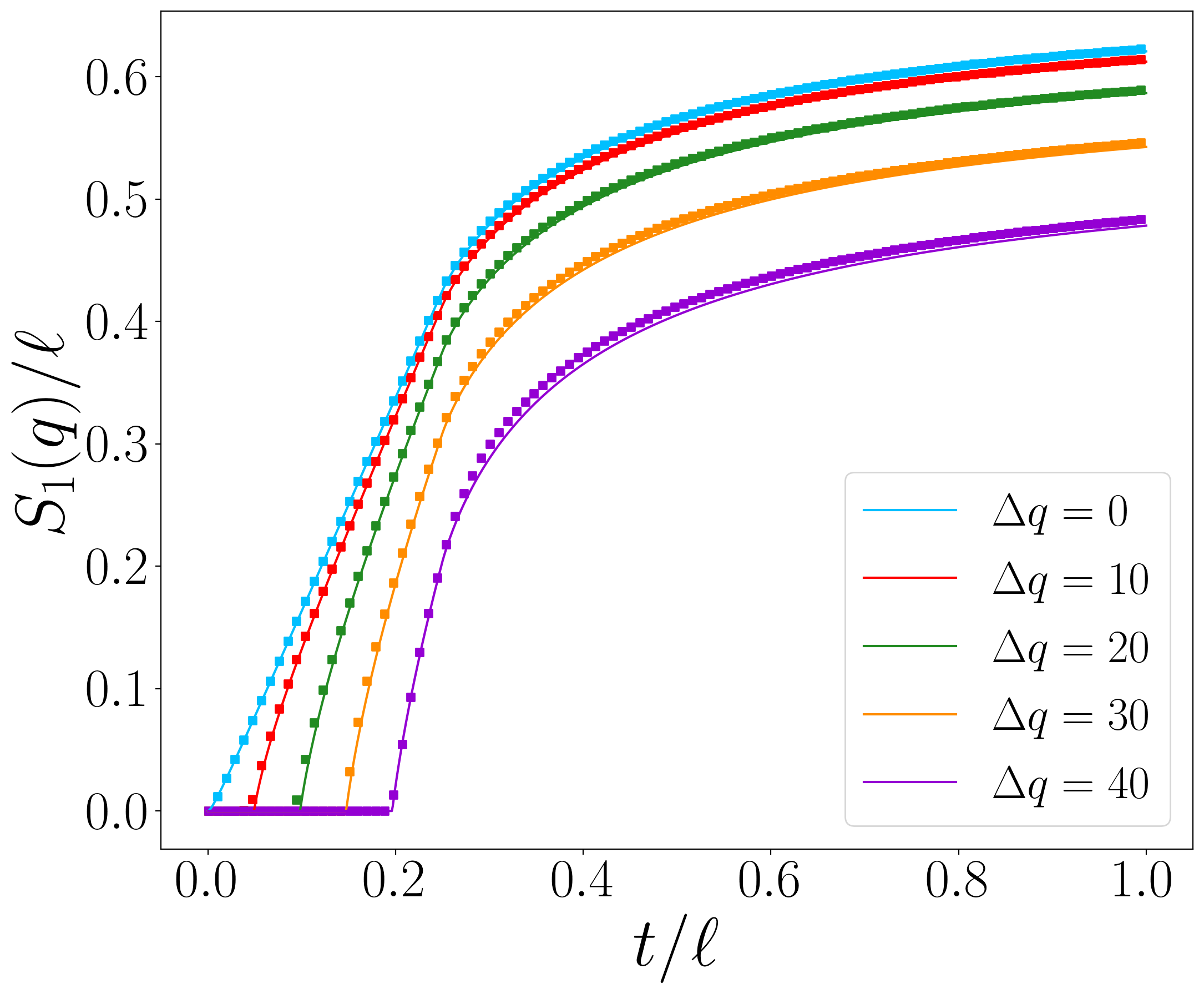} 
\includegraphics[width=0.45\textwidth]{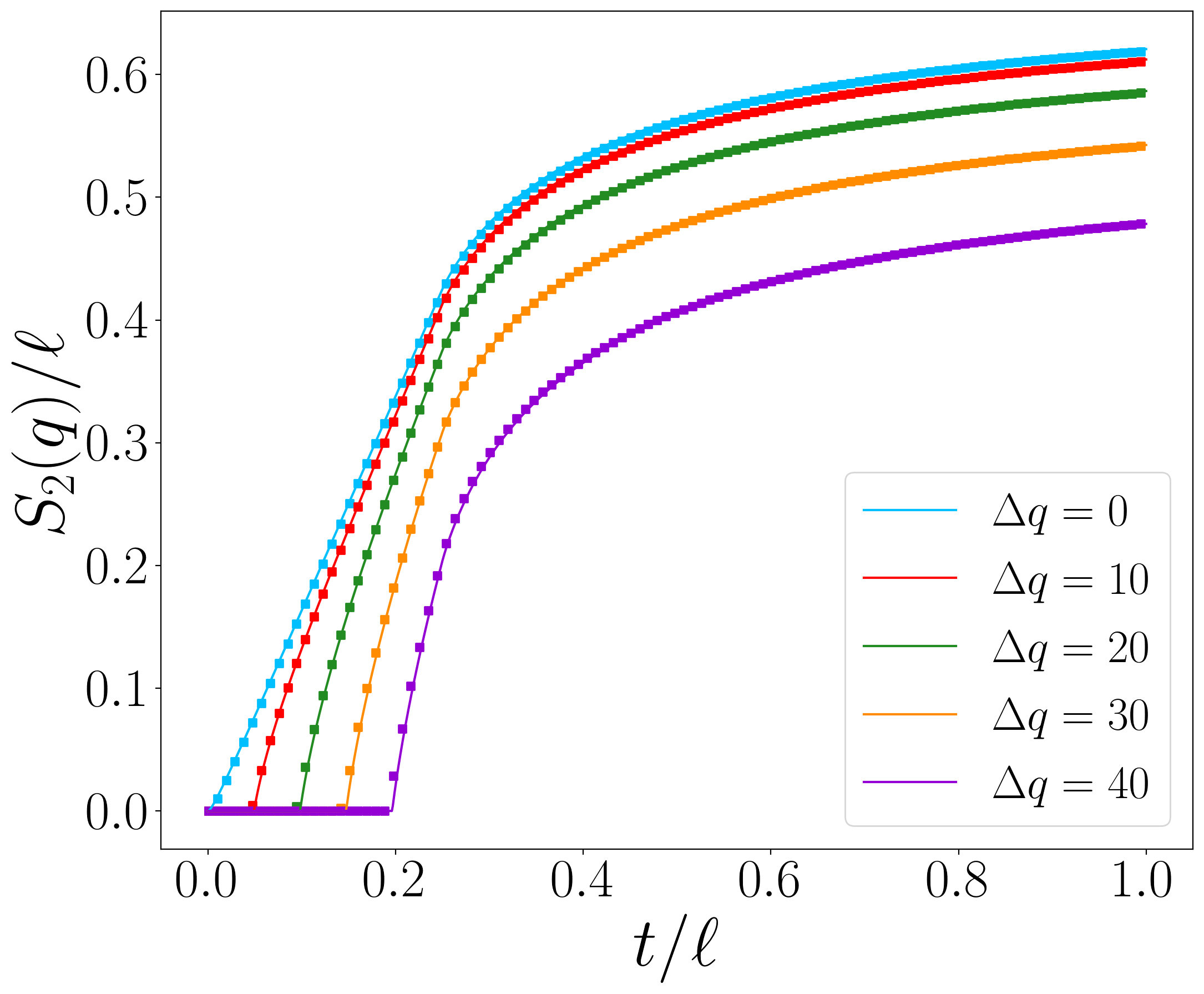} 
\includegraphics[width=0.45\textwidth]{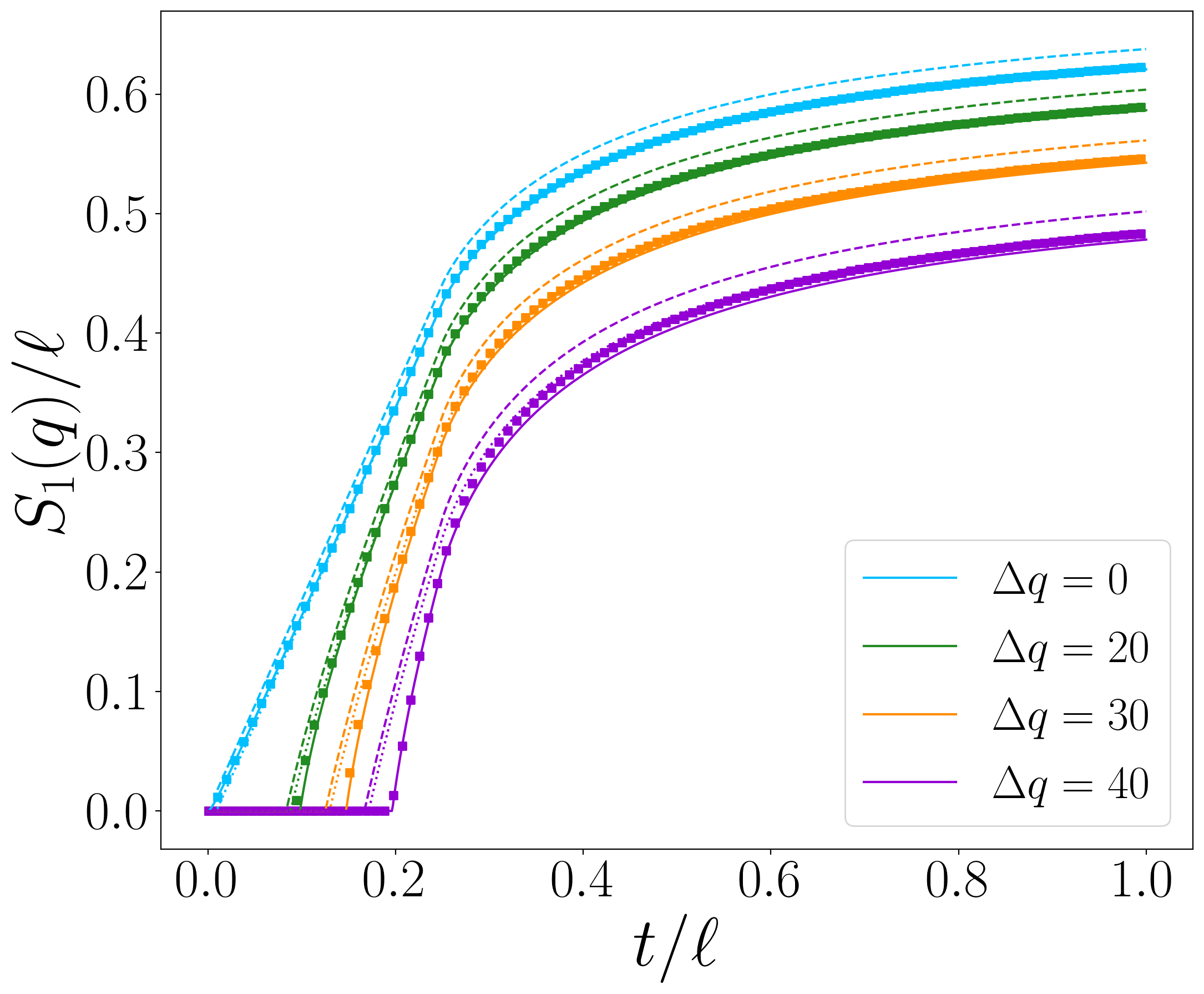} 
\includegraphics[width=0.45\textwidth]{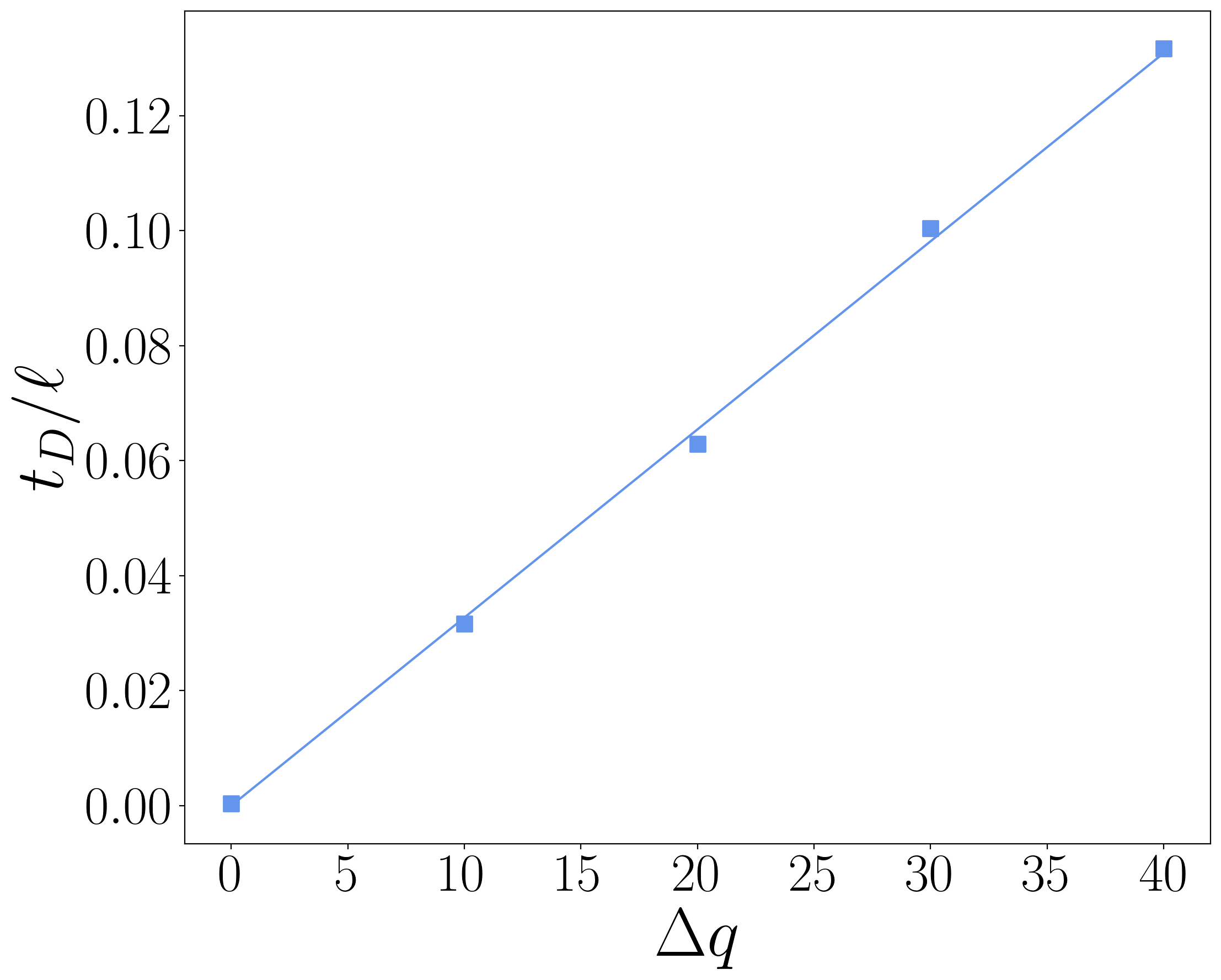}
\end{center}
\caption{ \textit{Top and bottom left:} Time evolution of the symmetry resolved entanglement entropies $S_n(q)$ after a quench from the N\'eel state in the tight-binding model \eqref{eq:Hfree}. The symbols are the exact numerical results for $\ell=160$ and various $\Delta q$. The solid lines are our prediction \eqref{eq:SnqIndep} that perfectly match the numerical data. 
\textit{Bottom left:} The dashed lines are the saddle-point approximation \eqref{eq:SqCorr} and the dotted ones are the same approximation with the logarithmic correction. 
\textit{Bottom right:} Delay time $t_D$ for various $\Delta q$ and $\ell=240.$ The solid line is the analytic prediction \eqref{eq:tDNeel} and the symbols are the numerical results. They are obtained as the time for which $S_1(q)/\ell=0.007$.} 
\label{Fig:Snq}
\end{figure}

Following the discussion of the previous section, in the scaling limit, the entropies start to grow for $t>t_D$. In the top panels of Fig. \ref{Fig:Snq}, we compare ab-initio calculations of $S_n(q)$ and the exact prediction  \eqref{eq:SnqIndep}, finding perfect agreement. In particular, Fig. \ref{Fig:Snq} convincingly shows that the numerical results for $S_n(q)$ do not depend on $n$. We thus focus on $S_1(q)$ for the rest of this subsection. 
We also compare the prediction~\eqref{eq:tDNeel} for the delay with the numerical results at a finite $\ell$ (we fix $\ell=240$ and we calculated the delay time as 
the time when $S_1(q)/\ell=0.007$) in the bottom right panel of Fig. \ref{Fig:Snq} 
and again find a very good agreement. 

If in the scaling limit with $\ell$, we also take  $\J \gg |\Delta q|$, we can use Eq. \eqref{eq:Z1qNeelbigL} and plug it into \eqref{eq:SnqIndep}.
Expanding the result in powers of $|\Delta q|/\J$, we find
\begin{equation}
\label{eq:SqCorr}
S_n(q)= \J \left( \log 2 - 2 \Big( \frac{\Delta q}{\mathcal{J}}\Big)^2\right).
\end{equation}
Eq. \eqref{eq:SqCorr} implies that for small $|\Delta q|$ there is an effective equipartition of entanglement between the various symmetry sectors, 
with violations of order $(\Delta q)^2/\ell$. We report this prediction as dashed lines in the bottom left panel of Fig. \ref{Fig:Snq}. 
There appears to be some discrepancies, close to the delay time, between this approximation and the numerical results, even for small $|\Delta q|$, where one would expect the approximation to be very accurate. 
The reason for such disagreement is, obviously, that at the delay time $\J(t_D)=0$ and so the subleading terms in \eqref{eq:SqCorr} cannot be neglected. 
Looking at Eq. \eqref{eq:Z1qNeelbigL} (or equivalently Eq. \eqref{eq:Z1qSPA}), we see that the first correction to $S_n(q)$ in the large-$\ell$ expansion is proportional to 
$\log \ell$. In the same panel of Fig. \ref{Fig:Snq}, we add this correction and report $\frac{\J}\ell \left( \log 2 - 2 \Big( \frac{\Delta q}{\mathcal{J}}\Big)^2\right) -1/(2\ell)\log \ell$ as the dotted lines.  
These curves get closer to the numerical data and fit them perfectly for $t \gg t_D$, i,e, in the region where $\J$ is significantly larger than $|\Delta q|$. 
This correction is subleading in the scaling limit and hence we conclude that Eq.~\eqref{eq:SqCorr} correctly reproduces the numerical data in the large-$\ell$ limit.

\subsubsection{Total and number entropy}\label{sec:totSnNeel}
A final question worth investigating is whether our results for the symmetry resolved entanglement entropy are coherent with the known results for the dynamics of the total entanglement entropy $S_1$. 
To this aim, we should also investigate the contribution of the number entropy $S^n$ to the total entanglement. 
The number entropy $S^{n}$ is defined in Eq. \eqref{decompositionSvN}, providing in our case 
\begin{equation}
\label{eq:SnumNeel}
    S^n = - \sum \limits_{q=0}^{\ell} \mathcal{Z}_1(q) \log \mathcal{Z}_1(q)
\end{equation}
and, using the same asymptotic calculations as in Sec. \ref{sec:FTNeel} and Eq. \eqref{eq:Z1qSPAAsympt}, we find
\begin{equation}
    S^n = \sqrt{\frac{2}{\J \pi}} \sum_{\Delta q=-\ell/2}^{\ell/2} \Big(\frac{2 \Delta q^2}{\J}+\frac 12 \log \frac{\J \pi }{2}\Big) \eE^{-\frac{2 \Delta q^2}{\J} }
\end{equation}
for $\J \gg |\Delta q|$. The latter condition is not met for the extreme values of $|\Delta q|$ in the sum, where instead we have $\J \sim |\Delta q|$. However, such values
of the subsystem charge have a very low probability and so this approximation is expected to give excellent results, as we will soon confirm numerically.
Since we are in the limit of large~$\ell$, we transform the sums into Gaussian integrals. Their evaluation is straightforward and we have
\begin{equation}
    \label{eq:SnumApproxNeel}
    S^n = \frac 12 \Big(1+\log \frac{\J \pi }{2} \Big),
\end{equation}
i.e. the number entropy grows logarithmically with time until saturation to a value which is proportional to $\log \ell$.
Such behaviour is reminiscent of what found in many other different contexts \cite{kusf,kusf2,kufs3}.

\begin{figure}[t]
\begin{center}
\includegraphics[width=0.45\textwidth]{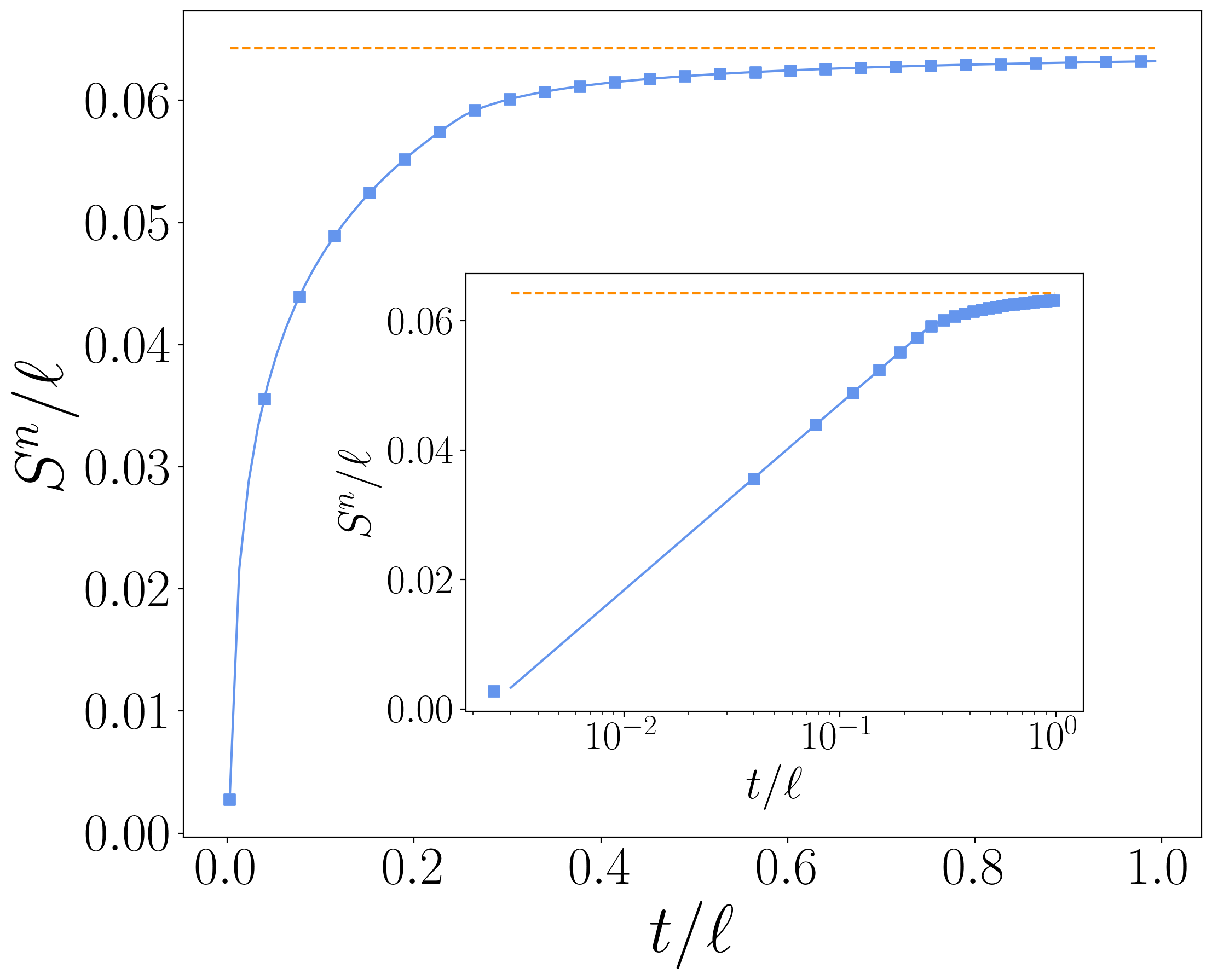} 
\includegraphics[width=0.45\textwidth]{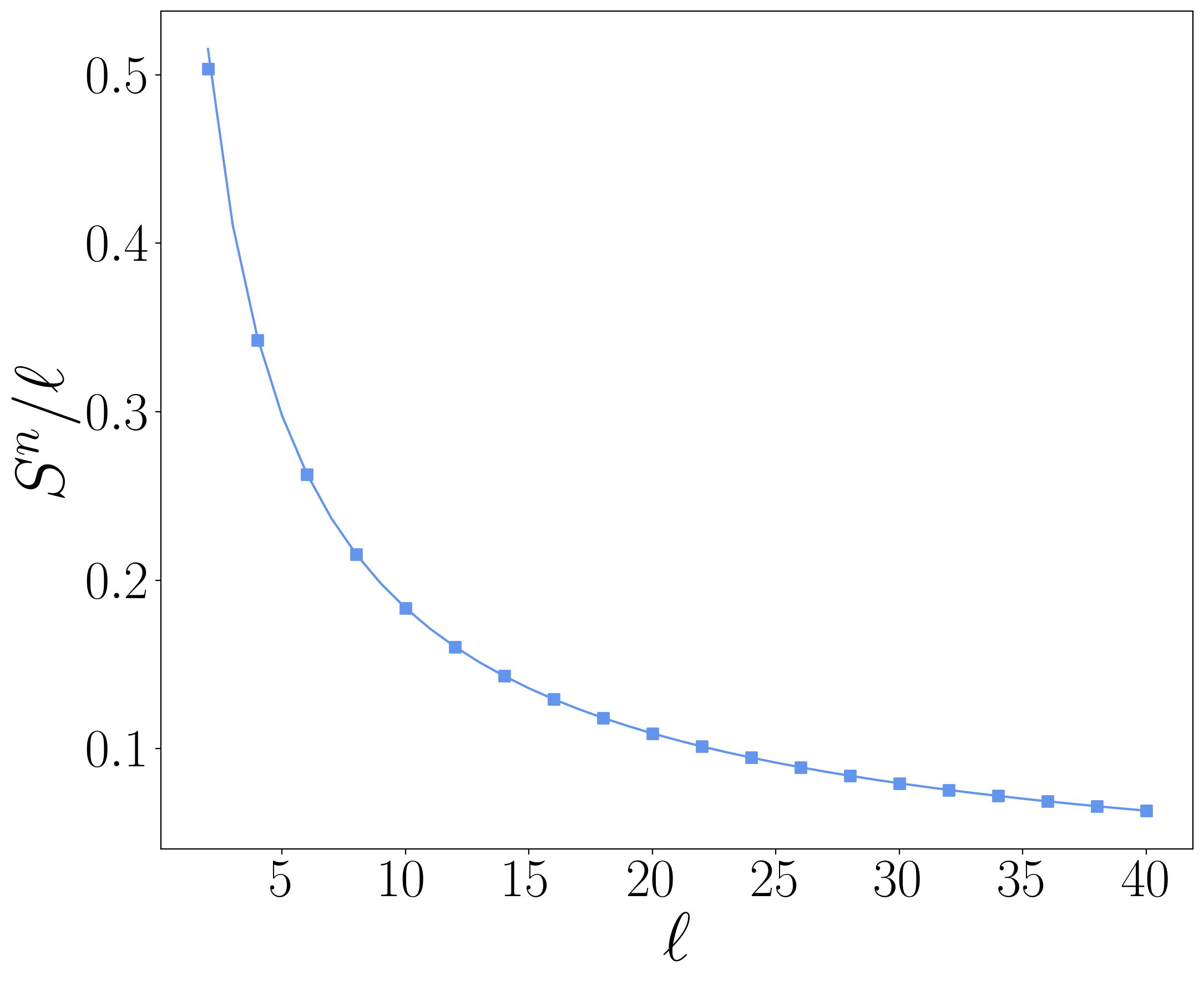} 
\end{center}
\caption{Evolution of $S^n$ after a quench from the N\'eel state in the tight-binding model \eqref{eq:Hfree} as a function of $t/\ell$ for $\ell=40$ (left panel), and as a function of $\ell$ with $t/\ell=1$ (right panel). The prediction \eqref{eq:SnumApproxNeel} (solid line) perfectly matches the numerical results (symbols). In the left panel, the horizontal dotted line is the asymptotic value $\frac {1}{2 \ell} \Big(1+\log \frac{\ell \pi }{2}\Big)$ for large times. In the same panel, the inset shows $S^n$ as a function of $t/\ell$ in a log-linear scale. }
\label{Fig:SnumNeel}
\end{figure}

In Fig. \ref{Fig:SnumNeel} we compare this result (solid lines) with ab-initio calculations of $S^n$ (symbols), and find perfect agreement. The inset in the left panel shows the time-evolution of $S^n$ in a log-linear scale and highlights the logarithmic growth in time before the saturation. It directly follows from Eq. \eqref{eq:SnumApproxNeel} that the large-time limit of the number entropy is $\displaystyle\lim_{t\to \infty}S^n =\frac 12 \Big(1+\log \frac{\ell \pi }{2} \Big)$, reported as a dotted horizontal line in the left panel of Fig. \ref{Fig:SnumNeel}. 
This analysis shows that in the scaling limit, the number entropy has a negligible contribution to the total entropy, 
$\displaystyle\lim_{t\to\infty}S^n/S_1 =\mathcal{O}(\ell^{-1}\log \ell)$.

For the total entropy, we plug Eq.~\eqref{eq:SnqIndep} into Eq.~\eqref{decompositionSvN} and find
\begin{equation}
    S_1 = \sum_{q=0}^{\ell} \mathcal{Z}_1(q) \J \log 2,
\end{equation}
where we used the relation $p(q)=\mathcal{Z}_1(q)$ from Eq.~\eqref{eq:pq}. 
The latter satisfies $\sum_{q=0}^{\ell} \mathcal{Z}_1(q)=1$, as it should. 
We thus recover the known result for the total entanglement entropy after a quench from the N\'eel state \cite{ac-17, ac-18}, namely
\begin{equation}
    S_1 = \log 2  \int \frac{\dd k}{2 \pi}\min(\ell,2 v_k t).
\end{equation}
For large times, we have $\displaystyle\lim_{t\to \infty}S_1 = \ell \log 2$.
We finally mention that it is straightforward to show that the number entropy exactly compensate for the $\log\ell$ subleading corrections present in the entanglement of 
each symmetry sector.

\subsection{Disjoint intervals}

In this subsection, we move to the analysis of the charged and symmetry resolved entropies for two disjoint intervals, as well as the symmetry resolved mutual information.

The starting point is the correlation matrix $C_{A_1\cup A_2}(t)$ that encodes correlations within the system $A=A_1 \cup A_2$ where the two subsystems have respective lengths 
$\ell_1$ and $\ell_2=\ell-\ell_1$, and are separated by $d$ lattice sites. The correlation matrix has the form
\begin{equation}
\label{eq:CADisjoin1}
C_{A_1\cup A_2}(t) = \begin{pmatrix}
C_{11}(t) & C_{12}(t) \\
C_{21}(t) & C_{22}(t)
\end{pmatrix}
\end{equation}
with
\begin{align}
\label{eq:CADisjoin2}
    [C_{11}(t)]_{j,k} &=  [C(t)]_{j,k}, \quad  &&j,k = 1,\dots \ell_1, \\
    [C_{22}(t)]_{j,k} &= [C(t)]_{j+d,k+d}, \quad  &&j,k = 1,\dots \ell_2, \\
    [C_{12}(t)]_{j,k} &= [C(t)]_{j,k+d}, \quad &&j = 1,\dots \ell_1, \quad k = 1,\dots \ell_2, \\
    [C_{21}(t)]_{j,k} &= [C(t)]_{j+d,k}, \quad  &&j= 1,\dots \ell_2, \quad k = 1,\dots \ell_1,
\end{align}
where the matrix elements are given in Eq. \eqref{eq:CNeel}.

\subsubsection{Charged moments}\label{sec:ZnadNeel}

\begin{figure}[t]
\begin{center}
\includegraphics[scale=0.3]{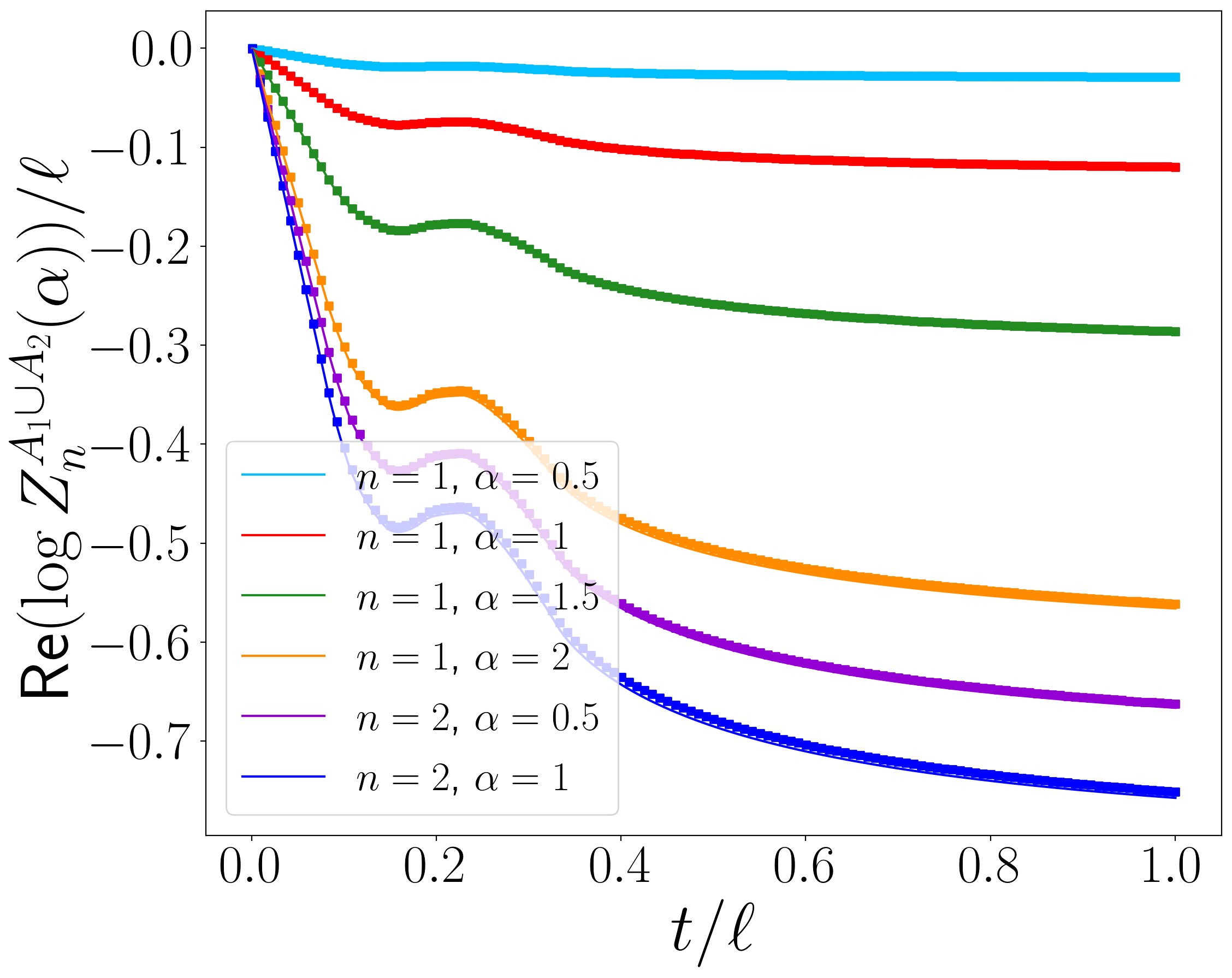}
\end{center}
\caption{Time evolution of the charged moments $Z_n^{A_1\cup A_2}(\alpha)$ after a quench from the N\'eel state in the tight-binding model \eqref{eq:Hfree} as a function of $t/\ell$ with $\ell_1=100$, $\ell_2=140$ and $d=80$. The analytical prediction of Eq.~\eqref{eq:ZnadExactNeel} (solid lines) perfectly matches the numerical data (symbols).}
\label{fig:Znalphadt}
\end{figure}

In order to compute the charged moments and the symmetry resolved entanglement measures, we compute the trace of arbitrary powers of the matrix $J_{A_1\cup A_2}(t) = 2 C_{A_1\cup A_2}(t)- \id_\ell$. We perform the computation of $\Tr J_{A_1\cup A_2}(t)^{m}$ in the scaling limit $t,\ell, \ell_1,\ell_2,d \to \infty$ where the various ratios are kept constant. The calculation uses the multidimensional stationary phase approximation, and generalises the methods used in \cite{fc-08} to the case of an arbitrary separation $d$, see \cite{pb-22}. We find
\begin{equation}
\label{eq:TrJdNeel}
\begin{split}
\Tr J_{A_1\cup A_2}(t)^{2j}&= \ell -  \int \frac{\dd k}{2 \pi}[\min(\ell_1,2 v_k t)+\min(\ell_2,2 v_k t)] \\
&+ \int \frac{\dd k}{2 \pi} [\max(d, 2v_kt)+ \max(d+\ell, 2v_kt)- \max(d+\ell_1, 2v_k t)- \max(d+\ell_2, 2v_k t)], \\
\Tr J_{A_1\cup A_2}(t)^{2j+1}&=0,
\end{split}
\end{equation}
with $v_k = 2 |\sin k|$. We note that the case $d=0$ reduces to the known result of Eq.~\eqref{eq:TrJNeeld0} for a single interval. The computation of the charged moments $Z_n^{A_1\cup A_2}(\alpha)$ proceeds exactly as for the case of a single interval discussed in Sec. \ref{sec:ZnaNeel}. We thus find
\begin{equation}
\label{eq:ZnadExactNeel}
Z_n^{A_1\cup A_2}(\alpha) = \left(\frac{\cos \frac{\alpha}{2}}{2^{n-1}} \right)^{\J_d} \eE^{\ir \ell \frac{\alpha}{2}}
\end{equation}
where we introduced
\begin{equation}
\label{eq:JdNeel}
\begin{split}
\J_d \equiv & \ell -\text{Tr}J_{{A_1\cup A_2}}(t)^2 \\
=& \int \frac{\dd k}{2 \pi}[\min(\ell_1,2 v_k t)+\min(\ell_2,2 v_k t)] \\
&-\int \frac{\dd k}{2 \pi} [\max(d, 2v_kt)+ \max(d+\ell, 2v_kt)- \max(d+\ell_1, 2v_k t)- \max(d+\ell_2, 2v_k t)].
\end{split}
\end{equation}
In Fig. \ref{fig:Znalphadt} we compare the prediction \eqref{eq:ZnadExactNeel} for $Z_n^{A_1\cup A_2}(\alpha)$ with ab-initio calculations and find perfect agreement. 

\subsubsection{Fourier transform and symmetry resolved R\'enyi entropies}
The calculations for the symmetry resolved moments and R\'enyi entropies are exactly the same as the computations of Secs. \ref{sec:FTNeel} and \ref{sec:SnqNeel} where $\J$ is systematically replaced by $\J_d$. We thus have 
\begin{equation}
\label{eq:Z1SnGammad}
\begin{split}
    S_n^{A_1\cup A_2}(q) = \J_d \log 2 + \log \mathcal{Z}_1^{A_1\cup A_2}(q), \quad \mathcal{Z}_1^{A_1\cup A_2}(q) \simeq 2^{-\J_d} \frac{\Gamma(\J_d+1)}{\Gamma\big(\frac{\J_d+2 \Delta q+2}{2}\big)\Gamma\big(\frac{\J_d-2 \Delta q+2}{2}\big)},
\end{split}
\end{equation}
and 
\begin{equation}
\label{eq:Z1SnAsymptd}
S_n^{A_1\cup A_2}(q)= \J_d \left( \log 2 - 2 \Big( \frac{\Delta q}{\mathcal{J}_d}\Big)^2\right), \quad \mathcal{Z}_1^{A_1\cup A_2}(q) \simeq  \sqrt{\frac{2}{\J_d \pi}} \eE^{-\frac{2 \Delta q^2}{\J_d} },
\end{equation}
in the limit where $\ell$ is large and $\J_d \gg |\Delta q|$. We do not report numerical tests for these symmetry resolved entropies because they do not to add 
any information compared to the single interval case.

\subsubsection{Symmetry resolved mutual information}\label{sec:SRMutNeel}

\begin{figure}[t]
\begin{center}
\includegraphics[scale=0.3]{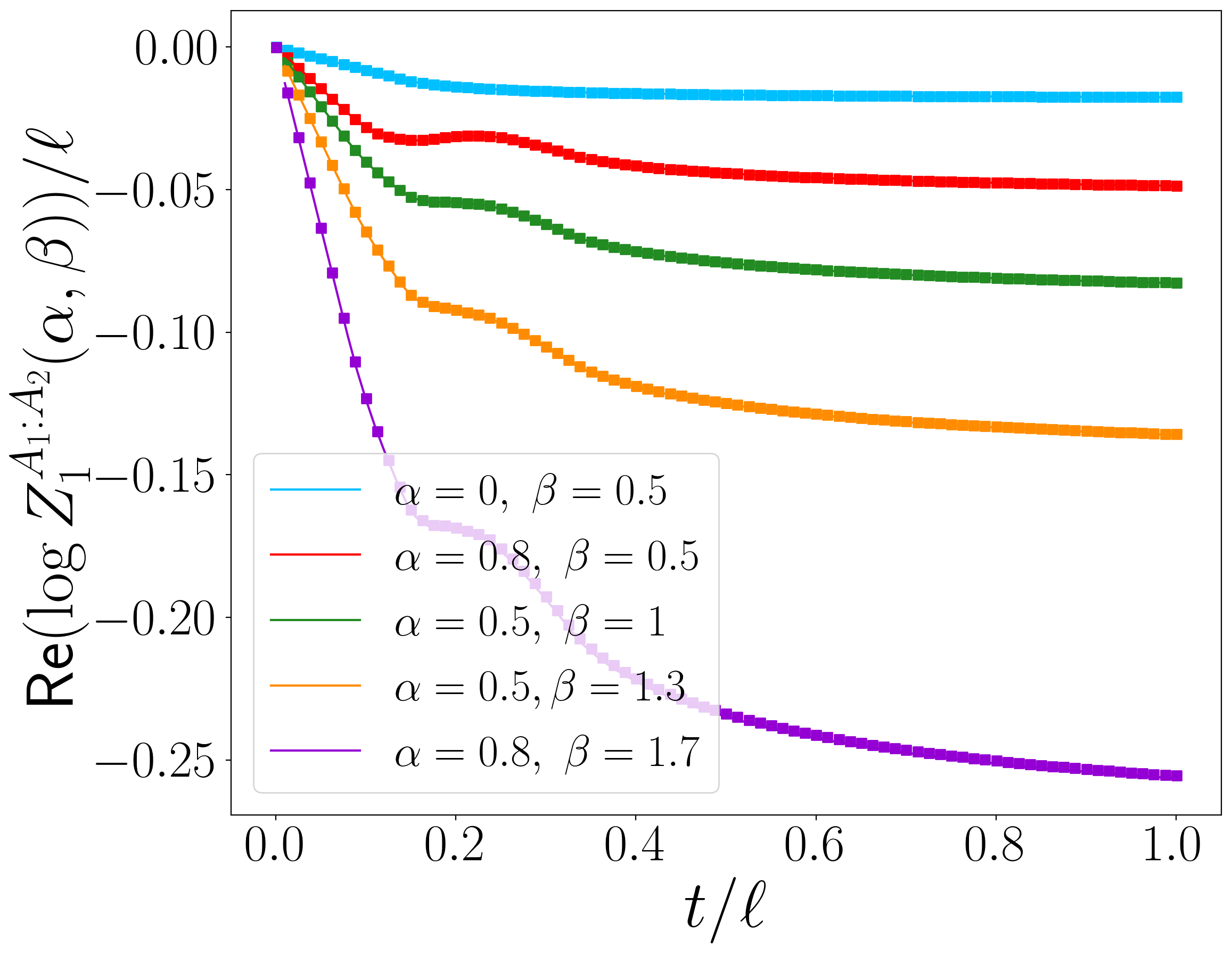}
\end{center}
\caption{Time evolution of $Z_1^{A_1: A_2}(\alpha,\beta)$ after a quench from the N\'eel state in the tight-binding model \eqref{eq:Hfree} as a function of $t/\ell$ with $\ell_1=100$, $\ell_2=140$ and $d=80$. The analytical prediction of Eq.~\eqref{eq:Z1abExactNeel} (solid lines) perfectly matches the numerical data (symbols).}
\label{fig:Z1abNeel}
\end{figure}

We now turn to the computation of the symmetry resolved mutual information defined as in Eq. \eqref{eq:SymResMut}. The first step is to compute $Z_1^{A_1:A_2} (\alpha,\beta)$. To do so, we compute the trace of powers of the matrix $J_{\alpha \beta} $ in Eq. \eqref{eq:TildeJA}. We adapt the calculations of Sec. \ref{sec:ZnadNeel} for the trace of $J_{A_1 \cup A_2}$ and find
\begin{equation}
\label{eq:TrTildeJdNeel}
\begin{split}
\Tr J_{\alpha \beta} (t)^{2j}&=\Big(\frac{\eE^{\ir \alpha}-1}{\eE^{\ir \alpha}+1}\Big)^{2j} \Big[\ell_1 -  \int \frac{\dd k}{2 \pi}\min(\ell_1,2 v_k t)\Big] \\
&+\Big(\frac{\eE^{\ir \beta}-1}{\eE^{\ir \beta}+1}\Big)^{2j} \Big[\ell_2-  \int \frac{\dd k}{2 \pi}\min(\ell_2,2 v_k t)\Big] \\
+(-1)^j\Big( \tan \frac{\alpha}{2} \tan \frac{\beta}{2}\Big)^j \int \frac{\dd k}{2 \pi} &[\max(d, 2v_kt)+ \max(d+\ell, 2v_kt)- \max(d+\ell_1, 2v_k t)- \max(d+\ell_2, 2v_k t)], \\[.5cm]
\Tr J_{\alpha \beta} (t)^{2j+1}&=0.
\end{split}
\end{equation}
The re-summation in Eq. \eqref{eq:Z1abSum} is direct and we find
\begin{equation}
\label{eq:Z1abExactNeel}
\begin{split}
\log Z_1^{A_1:A_2} (\alpha,\beta)&= \ir \frac{\ell_1 \alpha+\ell_2 \beta}{2}+ \log \Big(\cos \frac{\alpha}{2}\Big) \J_{A_1}+ \log \Big(\cos \frac{\beta}{2}\Big) \J_{A_2}\\
&+\frac 12 \log \Big( 1+\tan \frac{\alpha}{2} \tan \frac{\beta}{2}\Big) (\J_{A_1}+  \J_{A_2}-\J_d)
\end{split}
\end{equation}
where $\J_d$ is given in Eq. \eqref{eq:JdNeel} and the straightforward notation $\J_{A_{1,2}} = \J |_{\ell \to \ell_{1,2}}=\int \frac{\dd k}{2 \pi}\min(\ell_{1,2},2 v_k t)$. 
As a consistency check, $ Z_1^{A_1:A_2} (\alpha,\alpha)$ is equal to the charged moment $Z_1^{A_1 \cup A_2}(\alpha)$ given in Eq. \eqref{eq:ZnadExactNeel}. We compare the prediction of Eq. \eqref{eq:Z1abExactNeel} with numerical results in Fig. \ref{fig:Z1abNeel} and find a perfect agreement.

To perform the double Fourier transform in Eq. \eqref{eq:Z1q1q2} for $\mathcal{Z}_1^{A_1:A_2}(q_1,q-q_1)$, we exploit the saddle point approximation and 
expand $Z_1^{A_1:A_2} (\alpha,\beta)$ at quadratic order, obtaining
\begin{equation}
\label{eq:Z1abQuadrNeel}
\log Z_1^{A_1:A_2} (\alpha,\beta)= \ir \frac{\ell_1 \alpha+\ell_2 \beta}{2} -\frac{\alpha^2}{8} \J_{A_1} - \frac{\beta^2}{8} \J_{A_2}+ \frac{\alpha \beta}{8} (\J_{A_1}+  \J_{A_2}-\J_d).
\end{equation}
In this approximation, the Fourier transform is given as 
\begin{equation}
\label{eq:Z1qq1Neel}
\mathcal{Z}_1^{A_1:A_2}(q_1,q-q_1) = \frac{2}{\pi \sqrt{\J_{A_1}\J_{A_2}-\frac{\J_m^2}{4}}}\eE^{-\frac{8}{4\J_{A_1}\J_{A_2}-\J_m^2}\Big[\Delta q_1^2 \J_{A_2}+(\Delta q-\Delta q_1)^2 \J_{A_1}+\Delta q_1(\Delta q-\Delta q_1) \J_m\Big]}
\end{equation}
where we have introduced $\J_m=(\J_{A_1}+  \J_{A_2}-\J_d)$ to lighten the notations, and we have $\Delta q=q-\ell/2$ and $\Delta q_1=q_1-\ell_1/2$. Importantly, the weight $p(q_1,q-q_1)=\mathcal{Z}_1^{A_1:A_2}(q_1,q-q_1)/\mathcal{Z}_1^{A_1 \cup A_2}(q)$ defined in Eq. \eqref{eq:pqq1} satisfies $\sum_{q_1=0}^q p(q_1,q-q_1) =1$ also when the sum is approximated as an integral and the quantities are given by their Gaussian approximations of Eqs. \eqref{eq:Z1SnAsymptd} and \eqref{eq:Z1qq1Neel}.

Using the Gaussian expression for each term involved in the definition of Eq. \eqref{eq:SymResMut} and the same type of asymptotic calculations as in the previous sections, we find
\begin{equation}
\label{eq:IqAsymNeel}
\begin{split}
 I^{A_1 : A_2}_1(q) &= (\J_{A_1}+  \J_{A_2}-\J_d) \log 2 - \frac{1}{2} \Big( \log \frac{\J_{A_1}\J_{A_2} \pi}{2 \J_d} \Big)-\frac{4\J_{A_1}\J_{A_2}-\J_m^2}{8 \J_d}\Big(\frac{1}{\J_{A_1}}+\frac{1}{\J_{A_2}}\Big)\\
 &-2 \Delta q^2 \Big\{\Big(\frac{-\J_{A_1}+\J_{A_2}-\J_d}{2\J_d}\Big)^2\frac{1}{\J_{A_1}}+\Big(\frac{\J_{A_1}-\J_{A_2}-\J_d}{2\J_d}\Big)^2\frac{1}{\J_{A_2}}-\frac{1}{\J_{d}}\Big\}.
 \end{split}
\end{equation}

\begin{figure}[t]
\begin{center}
\includegraphics[scale=0.35]{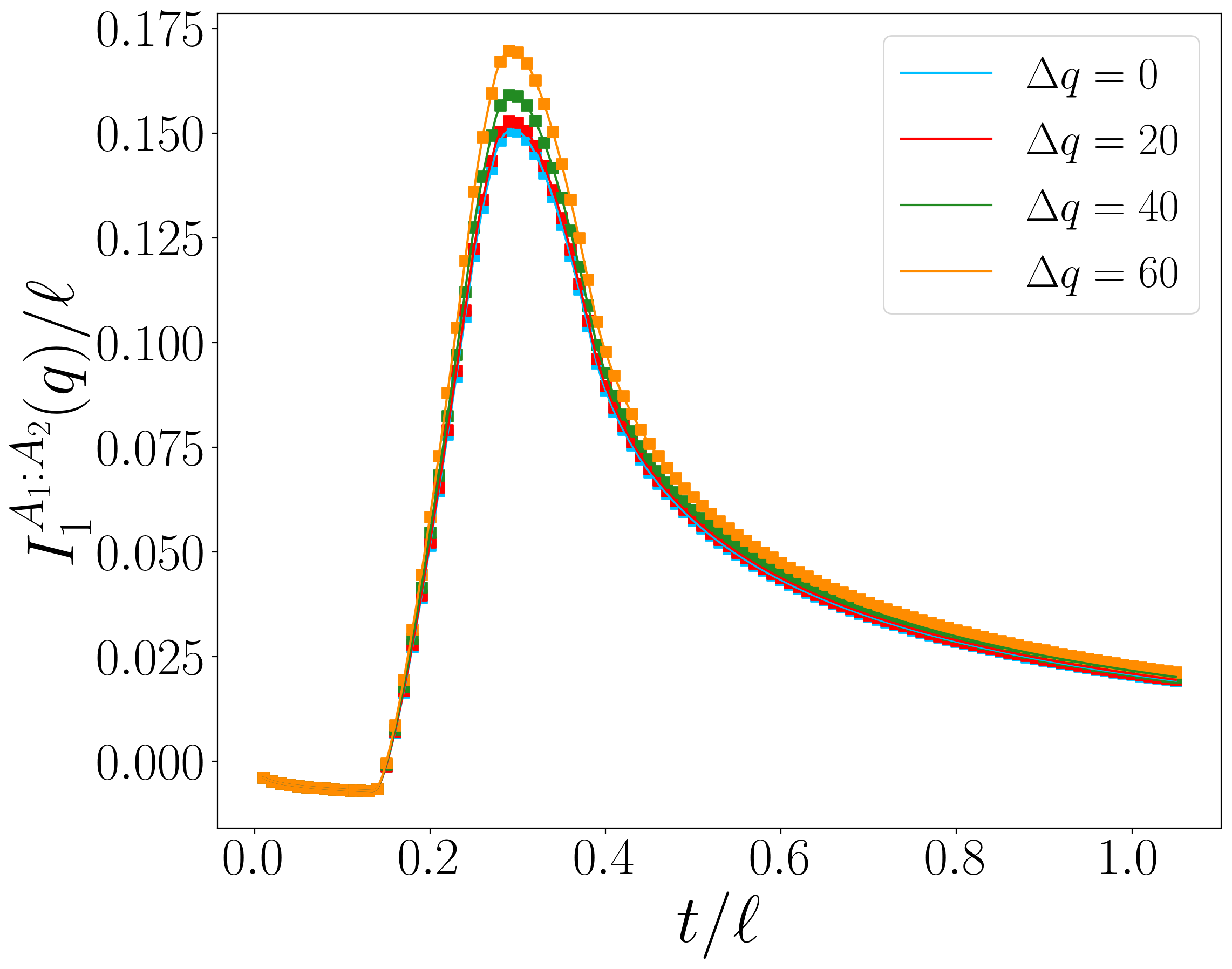}
\end{center}
\caption{Time evolution of $I^{A_1 : A_2}_1(q) $ after a quench from the N\'eel state in the tight-binding model \eqref{eq:Hfree} as a function of $t/\ell$ with $\ell_1=180$, $\ell_2=220$ and $d=220$ for various values of $\Delta q$. We compare the approximation of Eq. \eqref{eq:IqAsymNeel} (solid lines) with the exact result given by Eq. \eqref{eq:SymResMut} where the entropies are replaced by their exact formulas of the form of Eq. \eqref{eq:SnqIndep} and the weights are given by their Gaussian approximations (symbols).}
\label{fig:SRMutAsymNeel}
\end{figure}

Eq. \eqref{eq:IqAsymNeel} implies that at the leading order there is  equipartition of the symmetry resolved mutual information. 
Furthermore, for $t<d/4$ and in the limit $t \to \infty$, the subleading term vanishes, because $\J_d=\J_{A_1}+  \J_{A_2}$ in these two regimes, and the 
equipartition is broken at higher order. 
In Fig. \ref{fig:SRMutAsymNeel} we compare the prediction \eqref{eq:IqAsymNeel} (solid lines) with the ab-initio calculation of Eq. \eqref{eq:SymResMut} where the entropies 
have been replaced by their expressions in terms of Gamma functions, and the weights are replaced by their Gaussian expressions (symbols).  
We find a perfect agreement between asymptotic prediction and exact numerics. 
The slight negative values for early times arise from the term $- \frac{1}{2} \log \frac{\J_{A_1}\J_{A_2} \pi}{2 \J_d}$ in Eq.~\eqref{eq:IqAsymNeel}. 
This term is at most of order $\log \ell$ and is negligible in the scaling limit. For intermediate times, the equipartition is broken at order $\Delta q^2/\ell$.

The leading term in Eq. \eqref{eq:IqAsymNeel} is the total mutual information, 
\begin{equation}
\label{eq:ItotNeel}
\begin{split}
     I^{A_1:A_2}_1&= (\J_{A_1}+  \J_{A_2}-\J_d) \log 2\\
     &=\log 2\int \frac{\dd k}{2 \pi} [\max(d, 2v_kt)+ \max(d+\ell, 2v_kt)- \max(d+\ell_1, 2v_k t)- \max(d+\ell_2, 2v_k t)].
     \end{split}
\end{equation}
This immediately follow from the definition of Eq. \eqref{MutualInfo} and the exact results of Eqs. \eqref{eq:SnqIndep} and \eqref{eq:Z1SnAsymptd} for the entanglement entropies. This integral is precisely the form predicted by the quasiparticle picture for the spreading of entanglement between two non-complementary subsystems \cite{ac-19}. It is vanishing for times $t<d/4$ as well as in the large-time limit. To verify the consistency of our calculations, we must recover the total mutual information from the symmetry resolved ones with the combination given in Eq. \eqref{eq:IqTot}. Using the same Gaussian results as before, Eq. \eqref{eq:IqAsymNeel} and the asymptotic result for the number entropy of Eq. \eqref{eq:SnumApproxNeel}, we indeed find
\begin{equation}
\label{eq:eq:IqAsymNeelTot}
     \sum_{q=0}^{\ell} \mathcal{Z}_1^{A_1\cup A_2}(q) I^{A_1 : A_2}_1(q) + S^{A_1,n}+S^{A_2,n}-S^{A_1\cup A_2,n}= (\J_{A_1}+  \J_{A_2}-\J_d) \log 2.
\end{equation}

\section{Quench from the dimer state}
\label{Sect5}

For the quench from the dimer state $|D\rangle$ (cf. Eq. \eqref{eq:InitialStates}) in the tight binding model, the starting point is also the correlation matrix.
For this quench, it is convenient to write it in a block form as follows (see e.g. \cite{f-14}):
\begin{equation}
\label{eq:CtDimer}
    [C(t)]_{j,k} = \Big\langle \begin{pmatrix}c_{2j-1}^\dagger \\ c_{2j}^\dagger \end{pmatrix} \begin{pmatrix}c_{2k-1} & c_{2k} \end{pmatrix} \Big\rangle = \frac 12 \big( \delta_{j,k} \id_2 + \Pi_{k-j} \big)
\end{equation}
where $\Pi_m$ is
\begin{equation}
    \Pi_{m} =  \int_{-\pi}^{\pi}\frac{\dd k}{2 \pi}\eE^{-2\ir m k} \ \begin{pmatrix}
-f(k,t)&-g(k,t)\\
 -g(k,t)^*&f(k,t) \\
\end{pmatrix}
\end{equation}
with
\begin{equation}
\begin{split}
f(k,t) &= \sin k \sin(4 \cos(k) t), \\
g(k,t) &= \eE^{-\ir k}(\cos k + \ir \sin k \cos(4 \cos(k) t)).
\end{split}
\end{equation}

\subsection{Single interval}
Similarly to the N\'eel quench, we start by considering the case where $A$ is a block of $\ell$ consecutive sites. The correlation matrix $C_A(t)=\frac 12 (\id_\ell + J_A(t))$ is the $\ell\times \ell$ matrix whose entries are $[C(t)]_{j,k}$ given in Eq.~\eqref{eq:CtDimer} with $j,k \in \{1,\dots,\ell/2\}$.

\subsubsection{Charged moments}\label{sec:znaDimer}
\begin{figure}[t]
\begin{center}
\includegraphics[scale=0.24]{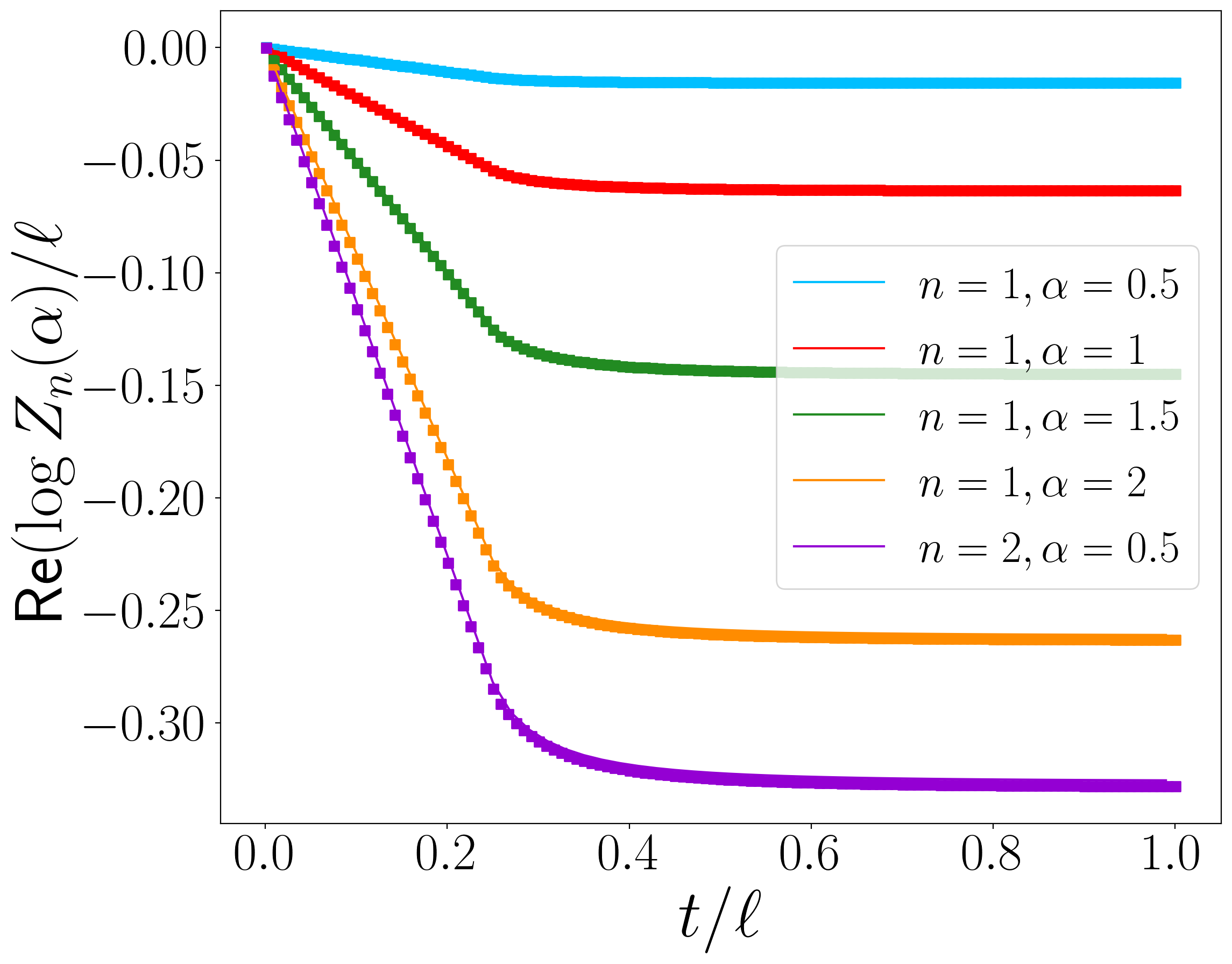}\quad
\includegraphics[scale=0.24]{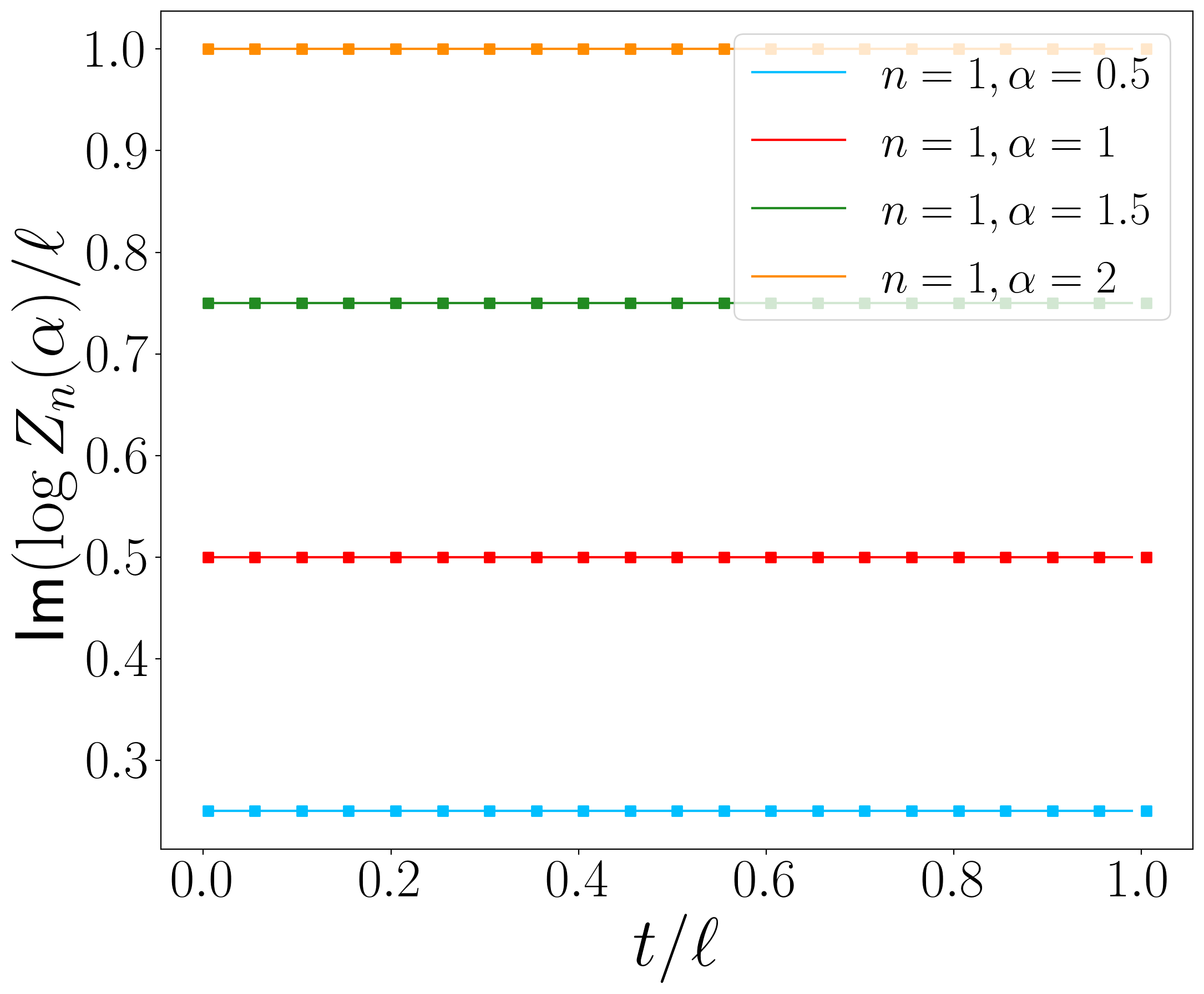} \\
\includegraphics[scale=0.24]{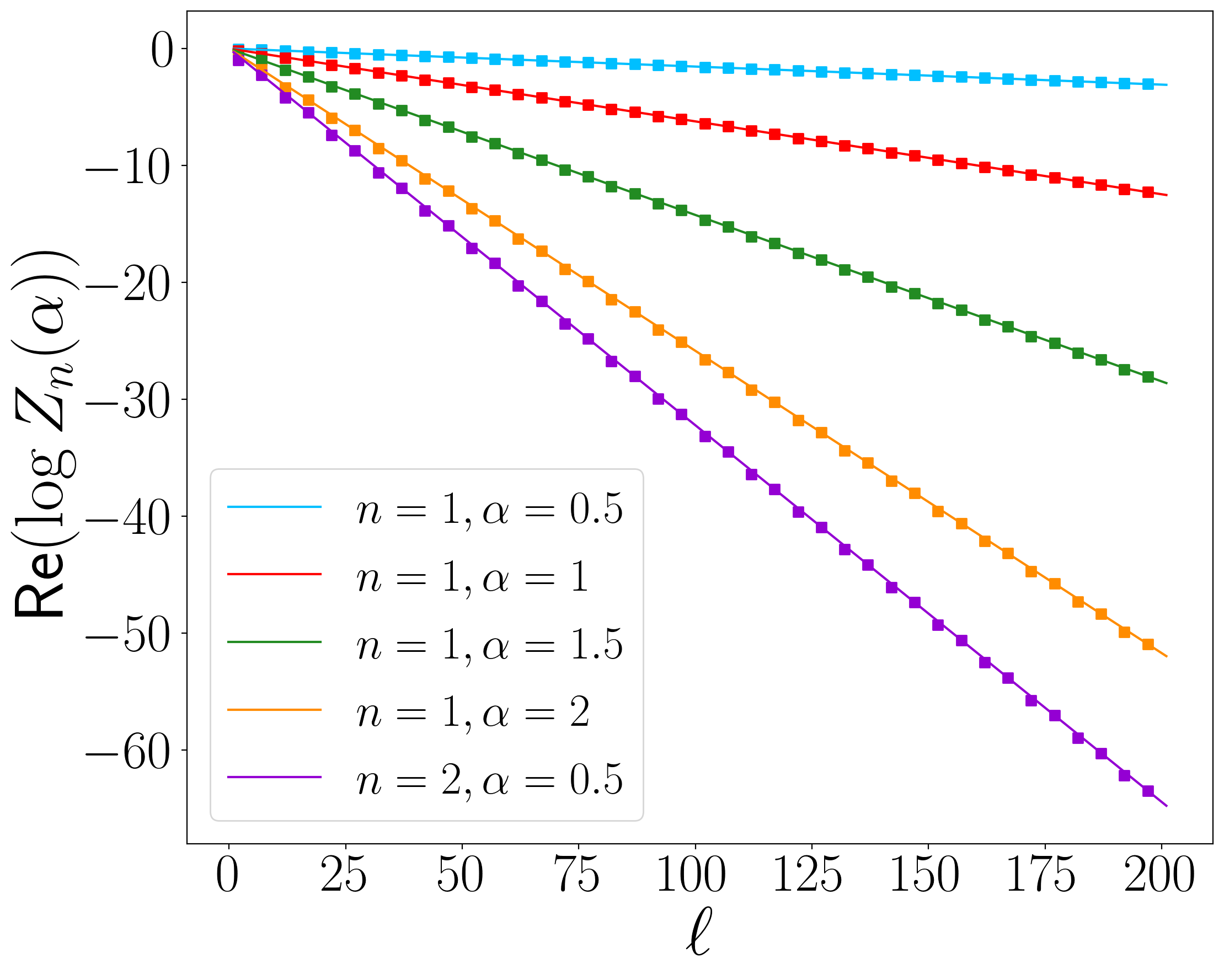} \quad 
\includegraphics[scale=0.24]{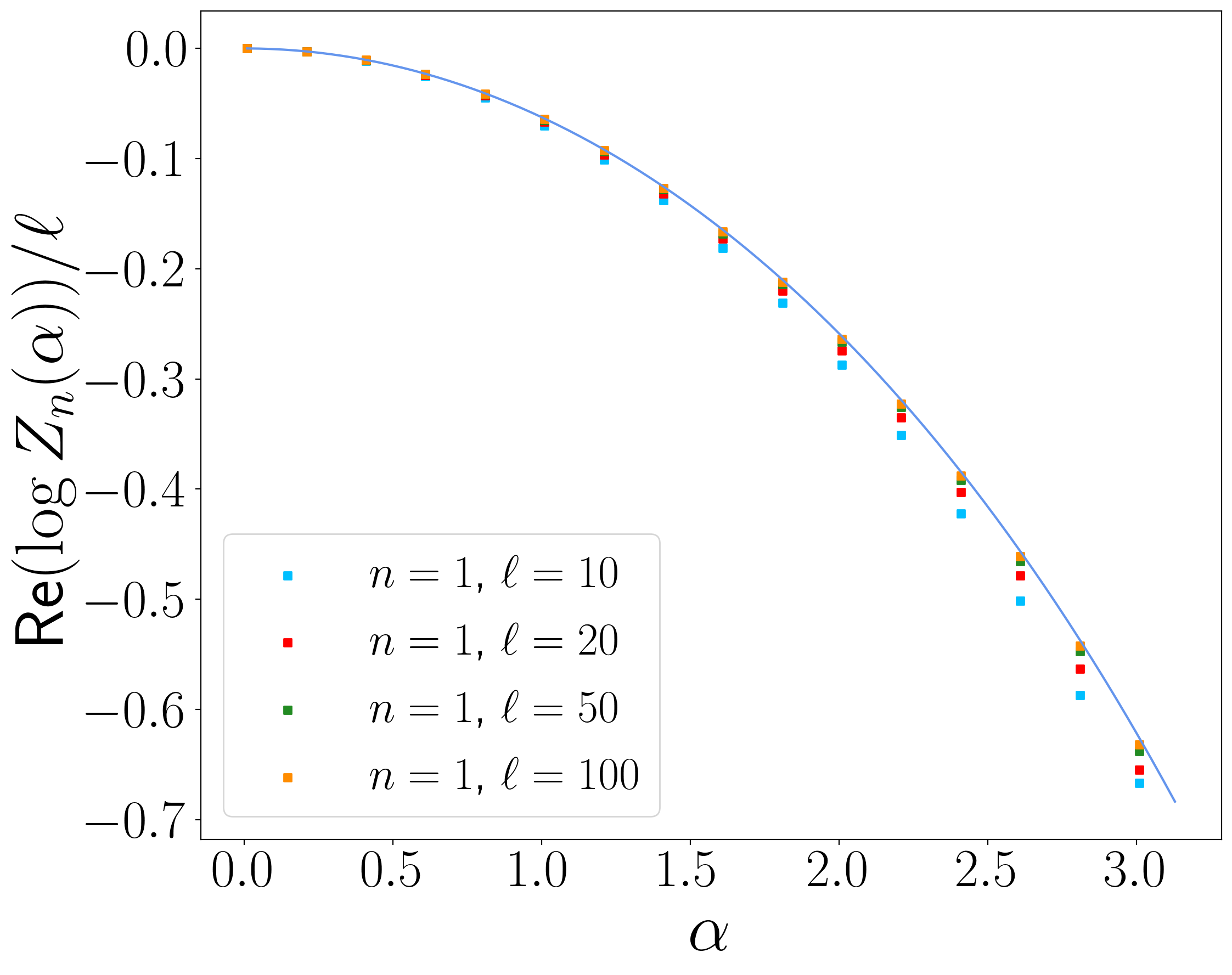}
\end{center}
\caption{Comparison between the exact result \eqref{eq:logZnaExactDimer} (solid lines) and numerical results (symbols) for $\log Z_n(\alpha)$ after a quench from the dimer state in the free fermion model \eqref{eq:Hfree}. \textit{Top left:} Time evolution of the real part of $\log Z_n(\alpha)$ as a function of $t/\ell$ with $\ell=120$. \textit{Top right}: Time evolution of the imaginary part of $\log Z_n(\alpha)$ as a function of $t/\ell$ with $\ell=20$ and $n=1$. \textit{Bottom left:} Real part of $\log Z_n(\alpha)$ as a function of $\ell$ with $t/\ell=0.5$. \textit{Bottom right:} Real part of $\log Z_n(\alpha)$ as a function of $\alpha$ with $t/\ell=0.5$ and $n=1$.
}
\label{fig:ZnalphaDimer}
\end{figure}

We compute the trace of $J_A(t)^m$ in the scaling limit where $t,\ell~\to~\infty$ with a fixed finite ratio. The computation is exactly the same as in \cite{fc-08} and we find
\begin{equation}
\label{eq:TrJDimerd0}
\Tr J_{A}(t)^{2j} = \ell - \int \frac{\dd k}{2 \pi}(1-[\cos k]^{2j})\min(\ell,2 v_k t), \quad \Tr J_{A}(t)^{2j+1} =0
\end{equation}
with $v_k = 2 |\sin k|$. As for the quench from the N\'eel state, the maximal velocity is $v_M=2$. We evaluate the sum \eqref{eq:chargedZTrK} with \eqref{eq:TrJDimerd0} and find 
\begin{equation}
\label{eq:logZnaExactDimer}
\log Z_n(\alpha) = \ir \ell \frac{\alpha}{2}+\int \frac{\dd k}{2 \pi}\mathrm{Re}[h_{n,\alpha}(\cos k)]\min(\ell,2 v_k t)
\end{equation}
where $h_{n,\alpha}(x)$ is given in Eq. \eqref{eq:hna}. This result is very similar to Eq. \eqref{eq:logZnaExactNeel} for the N\'eel quench, with the difference that the function $h_{n,\alpha}(x)$ is now evaluated in $x=\cos k$ and hence cannot be factorised out of the integral. 
In Fig. \ref{fig:ZnalphaDimer} we compare this prediction with the exact numerical data for the charged moments and find a very good agreement.  
As observed for the N\'eel quench and in many other cases, the corrections to the scaling are larger as $\alpha$ gets closer to  $\pm \pi$ and as $n$ increases. 

\subsubsection{Fourier transform and symmetry resolved R\'enyi entropies}\label{sec:FTandSRSn}

The symmetry resolved moments $\mathcal{Z}_n(q)$ with $q=\Delta q+ \langle Q_A \rangle$ are defined in Eq. \eqref{eq:ZnqFT}. With Eq. \eqref{eq:logZnaExactDimer} and $\langle Q_A \rangle=\ell /2$, we have
\begin{equation}
\label{eq:ZnqIntDimer}
\mathcal{Z}_n(q) = \int \frac{\dd \alpha}{2 \pi}\eE^{-\ir \alpha \Delta q+\int \frac{\dd k}{2 \pi}\mathrm{Re}[h_{n,\alpha}(\cos k)]\min(\ell,2 v_k t)}.
\end{equation}
Contrarily to the N\'eel quench, there is no closed-form expression similar to Eq. \eqref{eq:Z1Gamma} for this integral. 
We can nonetheless infer the validity of our two main results, namely the existence of a time delay that grows linearly in $|\Delta q|$ for small values of $|\Delta q|$, and an effective equipartition of entanglement broken at order $(\Delta q)^2/\ell$. 

For the delay, we analyse Eq. \eqref{eq:ZnqIntDimer} with the saddle point approximation in the regime where $2 v_M t < \ell$. The saddle point equation  is
\begin{equation}
\label{eq:SaddlePointDimer}
- \ir \Delta q + \ir t \int \frac{\dd k}{2 \pi} \partial_{\ir \alpha}\big( \mathrm{Re}[h_{n,\alpha}(\cos k)]\big) 2 v_k=0
\end{equation} 
and admits a solution for imaginary $\alpha$ only. In that case, the integral $\int \frac{\dd k}{2 \pi} \partial_{\ir \alpha}\big( \mathrm{Re}[h_{n,\alpha}(\cos k)]\big) 2 v_k$ is a monotonic function of $\ir \alpha$, going from $4/\pi$ at $\alpha = -\ir \infty$ to $-4/\pi$ at $\alpha = \ir \infty$. It thus follows that Eq. \eqref{eq:SaddlePointDimer} only admits a solution for $t> \frac{\pi |\Delta q|}{4}$. Hence we find precisely the same delay as for the N\'eel quench, with the same self-consistency relation $|\Delta q| < \ell /\pi$, as given in Eq. \eqref{eq:tDNeel}.

\begin{figure}[t]
\begin{center}
\includegraphics[width=0.45\textwidth]{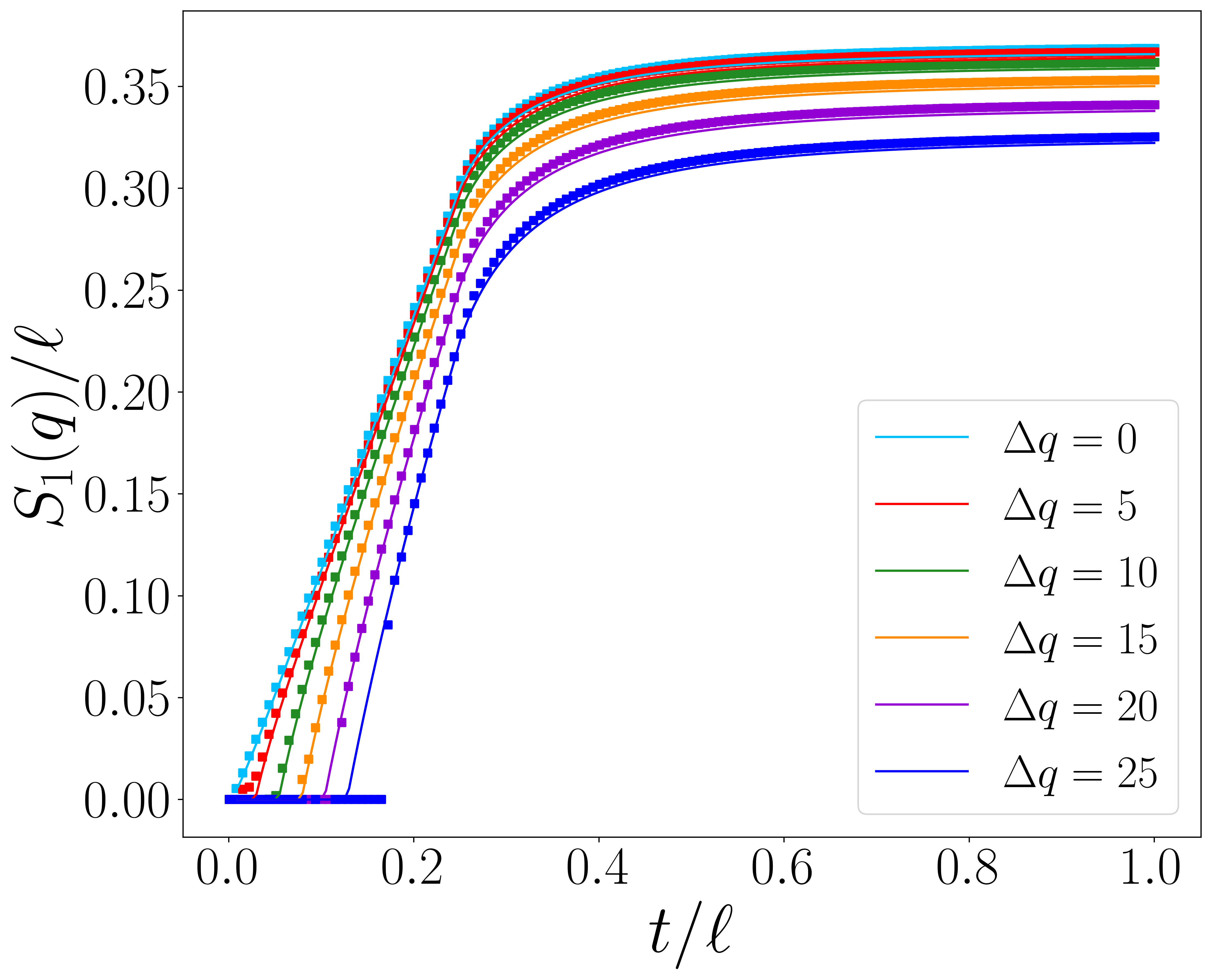} 
\includegraphics[width=0.45\textwidth]{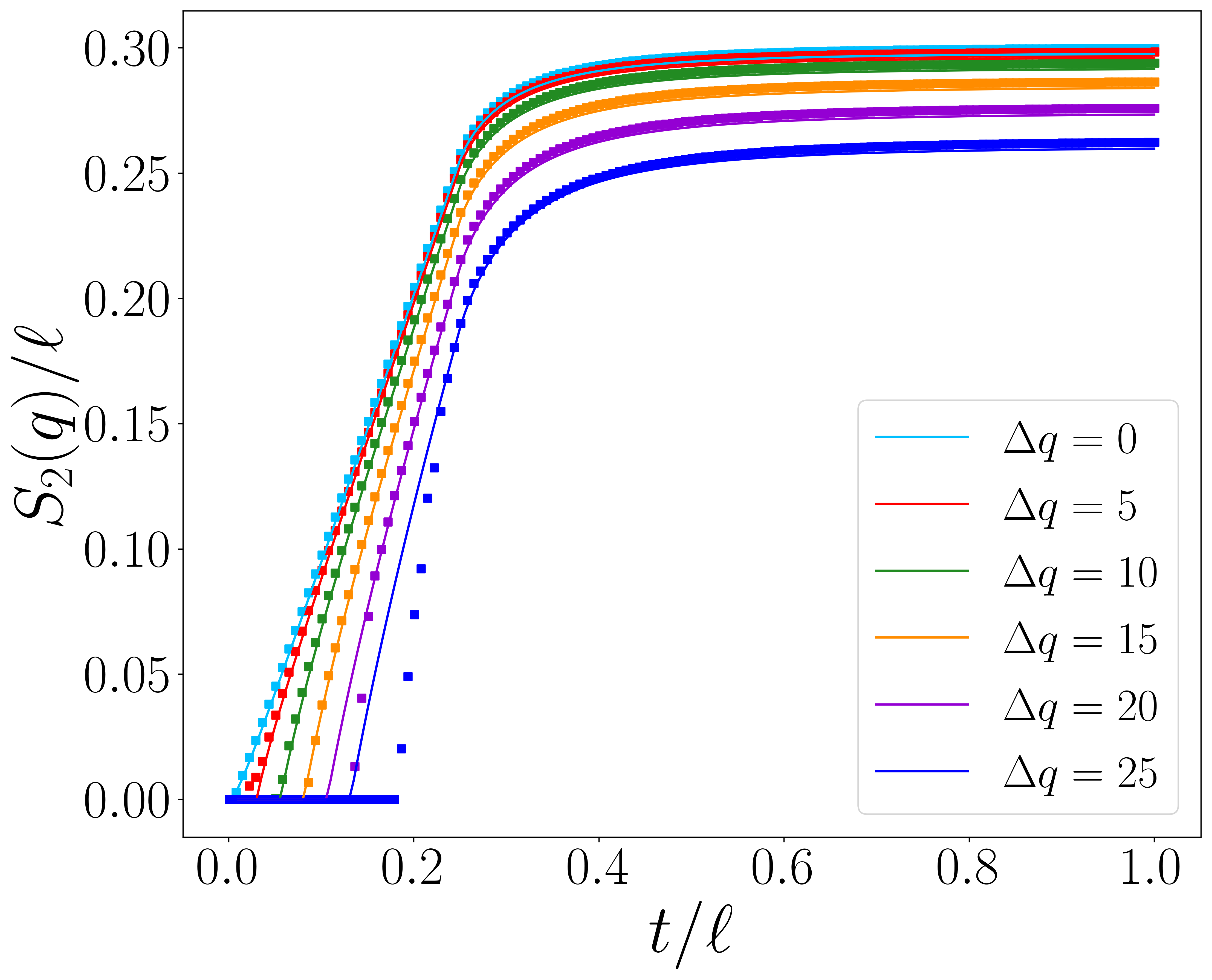} 
\includegraphics[width=0.45\textwidth]{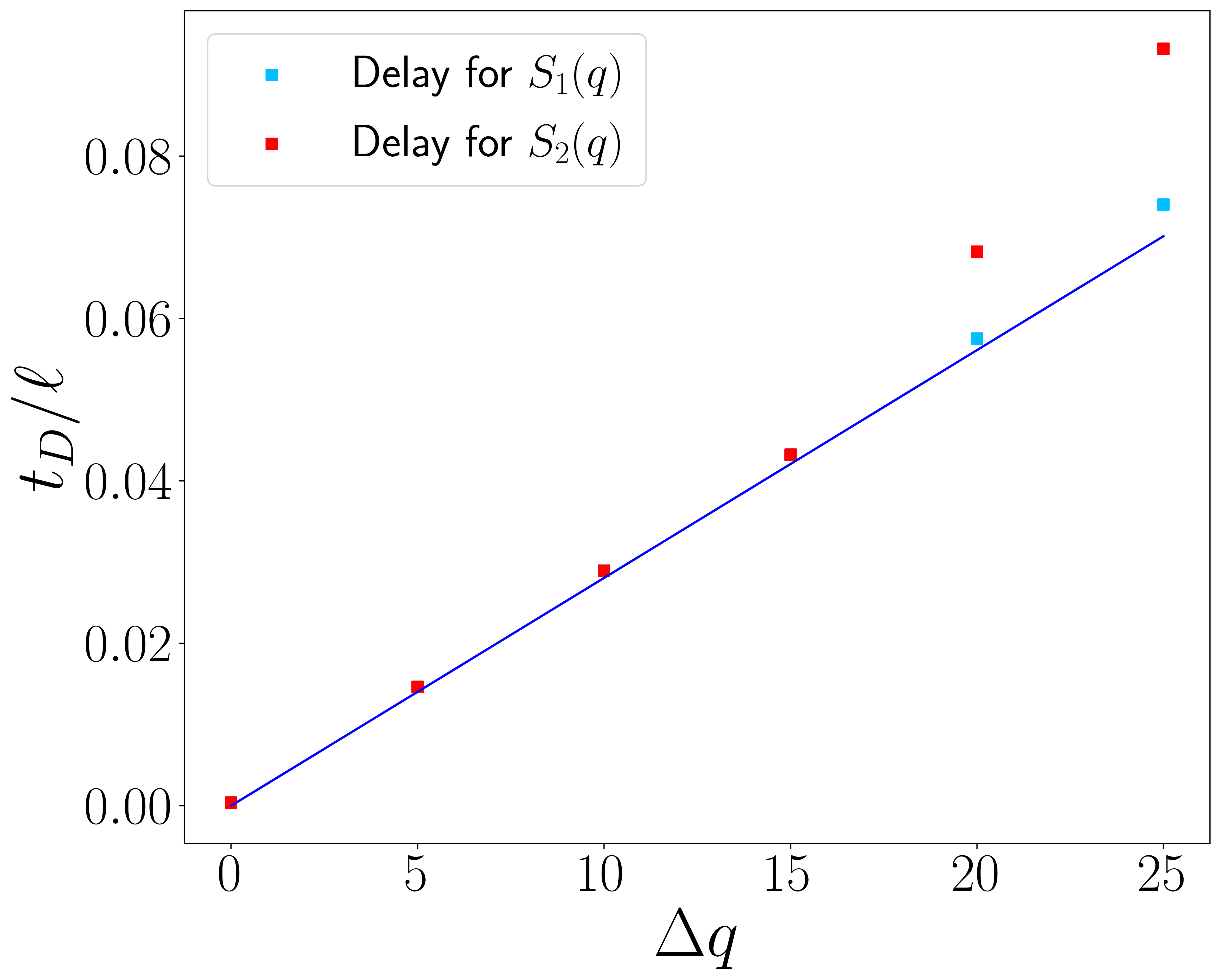} 
\end{center}
\caption{\textit{Top:} Time evolution of the symmetry resolved entropies $S_1(q)$ and $S_2(q)$ after a quench from the dimer state in the tight-binding model \eqref{eq:Hfree} for $\ell=140$ and various $\Delta q$. We compare the predictions of Eqs. \eqref{eq:SnqDimerApprox} and \eqref{eq:S1qDimerApprox} (solid lines) with numerical data (symbols). \textit{Bottom:} Delay time $t_D$ for various $\Delta q$ and $\ell=280.$ The solid line is the analytic prediction $t_D= \frac{\pi |\Delta q|}{4}$ and the symbols are the numerical results. They are obtained as the time for which $S_{1,2}(q)/\ell=0.01$.} 
\label{Fig:S2qDimer}
\end{figure}

For the equipartition, we expand the charged moments at quadratic order in $\alpha$, 
\begin{equation}
 \log Z_n(\alpha) =  \ir \ell \frac{\alpha}{2}+ \log Z_n(0) -\frac{\alpha^2}{8} \J^{(n)}
\end{equation}
with
\begin{equation}
\label{eq:JnDimer}
\J^{(n)} =  \int  \frac{\dd k}{2 \pi} j_n(k) \ \text{min}\big(\ell,2 v_k t \big)
\end{equation}
where 
\begin{equation}
\label{eq:jnk}
j_n(k)= \frac{4 \sin^{2n} k}{((1+\cos k)^n+(1-\cos k)^n)^2}.
\end{equation}
In the following, we also need the value of the derivative of $j_n(k)$ with respect to $n$ at $n=1$. It reads
\begin{equation}
(\partial_n j_n(k))|_{n=1}=[-(1-\cos k) \log(1-\cos k)-(1+\cos k) \log(1+\cos k)+2\log \sin k] \sin^2 k.
\end{equation}
These $\J^{(n)}$ integrals generalise $\J$ of Eq. \eqref{eq:JNeel}. In particular, we have $\J^{(0)}=\J$. With this quadratic expansion, we compute the Fourier transform and find
\begin{equation}
\label{eq:ZnqGaussDimer}
\mathcal{Z}_n(q) = Z_n(0)\eE^{-\frac{2\Delta q^2}{\J^{(n)}}}\sqrt{\frac{2}{\pi \J^{(n)}}}.
\end{equation}
Plugging this result in Eq.~\eqref{SvsZ}, we find at leading order
\begin{equation}
\label{eq:SnqDimer}
S_n(q) = S_n -\frac{2\Delta q^2}{(1-n)} \Big\{\frac{1}{\J^{(n)}}-\frac{n}{\J^{(1)}} \Big\}
\end{equation}
where we used that $\log Z_1(0)=0$ and the total entropy is $S_n = \frac{1}{1-n}\log Z_n(0)$. The symmetry resolved entanglement entropy is obtained as the limit $n \to 1$ of Eq. \eqref{eq:SnqDimer}, and reads
\begin{equation}
\label{eq:S1qDimer}
S_1(q) = S_1 -\frac{2\Delta q^2}{\J^{(1)}}\Big\{\frac{(\partial_n\J^{(n)})|_{n=1}}{\J^{(1)}}+1 \Big\}
\end{equation}
where $S_1 = -(\partial_n Z_n(0))|_{n=1}$. Equations \eqref{eq:SnqDimer} and \eqref{eq:S1qDimer} are physically equivalent to Eq. \eqref{eq:SqCorr}. Similarly to the case of a quench from the N\'eel state, the match with the numerical data is better when we include the sub-leading terms that arise from the square root in Eq. \eqref{eq:ZnqGaussDimer}. The result reads
\begin{equation}
\label{eq:SnqDimerApprox}
S_n(q) = S_n -\frac{2\Delta q^2}{(1-n)} \Big\{\frac{1}{\J^{(n)}}-\frac{n}{\J^{(1)}} \Big\}+\frac{1}{2} \log \frac{2}{\pi}-\frac{1}{2(1-n)}\log \frac{\J^{(n)}}{(\J^{(1)})^n}
\end{equation}
and the limit $n\to 1$ is 
\begin{equation}
\label{eq:S1qDimerApprox}
S_1(q) =S_1 -\frac{2\Delta q^2}{\J^{(1)}}\Big\{\frac{(\partial_n\J^{(n)})|_{n=1}}{\J^{(1)}}+1 \Big\}+\frac{1}{2} \log \frac{2}{\pi}+\frac{1}{2}\Big(\frac{(\partial_n\J^{(n)})|_{n=1}}{\J^{(1)}}-\log \J^{(1)} \Big).
\end{equation}
We compare these predictions with ab-initio results in Fig. \ref{Fig:S2qDimer}. As expected, the delay is not well reproduced by the approximation for large $|\Delta q|$, but the asymptotic values match convincingly.
The reason of this disagreement at intermediate times is the same as for the N\'eel quench and is not discussed again here.
 In the same figure we report the delays for $S_1(q)$ and $S_2(q)$ and compare them with the prediction $t_D=\frac{\pi |\Delta q|}{4}$. For $S_1(q)$, the match is perfect. We expect that finite-size effects play an increasing role for larger values of $n$ and $|\Delta q|$, and indeed the match is less precise for $S_2(q)$ at large $|\Delta q|$.
 
\subsubsection{Total and number entropy}
The total entropies are given by $S_n = \frac{1}{1-n}\log Z_n(0)$. With Eq. \eqref{eq:logZnaExactDimer} we have 
\begin{equation}
\label{eq:SntotDimer}
S_n = \frac{1}{1-n} \int \frac{\dd k}{2 \pi}h_{n,0}(\cos k)\min(\ell,2 v_k t)
\end{equation}
where 
\begin{equation}
h_{n,0}(\cos k) =  \log \left[ \left(\frac{1+\cos k}{2} \right)^n + \left( \frac{1 - \cos k}{2}\right)^n \right]
\end{equation}
is a specialisation of Eq. \eqref{eq:hna}. The term $n_k=\frac{1+\cos k}{2}$ is the mode occupation in the steady state. The limit $n\to 1$ yields
\begin{equation}
S_1 =\int \frac{\dd k}{2 \pi} (-n_k \log n_k -(1-n_k)\log(1-n_k))\min(\ell,2 v_k t)
\end{equation}
that matches the quasiparticle prediction \cite{ac-17,ac-18}. As in the case of a quench from the N\'eel state, we recover the total entanglement entropy $S_1$ from the symmetry resolved quantities using Eq. \eqref{decompositionSvN}. To see this, we plug the Gaussian approximations  \eqref{eq:ZnqGaussDimer} and \eqref{eq:S1qDimerApprox} into Eq. \eqref{decompositionSvN} and find
\begin{equation}
S_1= -(\partial_n Z_n(0))|_{n=1} + \frac{(\partial_n\J^{(n)})|_{n=1}}{\J^{(1)}} \Big(\frac 12 - \sqrt{\frac{2}{\pi \J^{(1)}}} \sum_{q=0}^\ell \frac{2\Delta q^2}{\J^{(1)}} \eE^{-\frac{2\Delta q^2}{\J^{(1)}}}  \Big)
\end{equation}
where we also used $\sum_q \mathcal{Z}_1(q)=1$. In the large-$\ell$ limit, the sum in the right-hand side becomes a Gaussian integral. The whole parenthesis then vanishes and we recover $S_1= -(\partial_n Z_n(0))|_{n=1}$, as desired.

The number entropy is given in Eq. \eqref{eq:SnumNeel} and only depends on $\mathcal{Z}_1(q)$. We use the quadratic approximation \eqref{eq:ZnqGaussDimer} 
and find that the calculations are exactly the same as in Sec. \ref{sec:totSnNeel} where $\J$ is replaced by $\J^{(1)} = \int \frac{\dd k}{2 \pi}(1-\cos^2k)\min(\ell,2 v_k t).$ We thus find 
\begin{equation}
    \label{eq:SnumApproxDimer}
    S^n = \frac 12 \Big(1+\log \frac{\J^{(1)} \pi }{2} \Big).
\end{equation}

\begin{figure}
\begin{center}
\includegraphics[width=0.45\textwidth]{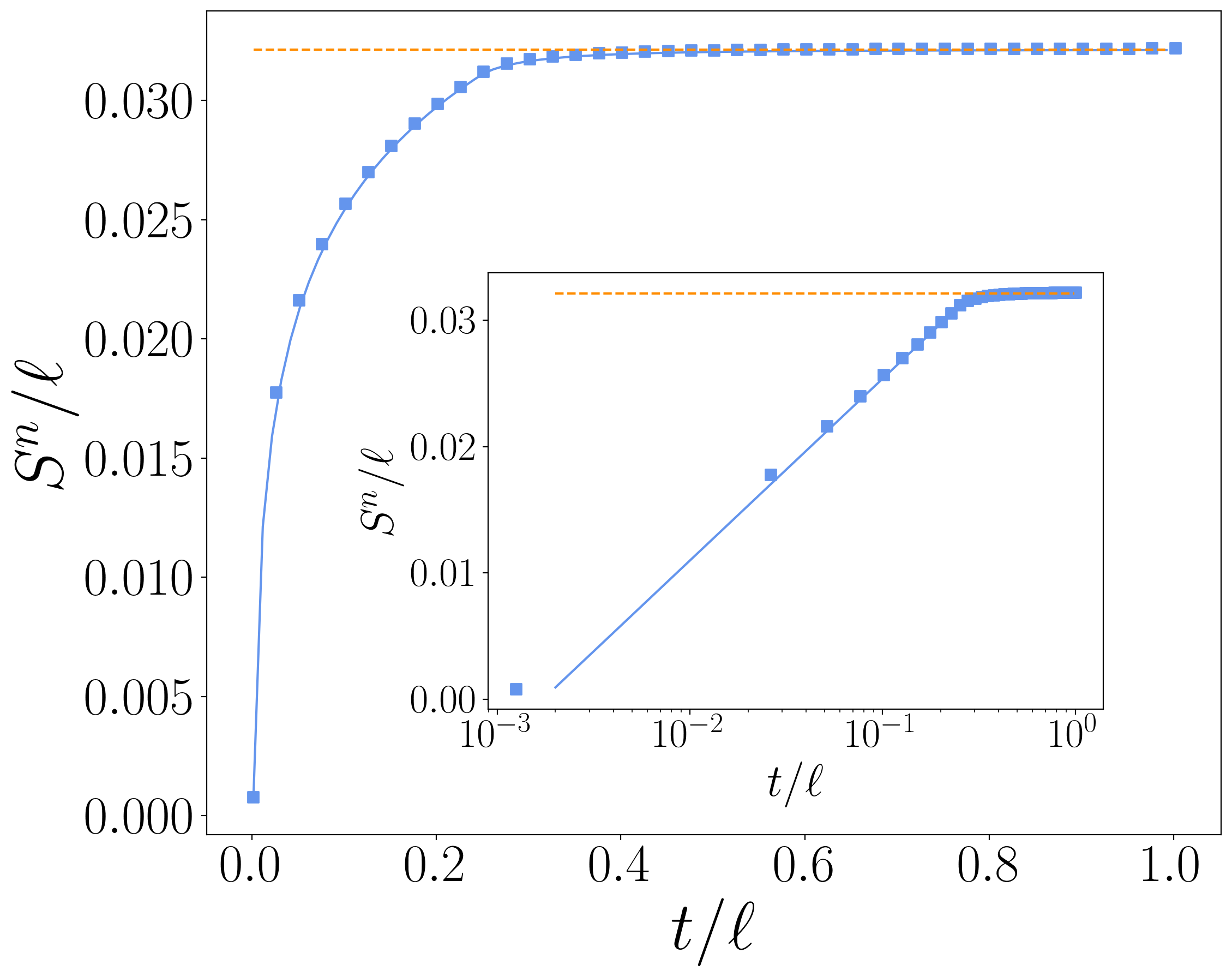} 
\includegraphics[width=0.45\textwidth]{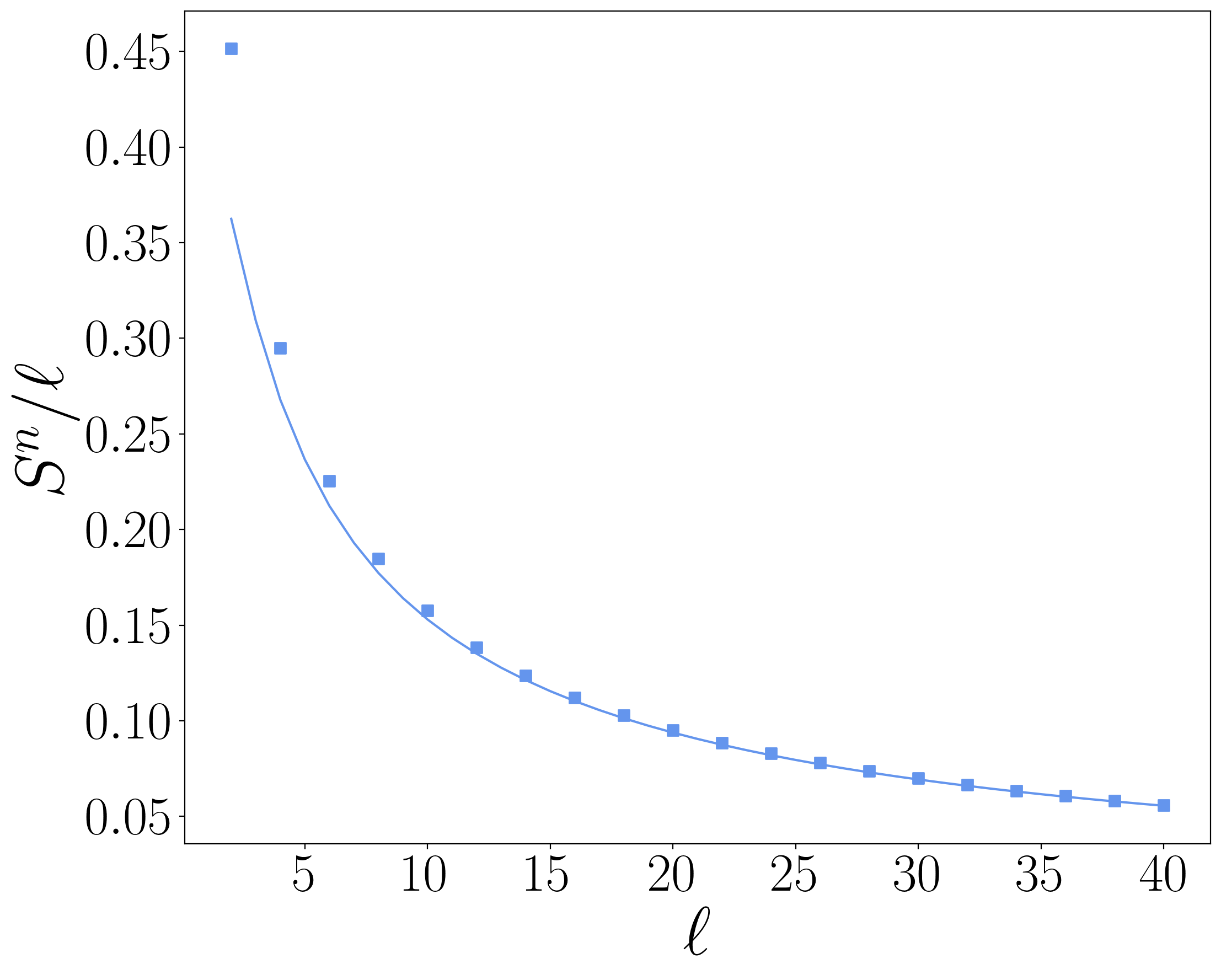} 
\end{center}
\caption{Evolution of $S^n$ after a quench from the dimer state in the tight-binding model \eqref{eq:Hfree} as a function of $t/\ell$ for $\ell=80$ (left panel), and as a function of $\ell$ with $t/\ell=1$ (right panel). The prediction \eqref{eq:SnumApproxDimer} (solid line) perfectly matches the numerical results (symbols). In the left panel, the horizontal dotted line is the asymptotic value $\frac {1}{2 \ell} \Big(1+\log \frac{\ell \pi }{4}\Big)$ for large times. In the same panel, the inset shows $S^n$ as a function of $t/\ell$ in a log-linear scale.}
\label{Fig:SnumDimer}
\end{figure}

In Fig. \ref{Fig:SnumDimer} we compare this result (solid lines) with ab-initio calculations of $S^n$ (symbols), and find perfect agreement. The inset in the left panel shows the time-evolution of $S^n$ in a log-linear scale. We note that $\J^{(1)}$ saturates to the value $\ell/2$ for large times, so that $\displaystyle \lim_{t\to \infty}S^n =\frac 12 \Big(1+\log \frac{\ell \pi }{4} \Big)$. This asymptotic value is reported as a dotted horizontal line in the left panel of Fig. \ref{Fig:SnumDimer}. Similarly to the quench from the N\'eel state, the number entropy grows logarithmically with time before saturation, and has negligible contribution to the total entropy in the large-time limit, $\lim_{t\to\infty}S^n/S_1 =\mathcal{O}(\ell^{-1}\log \ell)$. 

\subsection{Disjoint intervals}
We now turn to the case where the system $A$ is composed of two disjoint subsystems $A_1$ and $A_2$ of lengths $\ell_1$ and $\ell_2=\ell-\ell_1$ and separated by a distance $d$. The correlation matrix has the form of Eqs.~\eqref{eq:CADisjoin1} and \eqref{eq:CADisjoin2} with the entries given in Eq. \eqref{eq:CtDimer}.

\subsubsection{Charged moments}\label{sec:ZnadDimer}
\begin{figure}
\begin{center}
\includegraphics[scale=0.3]{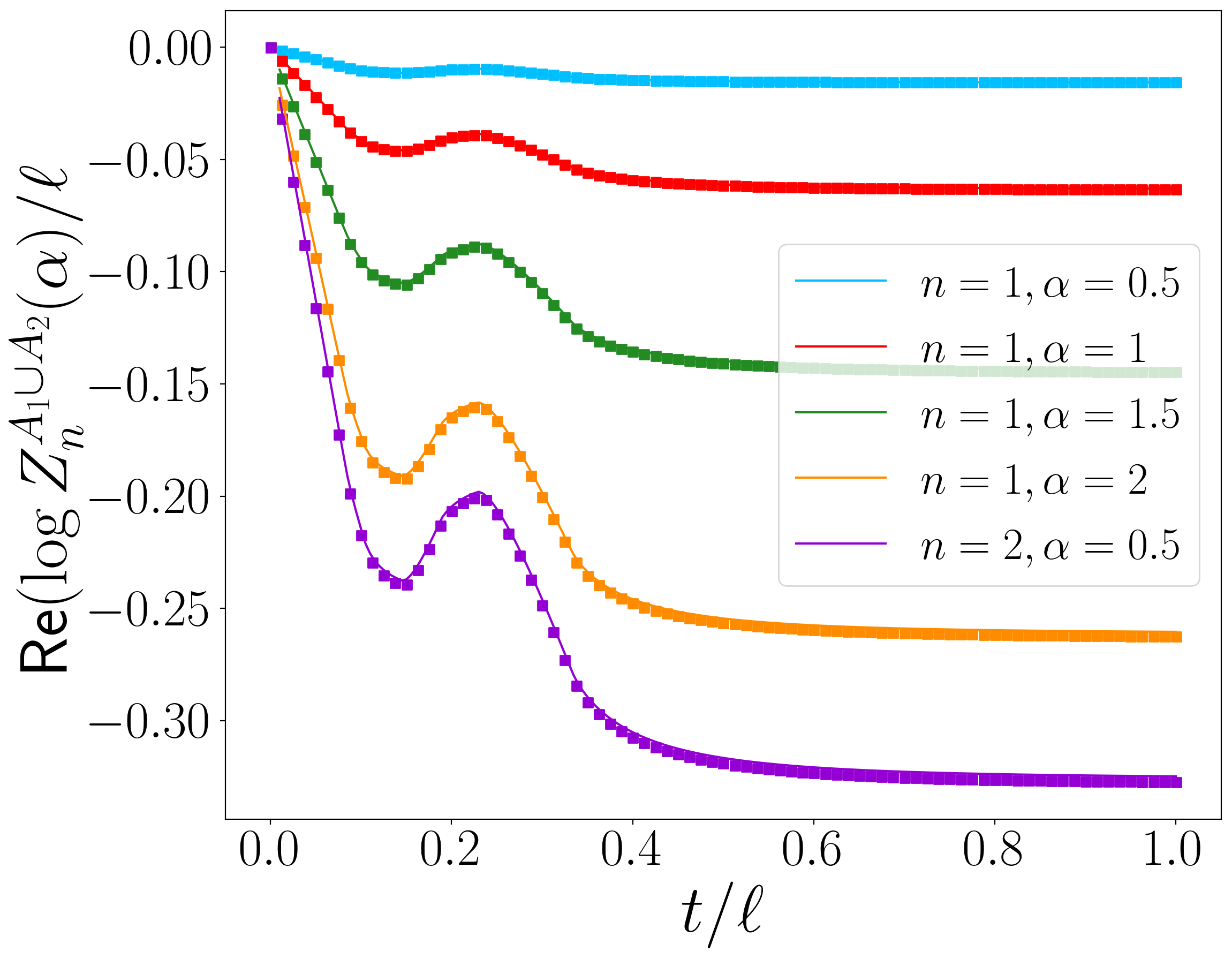}
\end{center}
\caption{Time evolution of the charged moments $Z_n^{A_1\cup A_2}(\alpha)$ after a quench from the dimer state in the tight-binding model \eqref{eq:Hfree} as a function of $t/\ell$ with $\ell_1=100$, $\ell_2=140$ and $d=80$. The analytical prediction of Eq.~\eqref{eq:logZnadExactDimer} (solid lines) perfectly matches the numerical data (symbols).}
\label{fig:ZnalphadtDimer}
\end{figure}

Similarly to the calculations of Sec. \ref{sec:ZnadNeel}, we compute the trace of arbitrary powers of the matrix $J_{A_1\cup A_2}(t) = 2 C_{A_1\cup A_2}(t)- \id_\ell$. We perform the computation of $\Tr J_{A_1\cup A_2}(t)^{m}$ in the scaling limit $t,\ell, \ell_1,\ell_2,d \to \infty$ where the various ratios are kept constant. We find
\begin{equation}
\label{eq:TrJdDimer}
\begin{split}
\Tr &J_{A_1\cup A_2}(t)^{2j}= \ell  -  \int \frac{\dd k}{2 \pi}(1-[\cos k]^{2j})[\min(\ell_1,2 v_k t)+\min(\ell_2,2 v_k t)] \\
&+\int \frac{\dd k}{2 \pi}(1-[\cos k]^{2j}) [\max(d, 2v_kt)+ \max(d+\ell, 2v_kt)- \max(d+\ell_1, 2v_k t)- \max(d+\ell_2, 2v_k t)],\\[.35cm]
\Tr &J_{A_1\cup A_2}(t)^{2j+1}=0,
\end{split}
\end{equation}
with $v_k = 2 |\sin k|$. We note that the case $d=0$ reduces to the known result \eqref{eq:TrJDimerd0} for a single interval. The calculation is similar to case of a single interval as presented in Sec. \ref{sec:znaDimer}, and we find 
\begin{equation}
\label{eq:logZnadExactDimer}
\begin{split}
\log Z&^{A_1 \cup A_2}_n(\alpha) = \ir \ell \frac{\alpha}{2}+\int \frac{\dd k}{2 \pi}\mathrm{Re}[h_{n,\alpha}(\cos k)](\min(\ell_1,2 v_k t)+\min(\ell_2,2 v_k t)) \\
&-\int \frac{\dd k}{2 \pi}\mathrm{Re}[h_{n,\alpha}(\cos k)](\max(d, 2v_kt)+ \max(d+\ell, 2v_kt)- \max(d+\ell_1, 2v_k t)- \max(d+\ell_2, 2v_k t)).
\end{split}
\end{equation}
We compare the prediction of Eq. \eqref{eq:logZnadExactDimer} for $Z_n^{A_1\cup A_2}(\alpha)$ with ab-initio calculations in Fig. \ref{fig:ZnalphadtDimer} and find perfect agreement. 

\subsubsection{Fourier transform and symmetry resolved R\'enyi entropies}
In order to perform analytical calculations, we approximate $\log Z^{A_1 \cup A_2}_n(\alpha)$ at quadratic order in $\alpha$, 
\begin{equation}
 \log Z^{A_1 \cup A_2}_n(\alpha) =  \ir \ell \frac{\alpha}{2}+ \log Z^{A_1 \cup A_2}_n(0) -\frac{\alpha^2}{8} \J_d^{(n)}
\end{equation}
with
\begin{equation}
\label{eq:JdnDimer}
\begin{split}
&\J_d^{(n)} =  \int  \frac{\dd k}{2 \pi} j_n(k)(\min(\ell_1,2 v_k t)+\min(\ell_2,2 v_k t))\\
&- \int  \frac{\dd k}{2 \pi} j_n(k)(\max(d, 2v_kt)+ \max(d+\ell, 2v_kt)- \max(d+\ell_1, 2v_k t)- \max(d+\ell_2, 2v_k t))
\end{split}
\end{equation}
where $j_n(k)$ is given in Eq. \eqref{eq:jnk}. As in the case of a single interval, these $\J_d^{(n)}$ integrals generalise $\J_d$ of Eq. \eqref{eq:JdNeel}. In particular, we have $\J_d^{(0)}=\J_d$. We follow Sec. \ref{sec:FTandSRSn} and find 
 \begin{equation}
\label{eq:ZnqdGaussDimer}
\mathcal{Z}^{A_1 \cup A_2}_n(q) = Z^{A_1 \cup A_2}_n(0)\eE^{-\frac{2\Delta q^2}{\J_d^{(n)}}}\sqrt{\frac{2}{\pi \J_d^{(n)}}},
\end{equation}
\begin{equation}
\label{eq:SnqdDimer}
S^{A_1 \cup A_2}_n(q) = S^{A_1 \cup A_2}_n -\frac{2\Delta q^2}{(1-n)} \Big\{\frac{1}{\J_d^{(n)}}-\frac{n}{\J_d^{(1)}} \Big\}+\frac{1}{2} \log \frac{2}{\pi}-\frac{1}{2(1-n)}\log \frac{\J_d^{(n)}}{(\J_d^{(1)})^n},
\end{equation}
for $|\Delta q| \ll \ell$. We used that $\log Z^{A_1 \cup A_2}_1(0)=0$ and the total entropy is $S^{A_1 \cup A_2}_n = \frac{1}{1-n}\log Z^{A_1 \cup A_2}_n(0)$. The symmetry resolved entanglement entropy is
\begin{equation}
\label{eq:S1qdDimer}
S_1^{A_1 \cup A_2}(q) = S^{A_1 \cup A_2}_1 -\frac{2\Delta q^2}{\J_d^{(1)}}\Big\{\frac{(\partial_n\J_d^{(n)})|_{n=1}}{\J_d^{(1)}}+1 \Big\}+\frac{1}{2} \log \frac{2}{\pi}+\frac{1}{2}\Big(\frac{(\partial_n\J_d^{(n)})|_{n=1}}{\J_d^{(1)}}-\log \J_d^{(1)} \Big).
\end{equation}

\subsubsection{Symmetry resolved mutual information}

We now turn to the computation of the symmetry resolved mutual information after a quench from the dimer state. Most of the results and computations are similar to those presented in Sec. \ref{sec:SRMutNeel} in the case of a quench from the N\'eel state. The first step is to compute $Z_1^{A_1:A_2} (\alpha,\beta)$ and hence the trace of powers of the matrix $J_{\alpha \beta} $ defined in Eq. \eqref{eq:TildeJA}. We adapt the calculations of Sec. \ref{sec:ZnadDimer} for the trace of $J_{A_1 \cup A_2}$ and find
\begin{equation}
\label{eq:TrTildeJdDimer}
\begin{split}
\Tr J_{\alpha \beta} (t)^{2j}&=\Big(\frac{\eE^{\ir \alpha}-1}{\eE^{\ir \alpha}+1}\Big)^{2j} \Big[\ell_1 -  \int \frac{\dd k}{2 \pi}(1-[\cos k]^{2j})\min(\ell_1,2 v_k t)\Big] \\
&+\Big(\frac{\eE^{\ir \beta}-1}{\eE^{\ir \beta}+1}\Big)^{2j} \Big[\ell_2-  \int \frac{\dd k}{2 \pi}(1-[\cos k]^{2j})\min(\ell_2,2 v_k t)\Big] \\
+ \int \frac{\dd k}{2 \pi} \xi(k,j)&[\max(d, 2v_kt)+ \max(d+\ell, 2v_kt)- \max(d+\ell_1, 2v_k t)- \max(d+\ell_2, 2v_k t)], \\[.5cm]
\Tr J_{\alpha \beta} (t)^{2j+1}&=0,
\end{split}
\end{equation}
where
\begin{equation}
\xi(k,j) =  \sum_{s=1}^j \sum_{t=0}^{2j-2s}(-1)^t \Big(\frac{\eE^{\ir \alpha}-1}{\eE^{\ir \alpha}+1}\Big)^{s+t} \Big(\frac{\eE^{\ir \beta}-1}{\eE^{\ir \beta}+1}\Big)^{2j-s-t} \gamma_{s,t,j}\cos k^{2j-2 s} \sin k^{2s} 
\end{equation}
and 
\begin{equation}
\label{eq:gammaSTJ}
\gamma_{s,t,j} = \frac{ s j}{(2j-s-t)(s+t)}\begin{pmatrix}
2j-s-t \\ s
\end{pmatrix} \begin{pmatrix}
s+t \\ s
\end{pmatrix}.
\end{equation}
Let us stress that the function $\gamma_{s,t,j}$ satisfies the crucial relation 
\begin{equation}
\sum_{t=0}^{2j-2s}(-1)^t \gamma_{s,t,j} = \begin{pmatrix}
j \\s
\end{pmatrix}.
\end{equation}

The re-summation in Eq. \eqref{eq:Z1abSum} yields the formal expression
\begin{equation}
\label{eq:Z1abExactDimer}
\begin{split}
\log Z_1^{A_1:A_2} (\alpha,\beta)&= \ir \frac{\ell_1 \alpha+\ell_2 \beta}{2}+\int \frac{\dd k}{2 \pi}\mathrm{Re}[h_{1,\alpha}(\cos k)]\min(\ell_1,2 v_k t) +\int \frac{\dd k}{2 \pi}\mathrm{Re}[h_{1,\beta}(\cos k)]\min(\ell_2,2 v_k t)\\
&+\int \frac{\dd k}{2 \pi}\Xi(k)(\max(d, 2v_kt)+ \max(d+\ell, 2v_kt)- \max(d+\ell_1, 2v_k t)- \max(d+\ell_2, 2v_k t))
\end{split}
\end{equation}
with the infinite sum
\begin{equation}
\Xi(k) = \sum_{j=1}^\infty \tilde c(2j)\xi(k,j),\quad \tilde c(2j)=-\frac{1}{2j},
\end{equation}
where we recall that the coefficients $\tilde c(j)$ are the Taylor coefficients of the function $\tilde h (x) = \log [ (1+x)/2]$. Despite its cumbersome form, the sum can be simplified. We find
\begin{equation}
\Xi(k) = -\frac{1}{2}\mathrm{Re}[h_{1,\alpha}(\cos k)+h_{1,\beta}(\cos k)-h_{1,\alpha-\beta}(\cos k)],
\end{equation}
and hence
\begin{multline}
\label{eq:Z1abExactDimer2}
\log Z_1^{A_1:A_2} (\alpha,\beta)= \ir \frac{\ell_1 \alpha+\ell_2 \beta}{2}+\int \frac{\dd k}{2 \pi}\mathrm{Re}[h_{1,\alpha}(\cos k)]\min(\ell_1,2 v_k t) +\int \frac{\dd k}{2 \pi}\mathrm{Re}[h_{1,\beta}(\cos k)]\min(\ell_2,2 v_k t)\\
-\frac{1}{2}\int \frac{\dd k}{2 \pi} \mathrm{Re}[h_{1,\alpha}(\cos k)+h_{1,\beta}(\cos k)-h_{1,\alpha-\beta}(\cos k)] (\max(d, 2v_kt)\\ + \max(d+\ell, 2v_kt) - \max(d+\ell_1, 2v_k t)- \max(d+\ell_2, 2v_k t)).
\end{multline}
%
%

\begin{figure}
\begin{center}
\includegraphics[scale=0.31]{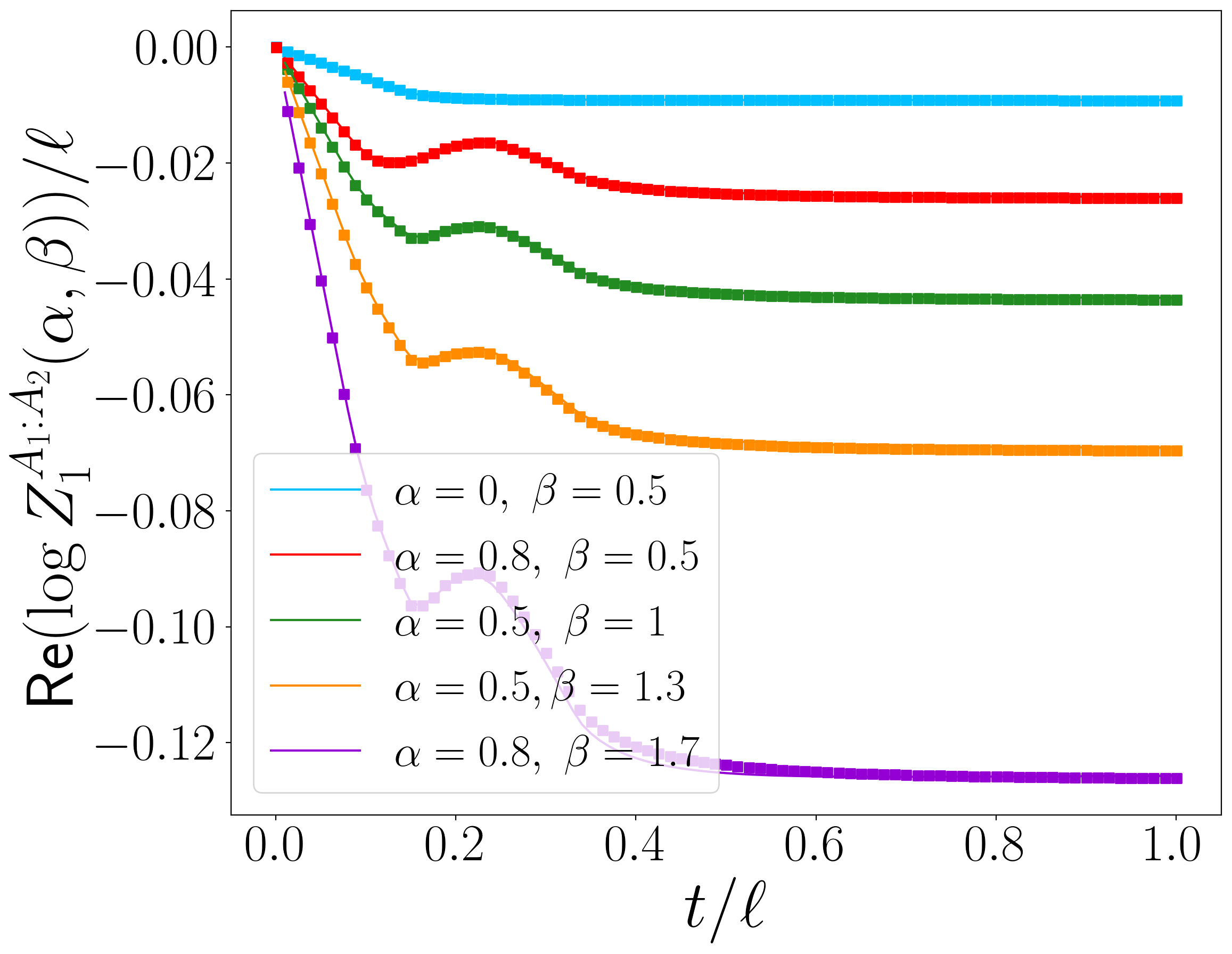}
\end{center}
\caption{Time evolution of $Z_1^{A_1: A_2}(\alpha,\beta)$ after a quench from the dimer state in the tight-binding model \eqref{eq:Hfree} as a function of $t/\ell$ with $\ell_1=100$, $\ell_2=140$ and $d=80$. The analytical prediction of Eq.~\eqref{eq:Z1abExactDimer2} (solid lines) perfectly matches the numerical data (symbols).}
\label{fig:Z1abDimer}
\end{figure}

It is direct to show that $Z_1^{A_1:A_2} (\alpha,\alpha)$ in Eq. \eqref{eq:Z1abExactDimer2} reduces to $Z_1^{A_1\cup A_2} (\alpha)$ given in Eq. \eqref{eq:logZnadExactDimer}, as expected. Moreover, replacing $\cos k$ by $0$ in Eq. \eqref{eq:Z1abExactDimer2} yields the result of Eq. \eqref{eq:Z1abExactNeel} for the quench from the N\'eel state. We compare the prediction of Eq. \eqref{eq:Z1abExactDimer2} with ab-initio calculations in Fig. \ref{fig:Z1abDimer} and find perfect agreement.

To proceed, we expand $Z_1^{A_1:A_2} (\alpha,\beta)$  at quadratic order in $\alpha$ and $\beta$.
We find
\begin{equation}
\label{eq:Z1abQuadrDimer}
\log Z_1^{A_1:A_2} (\alpha,\beta)= \ir \frac{\ell_1 \alpha+\ell_2 \beta}{2} -\frac{\alpha^2}{8} \J^{(1)}_{A_1} - \frac{\beta^2}{8} \J^{(1)}_{A_2}+ \frac{\alpha \beta}{8} (\J^{(1)}_{A_1}+  \J^{(1)}_{A_2}-\J^{(1)}_d),
\end{equation}
and the double Fourier transform is
\begin{equation}
\label{eq:Z1qq1Dimer}
\mathcal{Z}_1^{A_1:A_2}(q_1,q-q_1) = \frac{2}{\pi \sqrt{\J^{(1)}_{A_1}\J^{(1)}_{A_2}-\frac{(\J^{(1)}_m)^2}{4}}}\eE^{-\frac{8}{4\J^{(1)}_{A_1}\J^{(1)}_{A_2}-(\J^{(1)}_m)^2}\Big[\Delta q_1^2 \J^{(1)}_{A_2}+(\Delta q-\Delta q_1)^2 \J^{(1)}_{A_1}+\Delta q_1(\Delta q-\Delta q_1) \J^{(1)}_m\Big]}
\end{equation}
with $\J^{(1)}_m=(\J^{(1)}_{A_1}+  \J^{(1)}_{A_2}-\J^{(1)}_d)$ to lighten the notations, where $\J_{A_{1,2}}^{(n)}$ is simply $\J^{(n)}$ from Eq. \eqref{eq:JnDimer} where $\ell$ is replaced by $\ell_{1,2}$, and $\J_d^{(n)}$ is given in Eq. \eqref{eq:JdnDimer}. We also have $\Delta q=q-\ell/2$ and $\Delta q_1=q_1-\ell_1/2$. These two results are the same as Eqs. \eqref{eq:Z1abQuadrNeel} and \eqref{eq:Z1qq1Neel} where the $\J_s$ integrals are systematically replaced by $\J_s^{(1)}$. In a similar fashion as for the quench from the N\'eel state, the weight $p(q_1,q-q_1)=\mathcal{Z}_1^{A_1:A_2}(q_1,q-q_1)/\mathcal{Z}_1^{A_1 \cup A_2}(q)$ defined in Eq. \eqref{eq:pqq1} satisfies $\sum_{q_1=0}^q p(q_1,q-q_1) =1$. 

To compute the symmetry resolved mutual information, we use  Eq. \eqref{eq:SymResMut} and replace each quantity by its Gaussian approximation, namely Eqs.~\eqref{eq:ZnqGaussDimer}, \eqref{eq:Z1qq1Dimer}, \eqref{eq:S1qDimerApprox} and \eqref{eq:S1qdDimer}. The calculations are similar to those of Sec. \ref{sec:SRMutNeel}. We find
\begin{equation}
\label{eq:IqAsymDimer}
\begin{split}
& I^{A_1 : A_2}_1(q) =S_1^{A_1}+S_1^{A_2}-S_1^{A_1 \cup A_2} - \frac{1}{2} \Big( \log \frac{\J^{(1)}_{A_1}\J^{(1)}_{A_2} \pi}{2 \J^{(1)}_d} \Big)+\frac{1}{2}\Big(\frac{\partial_n(\J_{A_1}^{(n)})|_{n=1}}{\J^{(1)}_{A_1}}+\frac{\partial_n(\J_{A_2}^{(n)})|_{n=1}}{\J^{(1)}_{A_2}}-\frac{\partial_n(\J_{d}^{(n)})|_{n=1}}{\J^{(1)}_{d}}\Big)\\
&-\frac{4\J^{(1)}_{A_1}\J^{(1)}_{A_2}-(\J^{(1)}_m)^2}{8 \J^{(1)}_d}\Big\{\frac{1}{\J^{(1)}_{A_1}}\Big(\frac{\partial_n(\J_{A_1}^{(n)})|_{n=1}}{\J^{(1)}_{A_1}}+1\Big)+\frac{1}{\J^{(1)}_{A_2}}\Big(\frac{\partial_n(\J_{A_2}^{(n)})|_{n=1}}{\J^{(1)}_{A_2}}+1\Big)\Big\}\\
 &-2 \Delta q^2 \Big\{\Big(\frac{-\J^{(1)}_{A_1}+\J^{(1)}_{A_2}-\J^{(1)}_d}{2\J^{(1)}_d}\Big)^2\Big(\frac{\partial_n(\J_{A_1}^{(n)})|_{n=1}}{(\J^{(1)}_{A_1})^2}+\frac{1}{\J^{(1)}_{A_1}}\Big)+\Big(\frac{\J^{(1)}_{A_1}-\J^{(1)}_{A_2}-\J^{(1)}_d}{2\J^{(1)}_d}\Big)^2\Big(\frac{\partial_n(\J_{A_2}^{(n)})|_{n=1}}{(\J^{(1)}_{A_2})^2}+\frac{1}{\J^{(1)}_{A_2}}\Big)\Big\}\\
 &+2 \Delta q^2\Big(\frac{\partial_n(\J_{d}^{(n)})|_{n=1}}{(\J^{(1)}_{d})^2}+\frac{1}{\J^{(1)}_{d}}\Big).
  \end{split}
\end{equation}

\begin{figure}
\begin{center}
\includegraphics[scale=0.35]{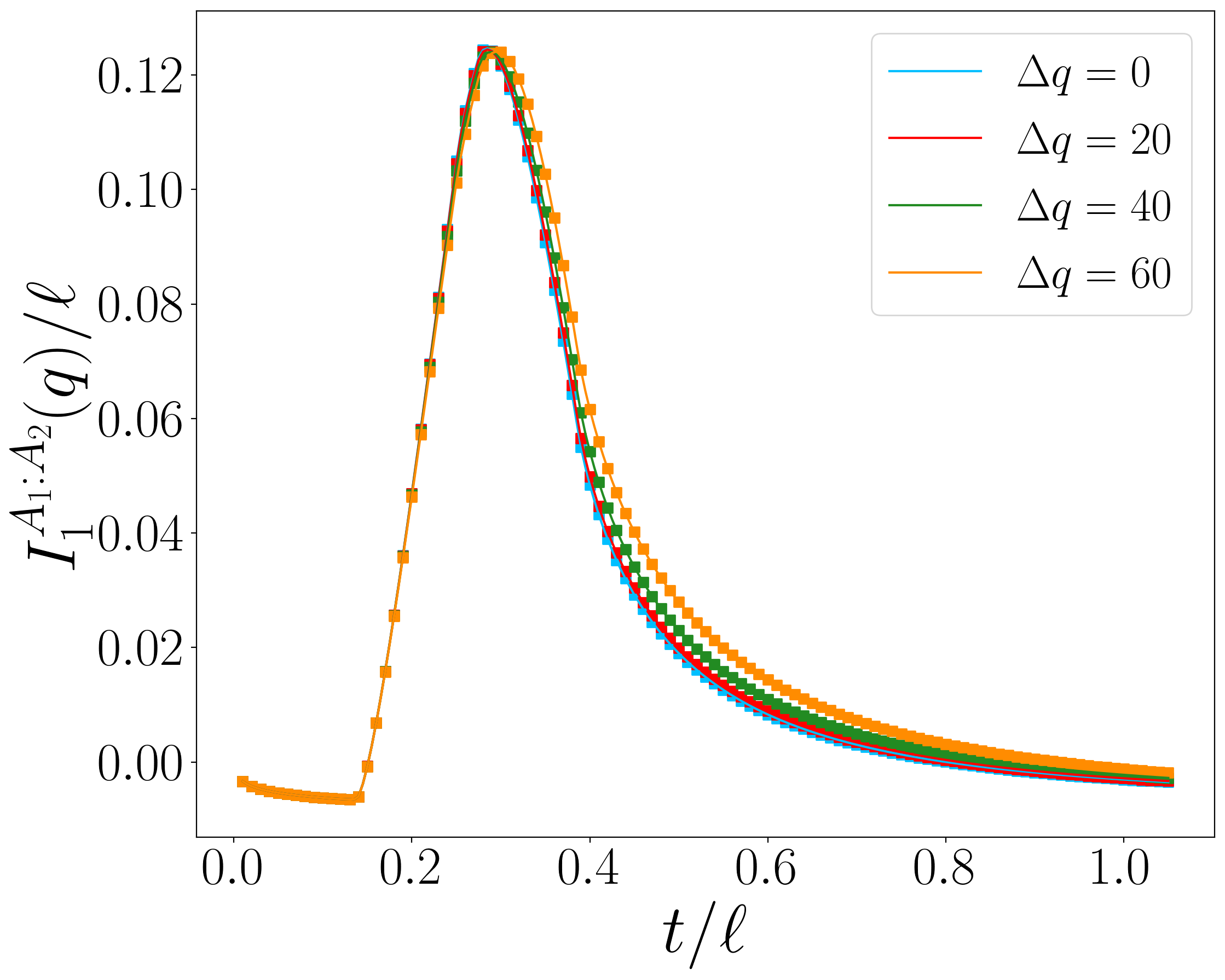}
\end{center}
\caption{Time evolution of $I^{A_1 : A_2}_1(q) $ after a quench from the dimer state in the tight-binding model \eqref{eq:Hfree} as a function of $t/\ell$ with $\ell_1=180$, $\ell_2=220$ and $d=220$ for various values of $\Delta q$. We compare the approximation of Eq. \eqref{eq:IqAsymDimer} (solid lines) with the definition of Eq. \eqref{eq:SymResMut} where the various quantities are replaced by their Gaussian approximations with the first corrections (symbols).}
\label{fig:SRMutAsymDimer}
\end{figure}

We compare this prediction with the definition of Eq. \eqref{eq:SymResMut} where the various quantities are replaced by their Gaussian approximations in Fig. \ref{fig:SRMutAsymDimer}, and find a perfect match. The result of Eq. \eqref{eq:IqAsymDimer} has the same physical interpretation as Eq. \eqref{eq:IqAsymNeel}, namely the first three terms of the right-hand side precisely yield the quasiparticle prediction for the total mutual information, equipartition is broken at order $\Delta q^2/\ell$ for intermediate times and at higher order for $t<d/4$ and $t \to \infty$, since $\J_d^{(n)}= \J^{(n)}_{A_1}+\J^{(n)}_{A_2}$ in those regimes. As for the quench from the N\'eel state, the combination of Eq. \eqref{eq:IqTot} yields the total mutual information, namely
\begin{equation}
    \sum_{q=0}^{\ell} \mathcal{Z}_1^{A_1\cup A_2}(q) I^{A_1 : A_2}_1(q) + S^{A_1,n}+S^{A_2,n}-S^{A_1\cup A_2,n}=I^{A_1:A_2}_1
\end{equation}
where
\begin{equation}
I^{A_1:A_2}_1=-\int \frac{\dd k}{2 \pi}\partial_n h_{n,0}(\cos k)|_{n=1}(\max(d, 2v_kt)+ \max(d+\ell, 2v_kt)- \max(d+\ell_1, 2v_k t)- \max(d+\ell_2, 2v_k t)).
\end{equation}

\section{Quasiparticle picture for the charged moments}
\label{Sect6}

In this section we use the quasiparticle picture to get general results for the dynamics  of the charged moments for the entanglement entropy.
On the one hand, this picture provides a physical explanation of the findings of the previous sections. 
On the other hand, it gives a generalisation valid for any free-fermion (or free-boson) model for arbitrary R\'enyi entropies.
It also suggests a conjecture for the charged entropies for arbitrary interacting models, but many quantitative details are still missing and its validity should eventually be tested 
against numerical simulations. 

\subsection{Dynamics of the entanglement entropies}

\begin{figure}
\begin{center}
\includegraphics[scale=0.65]{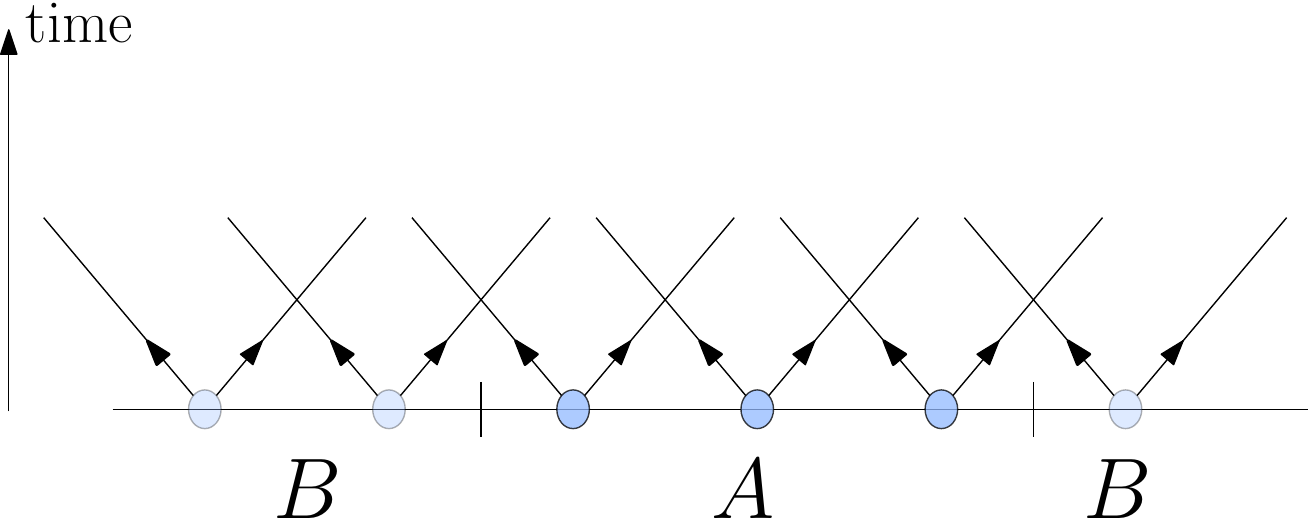}
\end{center}
\caption{Illustration of the quasiparticle picture for the entanglement spreading in a one-dimensional integrable system. The arrows indicate the propagation of the quasiparticles with the largest velocity $v_M$.}
\label{fig:QuasiPFi}
\end{figure}

Let us consider a many-body quantum system prepared in an initial state $|\psi_0\rangle$ with an extensive energy on top of the ground state of the Hamiltonian $H$ 
which governs its time evolution. 
The initial state can  be regarded as a source of quasiparticle excitations, which are assumed to be produced in independent pairs with opposite momenta $k$ and $-k$. 
(The assumption can be weakened in free  models, see e.g. \cite{btc-18,bc-18,bc-20b}, it is instead fundamental for interacting integrable systems \cite{PPV-17}.) 
This situation is illustrated in Fig.~\ref{fig:QuasiPFi}. 
We first suppose that the quasiparticles are all of the same species, so that we can characterise them only by their momentum $k$.
After being produced, each particle moves ballistically with velocity $v(k)$. 
The key assumption of this description is that the quasiparticles emitted from different points are incoherent, while those emitted from the same point are entangled and 
as they move far from each other they spread the entanglement and the correlations through the system. 
In particular, the entanglement of a subsystem $A$ with its complement after a time $t$ is proportional to the total number of entangled pairs shared by the two parts at that time.
Hence, just by counting these pairs, when the subsystem $A$ is an interval of length $\ell$ embedded in an infinite system, the entanglement entropy evolves as \cite{cc-05}
\begin{equation}
    \label{EntropyQuasi}
    S_1(t)= 2 t \int_{2 v(k)t<\ell}\dd k \ v(k) s(k)+ \ell \int_{2 v(k)t>\ell} \dd k \ s(k),
\end{equation}
where the function $s(k)$ depends on the rate of production of the quasiparticles with momentum $\pm k$ and their contribution to the entanglement entropy. 
The velocity $v(k)$ has a maximum value $v_M$, whose existence, for spin chains, is guaranteed by the Lieb-Robinson bound \cite{lr-72}. Hence, the two terms of Eq.~\eqref{EntropyQuasi} give two different regimes as the entropy evolves in time. For times $t\leqslant \ell/(2v_M)$ the domain of integration of the second integral vanishes, and the entropy grows linearly in time. For larger times, the entanglement growth slows down and for $t\gg \ell/2 v_M$ the second integral dominates and the entropy is finally  
extensive in the subsystem length. 

In the presence of different species of independent quasiparticles the result can be generalised  by summing over them \cite{ac-17},
\begin{equation}
    \label{EntropyQuasiSpecies}
     S_1(t)=\sum_m \left[ 2 t \int_{2 v_m(k)t<\ell}\dd k \ v_m(k) s_m(k)+ \ell \int_{2 v_m(k)t>\ell} \dd k \ s_m(k)\right].
\end{equation}
By properly characterising the functions $v_m(k)$ and $s_m(k)$, Eq. \eqref{EntropyQuasiSpecies}  can be applied to all integrable systems \cite{ac-17,ac-18,c-20}.
For what follows, it is important to stress that while the generalisation to the R\'enyi entropies is very easy for free models (in Eq. \eqref{EntropyQuasi} it is enough to replace 
$s(k)$ with an appropriate $s_n(k)$, see \cite{ac-17b} and below), it is still an open problem for interacting integrable models \cite{ac-17b,ac-17c,mac-18,clsv-19,kb-21}. 

The quasiparticle description can be used to derive analytical predictions for free fermions \cite{ac-17}. 
The main idea of Ref. \cite{ac-17} is that the kernel $s(k)$ in Eq. \eqref{EntropyQuasi} can be read off from the limit $t\to\infty$, 
where the density of entanglement entropy equals the density of the thermodynamic entropy in the stationary state, leading to 
\begin{equation}
 s(k) =\frac{1}{2 \pi} (-n_k \log n_k -(1-n_k)\log(1-n_k))
 \label{sk}
\end{equation}
and so
\begin{equation}
    \label{EvEntFree}
S_1 =\int \frac{\dd k}{2 \pi} (-n_k \log n_k -(1-n_k)\log(1-n_k))\min(\ell,2v_k t).
\end{equation}
Here $n_k$ is the probability of occupation of the mode $k$ in the steady state. 
For simplicity, we recast the sum of the two integrals as the integral where the function $\min(\ell,2 v_kt)$ is included in the integrand.
As anticipated, for free models, Eq. \eqref{EntropyQuasi} also holds for the R\'enyi entropy with the replacement \cite{ac-17b}
\begin{equation}
   s(k)\to s_n(k)=\frac{1}{2 \pi}\frac{1}{1-n} \log[n_k^n+(1-n_k)^n]\,.
\end{equation}
At this point, it is very natural to generalise the quasiparticle result to charged moments with the natural conjecture  
\begin{equation}
    \label{EvEntFree2}
\log Z_n(\alpha)   = \ir \alpha \langle Q_A\rangle  + \int \frac{\dd k}{2 \pi} \log[n_k^n \eE^{\ir\alpha}+(1-n_k)^n ]\min(\ell,2 v_k t).
\end{equation}
This conjecture matches the exact results for the two quenches considered here where $n_k=1/2$ (N\'eel case) and $n_k=(1+\cos k)/2$ (dimer case).

Actually, within free fermion models, the validity of the  quasiparticle result \eqref{EvEntFree2} can be inferred from the following argument.
For the R\'enyi entropy, the quasiparticles conjecture is 
\begin{equation}
    \label{EvRenFree}
    S_n(t)=\frac{1}{1-n} \int\frac{\dd k}{2 \pi}  h_{n,0}(x_k) \min(\ell,2 v_k t) =\frac{1}{1-n} \sum_{j=0}^{\infty} c_{n,0}(2j) \int\frac{\dd k}{2 \pi}  x_k^{2j} \min(\ell,2 v_k t) ,
\end{equation}
where $x_k$ is related to the probability of occupation of the mode $k$ by $n_k=(1+x_k)/2$ and $h_{n,0}(x)$ is given in Eq. \eqref{eq:hna} and has been expanded in $x$ (as in the rightmost side of Eq. \eqref{eq:hna}). In terms of the function $s_n(k)$, it is just $h_{n,0}(2n_k-1)=2 \pi (1-n) s_n(k)$. Similarly, the starting formula for the R\'enyi entropy in terms of the correlation matrix can be expanded as
\begin{equation}
    \label{RenyiFreeFerm}
    S_n(t)=\frac{1}{1-n}\sum_{i=1}^{\ell}\log\left[ \left(\frac{1+\nu_i}{2} \right)^n+\left(\frac{1-\nu_i}{2} \right)^n\right]=\frac{1}{1-n}\sum_{j=0}^{\infty} c_{n,0}(2j)\Tr J(t)^{2j}.
\end{equation}
Matching Eqs. \eqref{RenyiFreeFerm} and \eqref{EvRenFree} and properly keeping track of the normalisations (i.e. $\sum_{j=0}^{\infty}c_{n,0}(2j)=h_{n,0}(1)=0$ and  
$\Tr J(t)^0=\ell$), we get
\begin{equation}
    \label{TrJQuasi}
    \Tr J(t)^{2j}= \ell+ \int\frac{\dd k}{2 \pi}  (x_k^{2j}-1)\min(\ell,2 v_k t).
\end{equation}
Plugging Eq. \eqref{TrJQuasi} in Eq. \eqref{eq:chargedZTrK}, we obtain the desired result \eqref{EvEntFree2}. As for the total entropies, we conjecture that in the presence of different species of quasiparticles, the result of Eq. \eqref{EvEntFree2} generalises to a sum over the different species.

\subsection{Dynamics of the mutual information}

\begin{figure}
\begin{center}
\includegraphics[scale=0.65]{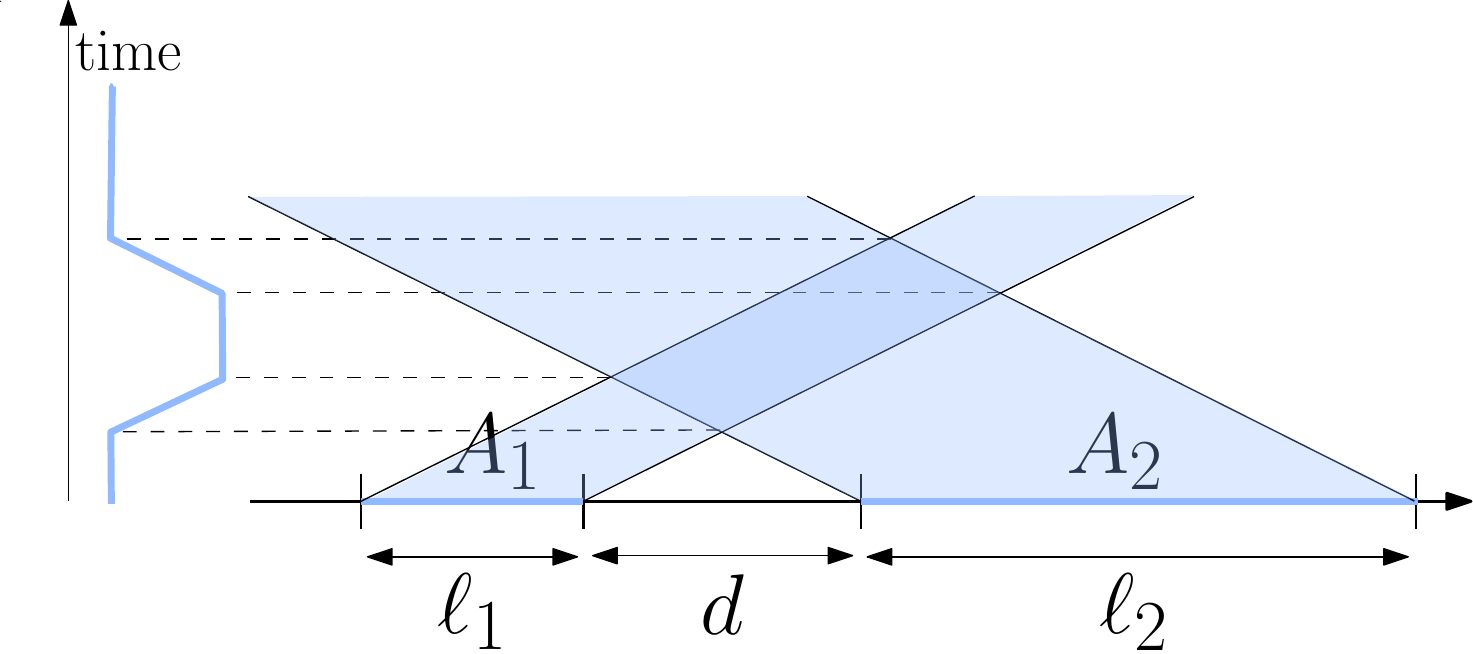}
\end{center}
\caption{Illustration of the entanglement and mutual information dynamics of two disjoint subsystems. On the left we report the shape of the mutual information 
with a single velocity of quasiparticles. In the main plot, the dark blue area counts the quasiparticles shared between the two disjoint sets.  }
\label{fig:QuasiPFiDisjoint}
\end{figure}

Let us consider the case in which the subsystem $A$ of length $\ell$ is formed by two disjoint intervals $A_1$ and $A_2$ of respective lengths $\ell_1$ and $\ell_2$ that are separated by a distance $d$. Without loss of generality we consider $\ell_1 \leqslant \ell_2$. 
Since the total mutual information is a combination of entanglement entropies for which the quasiparticle picture holds, it is natural that it must be related to the  
number of entangled pairs that, once emitted from the same point, are shared by the two subsystem at subsequent times.

This counting of quasiparticles is straightforward, but we report it here for completeness.   
Let us first assume that the quasiparticles possess a single velocity $v$. 
It follows that for short times $t\leqslant d/(2 v)$ there are no entangled quasiparticles shared by the two intervals and the mutual information is zero. It then increases linearly for $d/(2 v) \leqslant t \leqslant (d+\ell_1)/(2 v)$. There is a plateau for $(d+\ell_1)/(2 v) \leqslant t \leqslant (d+\ell_2)/(2 v)$, after which it decreases linearly in time up to $t=(d+\ell)/2v$. There are no longer any quasiparticles shared between the subsystems for larger times, so that the mutual information vanishes for $t\geqslant(d+\ell)/2v$. We illustrate this behaviour in Fig. \ref{fig:QuasiPFiDisjoint}. Accordingly, the resulting dynamics of the mutual information is proportional to $[\max(d,2vt)+ \max(d+\ell,2vt)- \max(d+\ell_1,2vt)-\max(d+\ell_2,2vt)]$\cite{ctc-14,ac-19}.

In the presence of a non-trivial dispersion (and hence different velocities $v(k)$), it is enough to integrate over $k$ to get  \cite{ac-17, ac-18}
\begin{equation}
    \label{MutualQuasi}
    I_1^{A_1:A_2}= \int \dd k \ s(k) [\max(d,2v(k)t)+ \max(d+\ell,2v(k)t)- \max(d+\ell_1,2v(k)t)-\max(d+\ell_2,2v(k)t)],
\end{equation}
with $s(k)$ in Eq. \eqref{sk} for free fermions.  

For the charged moment $\log Z_n^{A_1 \cup A_2}(\alpha)$ it is straightforward to adapt the argument of the previous subsection to get
\begin{multline}
    \label{eq:logZnaDisjointQPP}
  \log Z_n^{A_1 \cup A_2}(\alpha) = \ir \alpha \langle Q_A\rangle+\int \frac{\dd k}{2 \pi}\mathrm{Re}[h_{n,\alpha}(x_k)][\min(\ell_1,2 v_k t)+\min(\ell_2,2 v_k t)]\\
  - \int \frac{\dd k}{2 \pi}\mathrm{Re}[h_{n,\alpha}(x_k)][\max(d, 2v_kt)+ \max(d+\ell, 2v_kt)- \max(d+\ell_1, 2v_k t)- \max(d+\ell_2, 2v_k t)].
\end{multline}
This equation precisely matches our analytical results from Eqs. \eqref{eq:ZnadExactNeel} with $x_k=0$ for the quench from the N\'eel state, and \eqref{eq:logZnadExactDimer} with $x_k=\cos k$ in the case of the quench from the dimer state. 

To determine the symmetry resolved mutual information, we also need the moment $Z_1(\alpha,\beta)$ defined in Eq. \eqref{eq:Z1ab}. Our results of Eqs. \eqref{eq:Z1abExactNeel} and \eqref{eq:Z1abExactDimer2} suggest the general quasiparticle expression
\begin{multline}
\label{eq:Z1abQPP}
\log Z_1^{A_1:A_2} (\alpha,\beta)=  \ir \alpha \langle Q_{A_1}\rangle+ \ir \beta \langle Q_{A_2}\rangle+\int \frac{\dd k}{2 \pi}\mathrm{Re}[h_{1,\alpha}(x_k)]\min(\ell_1,2 v_k t) +\int \frac{\dd k}{2 \pi}\mathrm{Re}[h_{1,\beta}(x_k)]\min(\ell_2,2 v_k t)\\
-\frac{1}{2}\int \frac{\dd k}{2 \pi} \mathrm{Re}[h_{1,\alpha}(x_k)+h_{1,\beta}(x_k)-h_{1,\alpha-\beta}(x_k)]   (\max(d, 2v_kt)\\ + \max(d+\ell, 2v_kt) - \max(d+\ell_1, 2v_k t)- \max(d+\ell_2, 2v_k t)).
\end{multline}
 
As in the case of a single interval, we conjecture that in the presence of different species of quasiparticles, the results of Eqs. \eqref{eq:logZnaDisjointQPP} and \eqref{eq:Z1abQPP} generalise to a sum over the different species.


\section{Conclusions}
\label{Sect7}

In this manuscript, we study the symmetry resolved entanglement after a quantum quench in the one-dimensional tight binding model.
We start from the calculation of the charged moments $Z_n(\alpha)$ (cf. Eq. \eqref{ChargedMom}) that have been exactly determined by adapting the multidimensional
stationary phase calculation of Ref. \cite{fc-08}, see \cite{pb-22}. 
Our results for $Z_n(\alpha)$ are also valid when the subsystem $A$ consists of two (or more) disjoint intervals.
From the charged moments, the symmetry resolved entropy is obtained by Fourier transform and saddle point approximation in the charge flux.  
The structure of charged moments leads to two main physical results. 
First, the symmetry resolved entanglement with charge $\Delta q=q-\langle Q_A\rangle$ starts evolving only after a (calculable) delay time  proportional to $|\Delta q|$.
Second, for small $|\Delta q|$ there is an effective equipartition of entanglement broken at order $\Delta q^2/\ell$.
We also introduced a symmetry resolved mutual information \eqref{eq:SymResMut} that requires the computation of a charged moment with two independent 
flux insertions, see Eq. \eqref{eq:Z1ab}. We worked out exact predictions for this more complicated charged moment for two quenches in our models. 
We also found that the number entropy grows logarithmically with time before saturating to a value proportional to the logarithm of the subsystem size.

Our results for free fermions naturally lead to a conjecture based on the quasiparticle picture for $Z_n(\alpha)$ that can generically be written as 
\begin{equation}
\log Z_n(\alpha)   = \ir \alpha \langle Q_A\rangle  + \sum_m \int \frac{\dd k}{2 \pi} s_{n,\alpha}^m(k) \min(\ell,2 v_m(k) t)
\label{conj2}
\end{equation}
where the sum is over the independent quasiparticles' species (where the velocities $v_m(k)$ also depend on the state \cite{bonnes-2014}). 
To give predictive power to this conjecture we should still devise a way to fix the kernel $s_{n,\alpha}^m(k)$, but how to do this is still not know for
an interacting model, as already discussed for the R\'enyi entropy in Refs. \cite{ac-17b,ac-17c,mac-18,clsv-19,kb-21}. 
The solution is known only in the neighbourhood of $n=1$ where the kernel is the Yang-Yang entropy \cite{ac-17}.   
It would already be interesting to generalise this result at $n=1$ and small $\alpha\neq0$, that would be enough for the saddle-point approximation necessary to get the 
symmetry resolution. 
Independently of the details of the function $s_{n,\alpha}^m(k)$, the form of the conjecture  \eqref{conj2} is sufficient to deduce the two main 
physical results, i.e. existence of the delay time and equipartition for small $|\Delta q|$ for a generic integrable model. 
Moreover, the logarithmic growth of the number entropy follows from the behaviour of $Z_1(\alpha)$ close to $\alpha=0$ and so follows generically from ~Eq.~\eqref{conj2}.

Similarly to the charged moments, we conjectured that the doubly-charged moments $Z_1(\alpha,\beta)$ in Eq. \eqref{eq:Z1ab} needed to investigate the symmetry resolved mutual information can be described in the framework of the quasiparticle picture, see Eq. \eqref{eq:Z1abQPP}.
%

We conclude this manuscript by speculating on some generalisations of our results. 
A first question is whether symmetry resolved entropies can be obtained in inhomogeneous settings, as done for the total entropy in 
Refs. \cite{alba-inh,BFPC18,ABF19,a-19,dsvc-17,rcdd-19}.
Then, it is natural to wonder if some of the techniques developed for non-integrable 
models \cite{NRVH:17,nvh-18,zn-20,BKP:entropy,GoLa19,bc-20,PBCP20,mac-20,cdc-18,fcdc-19,rpv-19}
apply to the symmetry resolved quantities and whether our main physical findings, time delay and equipartition, are robust.

\section*{Acknowledgements}

All authors acknowledge support from ERC under Consolidator grant number 771536 (NEMO). GP thanks SISSA for hospitality and the Gustave B\"oel-Sofina Fellowships for financing his stay in Trieste. He is also supported by the Aspirant Fellowship FC 23367 from the F.R.S-FNRS and acknowledges support from the EOS contract O013018F.

\end{document}